\documentclass[12pt]{article}
\RequirePackage[authoryear]{natbib}

\RequirePackage{amsthm,amsmath,amsfonts,amssymb}
\RequirePackage[colorlinks,citecolor=blue,urlcolor=blue]{hyperref}
\RequirePackage{graphicx}

\usepackage{float}
\usepackage{enumitem}
\usepackage{url}
\usepackage{bbm}
\usepackage[nottoc,numbib]{tocbibind}
\usepackage{booktabs,caption,subcaption}
\usepackage[flushleft]{threeparttable}
\usepackage{multirow}

\usepackage[pdftex,dvipsnames]{xcolor}
\usepackage{comment}

\usepackage[margin=1in]{geometry}

\newcommand{\bA}{\boldsymbol{A}}

\newcommand{\bc}{\boldsymbol{c}}
\newcommand{\bC}{\boldsymbol{C}}

\newcommand{\bD}{\boldsymbol{D}}
\newcommand{\mD}{\mathcal{D}}
\newcommand{\E}{\mathbb{E}}

\newcommand{\bG}{\boldsymbol{G}}

\newcommand{\bI}{\boldsymbol{I}}

\newcommand{\bj}{\boldsymbol{j}}
\newcommand{\bJ}{\boldsymbol{J}}

\newcommand{\bM}{\boldsymbol{M}}
\newcommand{\mN}{\mathcal{N}}

\newcommand{\bP}{\boldsymbol{P}}
\newcommand{\bR}{\boldsymbol{R}}

\newcommand{\bs}{\boldsymbol{s}}
\newcommand{\bS}{\boldsymbol{S}}
\newcommand{\mS}{\mathcal{S}}
\newcommand{\mT}{\mathcal{T}}

\newcommand{\bU}{\boldsymbol{U}}
\newcommand{\V}{\mathbb{V}}
\newcommand{\bv}{\boldsymbol{v}}
\newcommand{\bV}{\boldsymbol{V}}

\newcommand{\bW}{\boldsymbol{W}}
\newcommand{\mW}{\mathcal{W}}

\newcommand{\bX}{\boldsymbol{X}}

\newcommand{\by}{\boldsymbol{y}}
\newcommand{\bY}{\boldsymbol{Y}}

\newcommand{\bz}{\boldsymbol{z}}
\newcommand{\bZ}{\boldsymbol{Z}}
\newcommand{\bzero}{\boldsymbol{0}}

\newcommand{\bbeta}{\boldsymbol{\beta}}

\newcommand{\bdelta}{\boldsymbol{\delta}}
\newcommand{\bDelta}{\boldsymbol{\Delta}}

\newcommand{\bKappa}{\boldsymbol{K}}

\newcommand{\bmu}{\boldsymbol{\mu}}

\newcommand{\bPsi}{\boldsymbol{\Psi}}
\newcommand{\bpsi}{\boldsymbol{\psi}}

\newcommand{\kl}{\text{\sc kl}}
\newcommand{\expit}{\mathrm{expit}}

\newtheorem{theorem}{Theorem}

\newtheorem{lemma}{Lemma}

\theoremstyle{definition}

\allowdisplaybreaks

\begin{document}

\title{\bf Turning the information-sharing dial: efficient inference from different data sources}
 \author{Emily C. Hector and Ryan Martin\hspace{.2cm}\\
  Department of Statistics, North Carolina State University}
  \date{}
 \maketitle
 
 \bigskip
\begin{abstract}
A fundamental aspect of statistics is the integration of data from different sources.  Classically, Fisher and others were focused on {\em how} to integrate homogeneous (or only mildly heterogeneous) sets of data.  More recently, as data are becoming more accessible, the question of {\em if} data sets from different sources should be integrated is becoming more relevant.  The current literature treats this as a question with only two answers: integrate or don't.  Here we take a different approach, motivated by information-sharing principles coming from the shrinkage estimation literature.  In particular, we deviate from the do/don't perspective and propose a dial parameter that controls the {\em extent} to which two data sources are integrated.  How far this dial parameter should be turned is shown to depend, for example, on the informativeness of the different data sources as measured by Fisher information.  In the context of generalized linear models, this more nuanced data integration framework leads to relatively simple parameter estimates and valid tests/confidence intervals.  Moreover, we demonstrate both theoretically and empirically that setting the dial parameter according to our recommendation leads to more efficient estimation compared to other binary data integration schemes. 
\end{abstract}

\noindent%
\vfill

\section{Introduction}
\label{s:intro}

Statistics aims to glean insights about a population by aggregating samples of individual observations, so {\em data integration} is at the core of the subject.  
In recent years, a keen interest in combining data---or statistical inferences---from multiple studies has emerged in both statistical and domain science research \citep{chatterjee2016constrained, jordan2018communication, yang2019combining, Michael-Thornton-Xie-Tian, lin2019race, chen2019distributed, Tang-Zhou-Song-2020,Cahoon-Martin}. While each sample is typically believed to be a collection of observations from one population, there is no reason to believe another sample would also come from the same population. This difficulty has given rise to a large body of literature for performing data integration in the presence of heterogeneity; see \cite{Claggett-Xie-Tian, Wang-Kim-Yang, duan2019heterogeneity, cai2019individual, Hector-Song-JMLR} and references therein for examples. These methods take an all-or-nothing approach to data integration: either the two sources of data can be integrated or not. This binary view of what can, or even should, be integrated is at best impractical and at worst useless when confronted with the messy realities of real data. Indeed, two samples are often similar enough that inferences can be improved with a joint analysis despite differences in the underlying populations.  It does us statistical harm to think in these binary terms when data sets come from similar but not identical populations: assuming data integration is wholly invalid leads to reduced efficiency whereas assuming it is wholly valid can unknowingly produce biased estimates \citep{Higgins-Thompson}. It is thus not practical to think of two data sets as being either entirely from the same population or not. This begs the obvious but non-trivial (multi-part) question: What and how much can be learned from a sample given additional samples from {\em similar} populations, and how to carry out this learning process?  

Towards answering this question, we consider a simple setup with two data sets: one for which a generalized linear model has already been fit, and another for which we wish to fit the same generalized linear model. (The case with more than two data sets is discussed in Section~\ref{ss:setup} below.) The natural question is if inference based on the second data set can be improved in some sense by incorporating results from the analysis of the first data set.  This problem has been known under many different names, of which ``transfer learning'' is the most recently popular \citep{Pan-Yang, Zhuang-etal}. The predominant transfer learning approach uses freezing and unfreezing of layers in a neural network to improve prediction in a target data set given a single source data set \citep{Tan-etal, Weiss-etal}. The prevailing insight into why deep transfer learning performs well in practice is that neural networks learn general data set features in the first few layers that are not specific to any task, while deeper layers are task specific \citep{Yosinski-etal, Dube-etal, Raghu-etal}. This insight, however, does not give any intuition into or quantification of the similarity of the source and target data sets, primarily because of a lack of model interpretability. More importantly, this approach fails to provide the uncertainty quantification required for statistical inference. Similarly, transfer learning for high-dimensional regression in generalized linear models aims to improve predictions and often does not provide valid inference \citep{Li-Cai-Li-2020, Li-Cai-Li-2022, Tian-Feng}. Another viewpoint casts this problem in an online inference framework \citep{Schifano-etal,Toulis-Airold,Luo-Song-2021} that assumes the true value of the parameter of interest in two sequentially considered data sets is the same.   

In contrast, we aim to tailor the inference in one data set according to the amount of relevant information in the other data set. Our goal is to provide a nuanced approach to data integration, one in which the integration tunes itself according to the amount of information available. In this sense, we do not view data sets as being from the same population or not, nor data integration as being valid or invalid, and we especially do not aim to provide a final inference that binarizes data integration into a valid or invalid approach. Rather, our perspective is that there is a continuum of more to less useful integration of information from which to choose, where we use the term ``useful'' to mean that we are minimizing bias and variance of model parameter estimates. 

This new and unusual perspective motivates our definition of {\em information-driven} regularization based on a certain Kullback--Leibler divergence penalty. This penalty term allows us to precisely quantify what and how much information is borrowed from the first data set when inferring parameters in the second data set. We introduce a so-called ``dial'' parameter that controls how much information is borrowed from the first data set to improve inference in the second data set. We prove that there exists a range of values of the dial parameter such that our proposed estimator has reduced mean squared error over the maximum likelihood estimator that only considers the second data set. This striking result indicates that there is always a benefit to integrating the data sets, but that the amount of information integrated depends entirely on the similarity between the source and target. Based on this result, we propose a choice of the dial parameter that calibrates the bias-variance trade-off to minimize the mean squared error, and show how to construct confidence intervals with our biased parameter estimator. Finally, we demonstrate empirically the superiority of our approach over alternative approaches, and show empirically that our estimator is robust to differences between the source and target data sets. Due to its disjointed nature, relevant literature is discussed throughout.

We describe the problem set-up and proposed information-driven shrinkage approach in Section \ref{s:formulation}. Section \ref{s:examples} gives the specific form of our estimator in the linear and logistic regression models. Theoretical properties of our estimator are established in Section \ref{s:theory}, demonstrating its efficiency compared to the maximum likelihood estimator. We investigate the empirical performance of our estimator through simulations in Section \ref{s:empirical} and a data analysis in Section \ref{s:data-analysis}. An R package implementing our methods is available at \verb|github.com/ehector/ISEDI|.

\section{A continuum of information integration}
\label{s:formulation}

\subsection{Problem setup}
\label{ss:setup}

Consider two populations, which we will denote as Populations~1 and 2.  Units are randomly sampled from the two populations and the features $\bX_{ij} \in \mathbb{R}^p$ and $y_{ij} \in \mathbb{R}$ are measured on unit $i$ from Population~$j$, with $i=1,\ldots,n_j$ and $j=1,2$.  Note that, while the values of these measurements will vary across $i$ and $j$, the features being measured are the same across the two populations.  Write $\mD_j = \{(\bX_{ij}, y_{ij}): i=1,\ldots,n_j\}$ for the data set sampled from Population~$j$, for $j=1,2$.  Note that we have not assumed any relationship between the two populations, only that we have access to independent samples consisting of measurements of the same features in the two populations. The reader can keep in mind the prototypical example where a well designed observational study or clinical trial has collected a set of high quality target data $\mD_2$ and the investigator has at his/her disposal a set of source data $\mD_1$ of unknown source and quality to use for improved inference.

While our focus is on the case of one source data set, our information-sharing method described below can be applied more generally.  Indeed, if an investigator has $M$-many source data sets from $M$ populations measuring the same outcome and features, as in Section~\ref{s:data-analysis} below, then they might consider creating a single source data set by concatenating these $M$ data sets.  The proposed information-sharing method can then be applied to the concatenated source data set.  This of course has advantages and disadvantages.  On the one hand, if all the data sets---source and target---are mostly homogeneous, then our proposed information-sharing leads to a substantial efficiency gain through concatenation; similarly, if the data sets are mostly heterogeneous, then there is effectively no risk since our proposed information-sharing procedure will down-weight the source data set and inference will rely mostly on the target.  On the other hand, if there are groups of source data sets that are homogeneous within and heterogeneous across, then the picture is far less clear: the concatenated source data set lacks some of the nuance of the individual source data sets which, depending on the circumstances, could improve or worsen the efficiency of the inference in the target data.  We investigate this potential mix of homogeneous and heterogeneous source data sets and make general recommendations in Appendix \ref{a:multi-source}, but leave it up to the individual investigator to determine if concatenation is justified in their particular application. Alternatives to concatenation include data-driven methods for detecting heterogeneous source data sets \citep{Li-Cai-Li-2022, Tian-Feng}, which are discussed in Section \ref{ss:t-GLM} in the context of ``negative transfer.''

Since the features measured in data sets $\mD_{1}$ and $\mD_2$ are the same, it makes sense to consider fitting identical generalized linear models to the two data sets.  These models assume the conditional probability density/mass function for $y_{ij}$, given $\bX_{ij}$, is of the form 
\begin{align}
\label{e:exp-family}
f_j(y_{ij}; \bbeta_j, \gamma_j) = c(y_{ij}; \gamma_j) \, \exp\bigl[ d(\gamma_j)^{-1} \{ y_{ij} \bX^\top_{ij} \bbeta_j - b(\bX^\top_{ij}\bbeta_j)\} \bigr],
\end{align}
$i=1,\ldots,n_j$, $j=1, 2$, where $b$, $c$, and $d$ are known functions, $\bbeta_j$ is the quantity of primary interest, and $\gamma_j$ is a nuisance parameter that will not receive much attention in what follows. Since inference of $\gamma_j$ is not of interest and we can appropriately estimate $\gamma_j$ based on $\mD_j$, we do not concern ourselves with how integration affects estimates of $\gamma_j$ and we assume $\gamma_{1} \neq \gamma_{2}$. The conditional distribution's dependence on $\bX_{ij}$ is implicit in the  ``$j$'' subscript on $f_j$.  We will not be concerned with the marginal distribution of $\bX_{ij}$ so, as is customary, we assume throughout that this marginal distribution does not depend on $(\bbeta_j, \gamma_j$), and there is no need for notation to express this marginal. We assume that $\bX_j = (\bX_{1j}, \ldots, \bX_{n_jj}) \in \mathbb{R}^{p \times n_j}$, $j=1, 2$, is full rank.

Of course, the two data sets can be treated separately and, for example, the standard likelihood-based inference can be carried out.  In particular, the maximum likelihood estimator (MLE) can be obtained as 
\[ (\widehat{\bbeta}_j, \widehat{\gamma}_j) = \arg\max_{\bbeta_j, \gamma_j} f_j^{(n_j)}(\by_j; \bbeta_j, \gamma_j), \quad j=1,2, \]
where $f_j^{(n_j)}(\by_j; \bbeta_j, \gamma_j) = \prod_{i=1}^{n_j} f_j(y_{ij}; \bbeta_j, \gamma_j)$ is the joint density/likelihood function based on data set $\mD_j$, $j=1,2$.  The MLEs, together with the corresponding observed Fisher information matrices, can be used to construct approximately valid tests and confidence regions for the respective $\bbeta_j$ parameters. This would be an appropriate course of action if the two data sets in question had nothing in common.  

Here, however, with lots in common between the two data sets, we have a different situation in mind.  We imagine that the analysis of data set $\mD_{1}$ has been carried out first, resulting in the MLE $\widehat{\bbeta}_{1}$, among other things.  Then the question of interest is whether the analysis of data set $\mD_{2}$ ought to depend in some way on the results of $\mD_{1}$'s analysis and, if so, how. While the problem setup is similar to the domain adaptation problem frequently seen in binary classification \citep{Cai-Wei,Reeve-Cannings-Samworth}, we emphasize that our interest lies in improving efficiency of inference of $\bbeta_{2}$. More specifically, can the inference of $\bbeta_{2}$ based on $\mD_{2}$ be improved through the incorporation of the results from $\mD_{1}$'s analysis?  

\subsection{Information-driven regularization}
\label{ss:proposal}

Our stated goal is to improve inference of $\bbeta_{2}$ in $\mD_{2}$ given the analysis of $\mD_{1}$. Further inspection of equation \eqref{e:exp-family} reveals that the primary difference between $f_{1}$ and $f_{2}$ is in the difference between $\widehat{\bbeta}_{1}$ and $\bbeta_{2}$. Thus, at first glance, the similarity between $\mD_{1}$ and $\mD_{2}$ is primarily driven by the closeness of $\widehat{\bbeta}_{1}$ and $\bbeta_{2}$. Intuitively, if $\widehat{\bbeta}_{1}$ and $\bbeta_{2}$ are close, say $\| \widehat{\bbeta}_{1} - \bbeta_{2} \|^2_{2} $ is small, then some gain in the inference of $\bbeta_{2}$ can be expected if the inference in $\mD_{1}$ is taken into account. This rationale motivates a potential objective to maximize $f^{(n_{2})}_{2}(\by_{2}; \bbeta_{2}, \gamma_j ) $ under the constraint $\| \bbeta_{2} - \widehat{\bbeta}_{1} \|^2 < c$ for some constant $c$. The choice of $c$ then reflects one's belief in the similarity or dissimilarity between $\mD_{1}$ and $\mD_{2}$, and data-driven selection of $c$ relies on the distance $\| \bbeta_{2} - \widehat{\bbeta}_{1} \|^2$. 

\begin{figure}[t]
\centering
\includegraphics[width=0.7\textwidth]{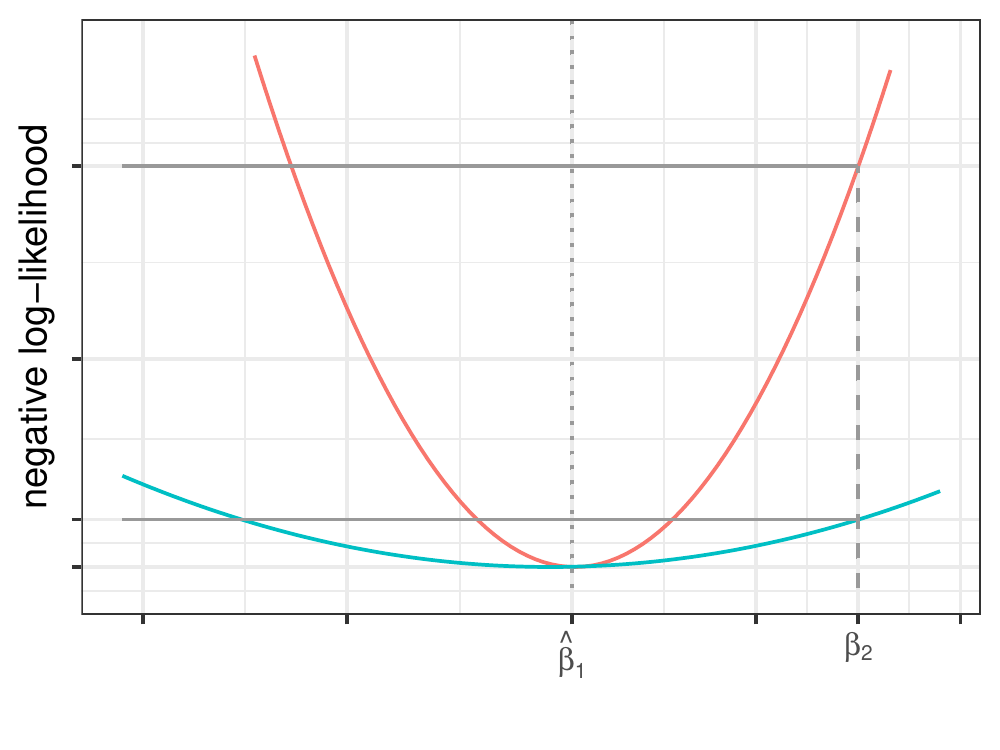}
\caption{An illustration of the likelihoods for more (red) and less (blue) informative data. \label{f:ll-motivation}}
\end{figure}

We must ask ourselves, however, if closeness between $\bbeta_{2}$ and $\widehat{\bbeta}_{1}$ is sufficient for us to integrate information from $\mD_{2}$ into estimation of $\bbeta_{2}$. For example, would we choose to constrain $\bbeta_{2}$ to be close to $\widehat{\bbeta}_{1}$ if $n_{1}$ were small? How about if $\bX_{1} \bX^\top_{1}$ is small, reflecting uninformative features and resulting in a large variance of $\widehat{\bbeta}_{1}$? These intuitive notions of what we consider to be {\em informative} highlight a gap in our argument so far: elements that are ``close'' in the parameter space may not be close in an information-theoretic sense, and vice-versa. This is best visualized by Figure~\ref{f:ll-motivation}, which plots two negative log-likelihood functions for $\bbeta_{1}$, one being based on a more informative data set than the other.  The true $\bbeta_{2}$ is also displayed there and, as expected, is different from the MLE $\widehat\bbeta_{1}$ based on $\mD_{1}$; for visualization purposes, we have arranged for $\widehat\bbeta_{1}$ to be the same for both the more and less informative data sets.  The plot reveals that the more informative data can better distinguish the two points $\bbeta_{2}$ and $\widehat\bbeta_{1}$, in terms of quality of fit, than can the less informative data---so it is not just the distance between $\bbeta_{2}$ and $\widehat\bbeta_{1}$ that matters! Intuitively, estimation of $\bbeta_{2}$ should account not only for its distance to $\widehat{\bbeta}_{1}$ but also the ``sharpness'' of the likelihood, or the amount of information in $\mD_{1}$. 

This motivates our proposal of a distance-driven regularization that takes both the ambient geometry of the parameter space and aspects of the statistical model into consideration.  That is, we propose a regularization based on a sort of ``statistical distance.'' The most common such distance, closely related to the Fisher information and the associated information geometry \citep{Amari-Nagaoka, Nielsen}, is the {\em Kullback--Leibler} (KL) divergence which, in our case, is given by 
\begin{align*}
&\kl_{n_{1}}(\bbeta; \widehat\bbeta_{1}, \gamma_{1}) = \int \log \frac{f_{1}^{(n_{1})}(\by; \widehat\bbeta_{1}, \gamma_{1})}{f_{1}^{(n_{1})}(\by; \bbeta, \gamma_{1})} \, f_{1}^{(n_{1})}(\by; \widehat\bbeta_{1}, \gamma_{1}) \, d\by \\
&\qquad= \frac{1}{d(\gamma_{1})} \sum \limits_{i=1}^{n_{1}} \bigl\{ b^\prime(\bX^\top_{i1} \widehat{\bbeta}_{1}) (\bX^\top_{i1} \widehat{\bbeta}_{1}-\bX^\top_{i1}\bbeta) + b(\bX^\top_{i1}\bbeta) - b(\bX^\top_{i1} \widehat{\bbeta}_{1}) \bigr\},
\end{align*}
where $b'(\bX_{i1}^\top \bbeta) = \E_{\bbeta, \gamma_{1}}(y_{i1} \mid \bX_{i1})$, a standard property satisfied by all natural exponential families.  Our proposal, then, is to use the aforementioned KL divergence to tie the estimates based on the two data sets together, i.e., to produce an estimator that solves the following constrained optimization problem:
\begin{equation}
\label{e:orig-opt}
\text{maximize $n_{2}^{-1} \log f_{2}^{(n_{2})}(\by_{2}; \bbeta, \gamma_{2})$ \; subject to \; $n_{1}^{-1} \kl_{n_{1}}(\bbeta; \widehat\bbeta_{1}, \gamma_{1}) \leq \epsilon$}, 
\end{equation}
where $\epsilon \geq 0$ is a constant to be specified by the user. The KL divergence measures the entropy of $f^{(n_{1})}_{1}(\by; \bbeta, \gamma_{1})$ relative to $f^{(n_{1})}_{1}(\by; \widehat{\bbeta}_{1}, \gamma_{1})$. As the Hessian of the KL divergence, the Fisher information describes the local shape of the KL divergence and quantifies its ability to discriminate between $f^{(n_{1})}_{1}(\by; \bbeta, \gamma_{1})$ and $f^{(n_{1})}_{1}(\by; \widehat{\bbeta}_{1}, \gamma_{1})$.

A key difference between what is being proposed here and other regularization strategies in the literature is that our penalty term is ``data-dependent'' or, perhaps more precisely, relative to $\mD_{1}$.  That is, it takes the {\em information} in $\mD_{1}$ into account when estimating $\bbeta_{2}$, hence the name information-driven regularization.  Of course, the extent of the information-driven regularization, i.e., how much information is shared between $\mD_{1}$ and $\mD_{2}$, is determined by $\epsilon$, so this choice is crucial.  We will address this, and the desired properties of the proposed estimator, in Section~\ref{s:theory} below.  

To solve the optimization problem in equation \eqref{e:orig-opt}, we propose the estimator 
\[ \widetilde{\bbeta}_{2}(\lambda) = \arg \max_{\bbeta} O(\bbeta; \lambda), \]
where the objective function is 
\begin{align}
O(\bbeta; \lambda)&= n_{2}^{-1} \log f_{2}^{(n_{2})}(\by_{2}; \bbeta, \gamma_{2}) - \lambda n_{1}^{-1} \kl_{n_{1}}(\bbeta; \widehat\bbeta_{1}, \gamma_{1}) \nonumber \\ 
& = \frac{1}{n_{2}} \sum \limits_{i=1}^{n_{2}} \left\{y_{i2} \bX^\top_{i2} \bbeta - b( \bX^\top_{i2} \bbeta) \right\} \label{e:const-opt} \\
& \qquad - \frac{ \lambda }{n_{1}} \sum \limits_{i=1}^{n_{1}} \{ b^\prime(\bX^\top_{i1} \widehat{\bbeta}_{1}) (\bX^\top_{i1} \widehat{\bbeta}_{1}-\bX^\top_{i1}\bbeta) + b(\bX^\top_{i1}\bbeta) - b(\bX^\top_{i1} \widehat{\bbeta}_{1}) \}, \nonumber
\end{align}
for a {\em dial parameter} $\lambda \geq 0$.  Thanks to the general niceties of natural exponential families, this objective function is convex and, therefore, is straightforward to optimize.  The addition of the KL penalty term in equation \eqref{e:const-opt} essentially introduces $\lambda$-discounted units of information from $\mD_{1}$ into the inference based on $\mD_{2}$. Note that we specifically avoid use of the term ``tuning parameter'' to make a clear distinction between our approach and the classical penalized regression approach. The addition of the KL divergence penalty intentionally introduces a bias in the estimator $\widetilde{\bbeta}_{2}(\lambda)$ for $\bbeta_{2}$ which we hope---and later show---will be offset by a reduction in variance.  This is the same basic bias--variance trade-off that motivates other shrinkage estimation strategies, such as ridge regression and lasso, but with one key difference: our motivation is information-sharing, so we propose to shrink toward a $\mD_{1}$-dependent target whereas existing methods' motivation is structural constraints (e.g., sparsity), so they propose to shrink toward a fixed target. With this alternative perspective comes new questions: does adding a small amount of $\mD_{1}$-dependent bias lead to efficiency gains?  Below we will identify a range of the dial parameter such that the use of the external information in $\mD_{1}$ reduces the variance of $\widetilde{\bbeta}_{2}(\lambda)$ enough so that we gain efficiency even with the small amount of added bias. The parameter $\lambda$ thus acts as a dial to calibrate the trade-off between bias and variance in $\widetilde{\bbeta}_{2}(\lambda)$.

The maximum of $O(\bbeta; \lambda)$ coincides with the root of the estimating function
\begin{align*}
\bPsi(\bbeta; \lambda)&=\frac{1}{n_{2}} \sum \limits_{i=1}^{n_{2}} \bX_{i2} \{ y_{i2} - b^\prime ( \bX^\top_{i2} \bbeta ) \} - \frac{\lambda}{n_{1}} \sum \limits_{i=1}^{n_{1}} \bX_{i1}\{ b^\prime(\bX^\top_{i1} \bbeta ) -b^\prime(\bX^\top_{i1} \widehat{\bbeta}_{1}) \} \\
&= n_{2}^{-1} \bX_{2} \{ \by_{2} - \bmu_{2}(\bbeta) \} - \lambda n_{1}^{-1} \bX_{1} \{ \bmu_{1}(\bbeta) - \bmu_{1} (\widehat{\bbeta}_{1}) \},
\end{align*}
where $\bmu_j (\bbeta)$ is the $n_j$-vector of (conditional) mean responses, given $\bX_j$, for $j=1,2$.  The root is the value of $\bbeta$ that satisfies
\begin{align*}
n_{2}^{-1} \bX_{2} \by_{2} + \lambda n_{1}^{-1} \bX_{1} \bmu_{1} (\widehat{\bbeta}_{1}) &= n_{2}^{-1} \bX_{2} \bmu_{2}(\bbeta) + \lambda n_{1}^{-1} \bX_{1} \bmu_{1}(\bbeta).
\end{align*}
From this equation, we see that a solution $\bbeta$ must be such that a certain linear combination of $\bmu_{1}(\bbeta)$ and $\bmu_{2}(\bbeta)$ agrees with the same linear combination of $\bmu_{1}(\widehat\bbeta_{1})$ and $\by_{2}$, two natural estimators of the mean responses in $\mD_{1}$ and $\mD_{2}$, respectively.  In Section~\ref{s:examples}, we give the specific form of the estimating function for two common generalized linear models: linear and logistic regression.

\subsection{Remarks}
\label{s:remarks}

\subsubsection{Comparison to Wasserstein distance}
\label{ss:wasserstein}

The Wasserstein distance has enjoyed recent popularity in the transfer learning literature; see for example \cite{Shen-etal, Yoon-etal, Cheng-etal, Zhu-Wang-Peng-Li}. An example illustrates why we prefer the KL divergence to the Wasserstein distance. Suppose $f_{1}(y_{i1}; \bbeta, \gamma_{1})$ is the Gaussian density with mean $\bX^\top_{i1} \bbeta$ and variance $\gamma_{1}=\sigma^2_{1}$. The KL divergence defined above is 
\begin{equation}
\label{e:gaussian.kl}
\kl_{n_{1}}(\bbeta; \widehat\bbeta_{1}, \gamma_{1}) = (2\sigma_{1}^2)^{-1} ( \widehat{\bbeta}_{1} - \bbeta )^\top \bX_{1} \bX^\top_{1} ( \widehat{\bbeta}_{1} - \bbeta ). 
\end{equation}
Clearly, this takes into consideration not only the distance between $\bbeta$ and $\widehat\bbeta_{1}$, but also other relevant information-related aspects of the data set $\mD_{1}$.  By contrast, following \cite{Olkin-Pukelsheim}, the $2$-Wasserstein distance between the two Gaussian joint densities  $f_{1}^{(n_{1})}(\cdot; \widehat{\bbeta}_{1}, \sigma^2_{1})$ and $f_{1}^{(n_{1})}(\cdot; \bbeta, \sigma^2_{1})$ is
\begin{align*}
\mW \{f_{1}^{(n_{1})}(\cdot; \widehat{\bbeta}_{1}, \sigma^2_{1}), & \, f_{1}^{(n_{1})}(\cdot; \bbeta, \sigma^2_{1}) \}\\
&= \|\widehat{\bbeta}_{1} - \bbeta \|^2_{2} + \mbox{trace} \{ \sigma^2_{1} \bI_{n_{1}} + \sigma^2_{1} \bI_{n_{1}} - 2 \sigma_{1}\sigma_{1} \bI_{n_{1}} \} \\
& =\| \widehat{\bbeta}_{1} - \bbeta \|^2_{2}.
\end{align*}
In this example, the Wasserstein distance is only a function of the distance between the means, and fails to take into account any other aspects of $\mD_{1}$, in particular, it does not depend on the observed Fisher information in $\mD_{1}$.  Replacing the KL divergence in our proposal above with the 2-Wasserstein distance would, therefore, reduce to a simple ridge regression formulated with $\mD_{1}$-dependent shrinkage target $\widehat\bbeta_{1}$.  As discussed above, such an approach would not satisfactorily achieve our objectives.  

\subsubsection{Connection to data-dependent penalties}
\label{ss:bayes}

The dependence of the constraint in equation \eqref{e:orig-opt} on the data $\mD_{1}$ is unusual in contrast to more common constraints on parameters. Our formulation leads to a penalty term in $O(\bbeta; \lambda)$ that depends on data. This approach allows an appealing connection to the empirical Bayes estimation framework, which allows us to treat information gained from analyzing $\mD_{1}$ as prior knowledge when analyzing $\mD_{2}$. Through this lens, $\bbeta_{2} \mapsto \exp\{-\lambda \kl_{n_{1}}(\bbeta_{2}; \widehat{\bbeta}_{1}, \gamma_{1})\}$ acts like a prior density and maximizing $O(\bbeta_{2}; \lambda)$ is equivalent to maximizing the corresponding posterior density for $\bbeta_{2}$, i.e., $\widehat{\bbeta}_{2}(\lambda)$ is the maximum {\em a posteriori} estimator of $\bbeta_{2}$. The dial parameter $\lambda$ adjusts the impact of the prior and reflects the experimenter's belief in the similarity of $\mD_{1}$ and $\mD_{2}$.

The prior term $ \kl_{n_{1}}(\bbeta; \widehat{\bbeta}_{1}, \gamma_{1})$ is itself a measure of the excess surprise \citep{Baldi-Itti} between the prior information $f^{(n_{1})}_{1}(\by_{1}; \widehat{\bbeta}_{1}, \gamma_{1})$ and the likelihood $f^{(n_{1})}_{1}(\by_{1}; \bbeta, \gamma_{1})$ in $\mD_{1}$. This observation leads to an intuitive understanding of our KL prior: if the prior $f^{(n_{1})}_{1}(\by_{1}; \widehat{\bbeta}_{1}, \gamma_{1})$ and the likelihood $f^{(n_{1})}_{1}(\by_{1}; \bbeta, \gamma_{1})$ are close for all values of $\bbeta$, then the prior probabilities are small for all values of $\bbeta$ and the prior is weakly informative. 

\subsubsection{Distinction from data fusion}
\label{ss:fusion}

Apart from the literature that considers (a subset of) effects to be {\em a priori} heterogeneous \citep[e.g.][]{Liu-Liu-Xie-2015, tang2020poststratification}, homogeneous \citep[e.g.][]{Xie-Singh-Strawderman} or somewhere in between \citep[e.g.][]{shen2020fusion}, some methods consider fusion of individual feature effects. Broadly, fusion approaches jointly estimate $\bbeta_{1}$ and $\bbeta_{2}$ by shrinking them towards each other. At first blush, these methods may appear similar to our approach, so we take the time to highlight two key differences. 

Primarily, these methods differ in that they do not consider estimation in $\mD_{1}$ to be fixed to $\widehat{\bbeta}_{1}$, and can jointly re-estimate both parameters for improved efficiency. This is quite different from our goal, which is to quantify the utility of a first, already analyzed data set $\mD_{1}$ in improving inference in a second data set $\mD_{2}$. Our approach has the advantage that we do not require the model to be correctly specified in $\mD_{1}$, which endows our method with substantial robustness properties which are lacking for fusion approaches. 

Moreover, data fusion's underlying premise is to exploit feature clustering structure across data sources, thereby determining which {\em features} should be combined \citep{Bondell-Reich,shen2010grouping, tibshirani2011solution, ke2015homogeneity,Tang-Song}. In particular, this leads to the integration of some feature effects but not others. Different sets of feature effects are estimated from different sets of data, which does not provide the desired quantification of the similarity between data sets.

\section{Examples}
\label{s:examples}

\subsection{Gaussian linear regression}
\label{ss:linear}

For the Gaussian linear model, the KL divergence is given in \eqref{e:gaussian.kl}. In this case, the objective function in \eqref{e:const-opt} is given by 
\begin{align*}
O(\bbeta; \lambda)
&=n_{2}^{-1} \bbeta^\top \bX_{2} \by_{2} - (2n_{2})^{-1} \bbeta^\top \bX_{2} \bX^\top_{2} \bbeta \\
&\qquad - \lambda (2n_{1})^{-1} ( \widehat{\bbeta}_{1} - \bbeta )^\top \bX_{1} \bX^\top_{1} ( \widehat{\bbeta}_{1} - \bbeta).
\end{align*}
Optimizing $O$ corresponds to finding the root of the  estimating function
\begin{align*}
\bPsi(\bbeta; \lambda)&= n_{2}^{-1} ( \bX_{2} \by_{2} - \bX_{2} \bX^\top_{2} \bbeta ) - \lambda n_{1}^{-1} ( \bX_{1} \bX^\top_{1} \bbeta - \bX_{1} \bX^\top_{1} \widehat{\bbeta}_{1} ).
\end{align*}
Let $\bG_j = n_j^{-1} \bX_j \bX^\top_j$, $j=1,2$, denote the scaled Gram matrices. Then the solution to the estimating equation $\bPsi(\bbeta; \lambda)=0$ is
\begin{equation}
\begin{split}
\label{e:beta2_linear}
\widetilde{\bbeta}_{2}(\lambda) &= \bigl( \bG_{2} + \lambda \bG_{1} \bigr)^{-1} \bigl( n_{2}^{-1} \bX_{2} \by_{2}+ \lambda \bG_{1} \widehat{\bbeta}_{1} \bigr)  \\
&= \bigl(\bG_{2} + \lambda \bG_{1} \bigr)^{-1} \bigl( \bG_{2} \widehat{\bbeta}_{2} + \lambda \bG_{1} \widehat{\bbeta}_{1} \bigr),
\end{split}
\end{equation}
where $\widehat\bbeta_{2} = (\bX_{2} \bX_{2})^{-1} \bX_{2} \by_{2}$ is the MLE based on $\mD_{2}$ only. The estimator $\widetilde{\bbeta}_{2}(\lambda)$ in equation \eqref{e:beta2_linear} progressively grows closer to $\widehat{\bbeta}_{1}$ as more weight (through $\lambda$) is given to $\mD_{1}$. This is desirable behavior: $\lambda$ acts as a ``dial'' allowing us to tune our preference towards $\mD_{1}$ or $\mD_{2}$. When $\bbeta_{1}=\bbeta_{2}$, we recognize $\widetilde{\bbeta}_{2}(\lambda)$ as a generalization of the best linear unbiased estimator. The estimator $\widetilde{\bbeta}_{2}(\lambda)$ does not rely on individual level data from the first data set, $\mD_{1}$; all that is needed is the sample size, the estimate, and the observed Fisher information. Thus, our proposed procedure is privacy preserving and can be implemented in a meta-analytic setting where only summaries of $\mD_{1}$ and $\mD_{2}$ are available. 

This estimator also takes the familiar form of a generalized ridge estimator \citep{Hoerl-Kennard, Hemmerle} and benefits from many well-known properties; see the excellent review of \cite{vanWieringen}. The estimator $\widetilde{\bbeta}_{2}(\lambda)$ is a convex combination of the MLEs from $\mD_{1}$ and $\mD_{2}$, with weights determined by the corresponding observed Fisher information matrices. This formulation allows us to identify the crucial role that $\lambda$ plays in balancing not only the bias but the variance of $\widehat{\bbeta}_{2}$, a role reminiscent of that played by the tuning parameter for ridge regression \citep{Theobald, Farebrother}.

Our estimator $\widetilde{\bbeta}_{2}(\lambda)$ can be rewritten as $\widetilde{\bbeta}_{2}(\lambda)= \widetilde{\bW}_{\lambda} \widehat{\bbeta}_{2} + ( \bI_p - \widetilde{\bW}_{\lambda}) \widehat{\bbeta}_{1}$ with
\begin{align*}
\widetilde{\bW}_{\lambda}&= n^{-1}_{2} \bigl(n^{-1}_{2} \bX_{2} \bX^\top_{2} + \lambda n^{-1}_{1} \bX_{1} \bX^\top_{1} \bigr)^{-1} \bX_{2} \bX^\top_{2}
\end{align*}
In contrast, \cite{Chen-Owen-Shi} consider pooling $\mD_{1}$ and $\mD_{2}$ to jointly estimate $\bbeta_{2}$ and $\bbeta_{1}=\bbeta_{2} + \bdelta_{2}$ with some penalty $\| \bX_{2} \bdelta_{2} \|^2_{2}$, where $\bdelta_{2}=\bbeta_1 - \bbeta_2$. Their estimator is given by $\widehat{\bbeta}_{2}(\bW_{\lambda})=\bW_{\lambda} \widehat{\bbeta}_{2} + (\bI_p - \bW_{\lambda}) \widehat{\bbeta}_{1}$, where
\begin{align*}
\bW_{\lambda}&=\bigl (\bX_{1} \bX^\top_{1} + \lambda \bX_{2} \bX^\top_{2} + \lambda \bX_{1} \bX^\top_{1}\bigr)^{-1} \bigl( \bX_{1} \bX^\top_{1} + \lambda \bX_{2} \bX^\top_{2}\bigr).
\end{align*}
When $n_{1}=n_{2}$, $\widetilde{\bW}_{\lambda} \preceq \bW_{\lambda}$ if and only if $(1-\lambda) \bI_p \preceq (\bX_{2} \bX^\top_{2})^{-1}\bX_{1} \bX^\top_{1}$. This is clearly always true when $\lambda \geq 1$. When $\lambda <1$, $\widetilde{\bW}_{\lambda} \preceq \bW_{\lambda}$ if $\mD_{1}$ is informative relative to $\mD_{2}$, and $\widetilde{\bW}_{\lambda} \succeq \bW_{\lambda}$ if $\mD_{1}$ is uninformative relative to $\mD_{2}$. Therefore, our estimator assigns more weight to $\mD_{1}$ than \cite{Chen-Owen-Shi}'s estimator when $\mD_{1}$ is more informative, and less when it is uninformative, as desired.

\subsection{Bernoulli logistic regression}
\label{ss:logistic}

For the standard logistic regression model, the KL divergence is given by
\begin{equation*}
\kl_{n_{1}}(\bbeta; \widehat\bbeta_{1}) =\sum \limits_{i=1}^{n_{1}} \left[ \frac{\exp ( \bX^\top_{i1} \widehat{\bbeta}_{1} )}{1+\exp ( \bX^\top_{i1} \widehat{\bbeta}_{1})} ( \bX^\top_{i1} \widehat{\bbeta}_{1} - \bX^\top_{i1} \bbeta) + \log \Bigl\{ \frac{1+\exp( \bX^\top_{i1} \bbeta ) }{ 1+\exp ( \bX^\top_{i1} \widehat{\bbeta}_{1} ) } \Bigr\} \right].
\end{equation*}
Then the objective function in \eqref{e:const-opt} can be written as
\begin{align*}
O(\bbeta; \lambda)&= \frac{1}{n_{2}} \sum \limits_{i=1}^{n_{2}} \bigl[ y_{i2} \bX^\top_{i2} \bbeta - \log \{ 1+\exp ( \bX^\top_{i2} \bbeta ) \} \bigr] \\
& \quad -\frac{\lambda}{n_{1}} \sum \limits_{i=1}^{n_{1}} \left[ \frac{\exp ( \bX^\top_{i1} \widehat{\bbeta}_{1} )}{1+\exp ( \bX^\top_{i1} \widehat{\bbeta}_{1})} ( \bX^\top_{i1} \widehat{\bbeta}_{1} - \bX^\top_{i1} \bbeta) + \log \Bigl\{ \frac{1+\exp( \bX^\top_{i1} \bbeta ) }{ 1+\exp ( \bX^\top_{i1} \widehat{\bbeta}_{1} ) } \Bigr\} \right].
\end{align*}
To optimize $O$, we find the root of the estimating function  given by
\begin{align*}
\bPsi(\bbeta; \lambda) &= n_{2}^{-1} \bX_{2} \bigl\{ \by_{2} - \expit(\bX_{2}^\top \bbeta) \bigr\} - \lambda n_{1}^{-1} \bX_{1} \bigl\{ \expit(\bX_{1}^\top \bbeta) - \expit(\bX_{1}^\top \widehat\bbeta_{1}) \bigr\}, 
\end{align*}
where $\expit(\bz)$ is the vector obtained by applying $z \mapsto e^z/(1 + e^z)$ to each component of the vector $\bz$.  The estimator $\widetilde{\bbeta}_{2}(\lambda)$ is the solution to $\bPsi(\bbeta; \lambda)=\bzero$.  Of course, there is no closed-form solution, but the solution can be obtained numerically. 

\section{Theoretical support}
\label{s:theory}

\subsection{Objectives}
\label{ss:objectives}

A distinguishing feature of our perspective here is that we treat $\mD_{1}$ as fixed.  In particular, we treat $\widehat\bbeta_{1}$ as a known constant.  A relevant feature in the analysis below is the difference $\bdelta = \bbeta_{2} - \widehat\bbeta_{1}$, which is just one measure of the similarity/dissimilarity between $\mD_{1}$ and $\mD_{2}$.  Of course, $\bdelta$ is unknown because $\bbeta_{2}$ is unknown, but $\bdelta$ can be estimated if necessary.  Note also that we do not assume $\|\bdelta\|_2$ to be small.

Thanks to well-known properties of KL divergence, $\bbeta \mapsto \E\{O(\bbeta; \lambda)\}$ is concave, so it has a unique maximum, denoted by $\bbeta^\star_{2}(\lambda)$.  If the true value of $\bbeta$ in $\mD_{2}$ is $\bbeta_{2}$, then clearly $\bbeta^\star_{2}(\lambda) \neq \bbeta_{2}$ when $\lambda \neq 0$ and $\bdelta \neq \bzero$. On the other hand, it is easy to verify that, as $\lambda \to 0^+$, the objective function satisfies $\E\{O(\bbeta; \lambda)\} \to \E\{O(\bbeta;0)\}$ uniformly on compact subsets of the parameter space of $\bbeta$, so, as expected, $\lim_{\lambda \rightarrow 0^+} \bbeta^\star_{2}(\lambda) = \bbeta_{2}$. As stated in Section \ref{ss:proposal}, our objective is to find a range of the dial parameter $\lambda$ --- depending on $\bdelta$, etc. --- such that the use of the external information in $\mD_{1}$ sufficiently reduces the variance of $\widetilde{\bbeta}_{2}(\lambda)$ so as to overcome the addition of a small amount of bias to $\widetilde{\bbeta}_{2}(\lambda)$. 

We first consider the Gaussian linear model for its simplicity, which allows for exact (non-asymptotic) results.  The ideas extend to generalized linear models, but there we will need asymptotic approximations as $n_{2} \to \infty$. Throughout, we use the notational convention $\bM^2=\bM^\top \bM$ for a matrix $\bM$.

Proofs of all the results are given in Appendix~\ref{S:proofs}. 

\subsection{Exact results in the Gaussian linear model}
\label{ss:t-linear}

In the Gaussian case, the expected estimating function evaluated at $\bbeta^\star_{2}(\lambda)$ is
\begin{align*}
\bzero&=\E_{\bbeta_{2}} \bigl\{ \bPsi(\bbeta^\star_{2}(\lambda); \lambda) \bigr\}\\
&=n_{2}^{-1} \bX_{2} \bigl\{ \E_{\bbeta_{2}} (\by_{2}) - \bX^\top_{2} \bbeta^\star_{2}(\lambda) \bigr\} - \lambda n_{1}^{-1} \bX_{1} \bigl\{ \bX_{1} \bbeta^\star_{2}(\lambda) - \bX_{1} \widehat{\bbeta}_{1} \bigr\}\\
& = \bG_{2} \{ \bbeta_{2} - \bbeta_{2}^\star(\lambda) \} - \lambda \bG_{1} \{ \bbeta^\star_{2}(\lambda) -  \widehat{\bbeta}_{1} \}.
\end{align*}
Denote $\bS(\lambda) = \bG_{2} + \lambda \bG_{1} $. Rearranging, we obtain
\begin{align*}
\bbeta^\star_{2}(\lambda)&= \bS^{-1}(\lambda) \bigl( \bG_{2} \bbeta_{2} + \lambda \bG_{1} \widehat{\bbeta}_{1} \bigr)=\bbeta_{2} - \lambda \bS^{-1}(\lambda) \bG_{1} \bdelta .
\end{align*}
Thus, $\bbeta^\star_{2}(\lambda) = \bbeta_{2}$ when $\lambda =0$ or $\bdelta = \bzero$.  In general, however, our proposed estimator $\widetilde\bbeta_{2}(\lambda)$ is estimating $\bbeta_{2}^\star(\lambda) \neq \bbeta_{2}$, so a practically important question is, if we already have an unbiased estimator, $\widehat\bbeta_{2}$, of $\bbeta_{2}$, then why would we introduce a biased estimator?  Theorem \ref{t:MSE} below establishes that there exists a range of $\lambda>0$ for which the mean squared error (MSE) of $\widetilde{\bbeta}_{2}(\lambda)$ is strictly less than that of $\widehat{\bbeta}_{2}$.  Details on the $\lambda$ range over which the efficiency gain is achieved are given in the third paragraph following the theorem statement.

\begin{theorem}
\label{t:MSE}
There exists a range of $\lambda>0$ on which the mean squared error of $\widetilde{\bbeta}_{2}(\lambda)$ is strictly less than the mean squared error of $\widehat{\bbeta}_{2}$.
\end{theorem}

This result is striking when considered in the context of data integration: we have shown that it is always ``better'' to integrate two sources of data than to use only one, even when the two are substantially different. We also claim that our estimator's gain in efficiency is robust to heterogeneity between the two data sets; we will return to this point in Section \ref{ss:t-GLM} to make general remarks in the context of transfer learning. 

Of course, the reader familiar with Stein shrinkage \citep{Stein,James-Stein} may not be surprised by our Theorem~\ref{t:MSE}. Our result has a similar flavor to Stein's paradox \citep{Efron-Morris-1977}, i.e., that some amount of shrinkage always leads to a more efficient estimator compared to a (weighted) sample average, here $\widehat{\bbeta}_{2}$. In their presciently titled paper, ``Combining possibly related estimation problems,'' \cite{Efron-Morris-JRSSB} anticipated the ubiquity of results like our Theorem~\ref{t:MSE} that show estimation efficiency gains when combining different but related data sets. 

The proof of Theorem \ref{t:MSE} shows that $\mbox{MSE}\{\widetilde{\bbeta}_{2}(\lambda)\} < \mbox{MSE}(\widehat{\bbeta}_{2})$ for all $\lambda >0$ when $\bdelta=\bzero$; when $\bdelta \neq \bzero$, it proves that $\mbox{MSE}\{\widetilde{\bbeta}_{2}(\lambda)\}$ is monotonically decreasing over the range $[0,\lambda^\star]$ and monotonically increasing over the range $[\lambda^\star, \infty)$ for some $\lambda^\star>0$ that does not have a closed-form. The proof does, however, provide a bound for $\lambda^\star$:
\begin{align}
\lambda^\star > \frac{\sigma^2_{2}}{n_{2}} \frac{\min_{r=1, \ldots, p} \kappa_{r} g^{-1}_{r2} }{ \max_{r=1,\ldots,p} \delta_r^2},
\label{e:lambda-range}
\end{align}
where $g_{r2}>0$, $r=1, \ldots, p$, the eigenvalues of $\bG_{2}$ in increasing order and $\kappa_{r}$, $r=1, \ldots, p$, the eigenvalues of $\bG^{1/2}_{2} \bG^{-1}_{1} \bG^{1/2}_{2}$ in decreasing order. That is, the MSE of $\widetilde\bbeta_{2}(\lambda)$ will be less than that of $\widehat\bbeta_{2}$ if $\lambda$ is no more than the right-hand side of \eqref{e:lambda-range}. From the above expression, we see that if elements of $\bdelta$ are large in absolute value, then the improvement in MSE only occurs for a small range of $\lambda$. Moreover, if $n^{-1}_{2}$, $\kappa_{r}$ or $g^{-1}_{r2}$ are small then the range of $\lambda$ such that $\mbox{MSE}\{\widetilde{\bbeta}_{2}(\lambda)\} < \mbox{MSE}(\widehat{\bbeta}_{2})$ is small: in other words, if $\mD_{2}$ is highly informative then very little weight should be given to $\mD_{1}$, regardless of how informative $\mD_{1}$ is. This is intuitively appealing and practically useful because it provides a loose guideline based on the informativeness of $\mD_{2}$ for how much improvement can be obtained using $\mD_{1}$.

In practice, we propose to find a data-driven version of the minimizer, $\lambda^\star$, of the MSE.  For this, we minimize an empirical version of the MSE based on plug-in estimators, i.e., 
\begin{align*}
\tilde\lambda &=\arg \min \limits_{\lambda > 0} \bigl[ n_{2}^{-1} \widehat{\sigma}^2_{2} \, \mbox{trace} \{ \bS^{-2}(\lambda) \bG_{2} \} + \lambda^2 \mbox{trace} \{ \bG_{1} \bS^{-2}(\lambda) \bG_{1} \widehat{\bdelta^2} \} \bigr],
\end{align*}
the minimizer of the estimated mean squared error of $\widetilde{\bbeta}_{2}(\lambda)$, where  
\[ \widehat{\sigma}^2_{2}=\frac{1}{n_{2}-p} \sum_{i=1}^{n_{2}} (y_{i2} - \bX^\top_{i2} \widehat{\bbeta}_{2})^2, \]
is the usual estimate of the error variance $\sigma^2_{2}$, and $\widehat{\bdelta^2}$ is a bias-adjusted estimate of $\bdelta \bdelta^\top$ computed as follows. If we define $\widehat{\bdelta}_p=\widehat{\bbeta}_{2} - \widehat{\bbeta}_{1}$, then the equality $\E_{\bbeta_{2}}(\widehat{\bdelta}_p \widehat{\bdelta}^\top_p)=\bdelta \bdelta^\top + \sigma^2_{2} n_{2}^{-1} \bG^{-1}_{2}$ implies that $\widehat{\bdelta}_p \widehat{\bdelta}^\top_p$ is a positively biased estimator of $\bdelta \bdelta^\top$. Similar to \cite{Vinod} and \cite{Chen-Owen-Shi}, we estimate $\bdelta \bdelta^\top$ with
\begin{align*}
\widehat{\bdelta^2}&= \bigl(\widehat{\bdelta}_p \widehat{\bdelta}^\top_p - \widehat{\sigma}^2_{2} n_{2}^{-1} \bG^{-1}_{2} \bigr)_+,
\end{align*} 
where $(\bA)_+=\bP \mathbf{diag} \{ \max(\lambda_i,0)\}_{i=1}^p \bP^\top$ for a symmetric matrix $\bA \in \mathbb{R}^{p\times p}$ with eigendecomposition $\bP \, \mathbf{diag} (\lambda_i)_{i=1}^p \, \bP^\top$. If all eigenvalues are negative, we use $\widehat{\bdelta^2}=\widehat{\bdelta}_p \widehat{\bdelta}^\top_p$. This bias-adjusted approach to estimating MSE is related to Stein's unbiased risk estimator \citep[SURE,][]{Stein-1981}; see also \cite{Vinod} for a discussion in the ridge regression setting. The minimizer $\tilde \lambda$ (almost) always exists since $\widehat{\bdelta^2}$ is non-zero with probability 1. That the minimizer is strictly positive, even if $\mD_{1}$ is minimally- or non-informative, might be counter-intuitive; but this is a consequence of Theorem~\ref{t:MSE}, which shows that we need only consider the set of strictly positive $\lambda$ values to improve inference in $\mD_{2}$. Finally, we show that constructing exact confidence intervals for $\bbeta_{2}$ using a debiased version of $\widetilde{\bbeta}_{2}(\lambda)$ reduces to the traditional inference using the MLE. It follows from the proof of Theorem \ref{t:MSE} that, 
\begin{align}
\widetilde{\bbeta}_{2}(\lambda) + \lambda \bS^{-1}(\lambda) \bG_{1} (\bbeta_{2} - \widehat{\bbeta}_{1}) \sim \mN \{ \bbeta_{2}, n_{2}^{-1} \sigma^2_{2} \bS^{-1}(\lambda) \bG_{2} \bS^{-1}(\lambda) \}, 
\label{e:tilde-dist}
\end{align}
$\lambda \geq 0$, and therefore, using the definition of $\widetilde{\bbeta}_{2}(\lambda)$ in equation \eqref{e:beta2_linear}, 
\[ \{ \bI_p - \lambda \bS^{-1}(\lambda) \bG_{1} \}^{-1} \bS^{-1}(\lambda) \bG_{2} \widehat{\bbeta}_{2} \sim \mN \{ \bbeta_{2}, (\sigma^2_{2}/n_{2}) \bD(\lambda) \}, \]
with 
\[ \bD(\lambda)=\{ \bI_p - \lambda \bS^{-1}(\lambda) \bG_{1} \}^{-1} \bS^{-1}(\lambda) \bG_{2} \bS^{-1}(\lambda) \{ \bI_p - \lambda \bG_{1} \bS^{-1}(\lambda) \}^{-1}. \]
Algebra reveals that $\{ \bI_p - \lambda \bS^{-1}(\lambda) \bG_{1} \}^{-1}\bS^{-1}(\lambda) \bG_{2} = \bI_p$, which finally yields the familiar result: $\widehat{\bbeta}_{2} \sim \mN\{ 0, n_{2}^{-1} \sigma^2_{2} \bG^{-1}_{2}\}$. Therefore, confidence intervals based on equation \eqref{e:tilde-dist} reduce to confidence intervals obtained from the MLE $\widehat{\bbeta}_{2}$ and familiar Gaussian sampling distribution. That is, the debiasing that ensures the confidence interval is centered (on average) at $\bbeta_{2}$ effectively negates the gain in efficiency of our estimator. As remarked by \cite{Obenchain} in ridge regression, our shrinkage does not produce ``shifted'' confidence regions, and the MLE is the most suitable choice if one desires valid confidence intervals. Nonetheless, we will consider in Sections \ref{ss:t-GLM} and \ref{s:empirical} if the biased estimator $\widetilde{\bbeta}_{2}(\lambda)$ can be used to derive confidence intervals with \emph{asymptotic} nominal coverage as $n_{2} \to \infty$.

\subsection{Asymptotic results in generalized linear models}
\label{ss:t-GLM}

Next we investigate the asymptotic properties of $\widetilde{\bbeta}_{2}(\lambda)$ in generalized linear models as $n_{2} \rightarrow \infty$. For $j=1,2$, let
\begin{align*}
\bA(\bX^\top_j \bbeta)=\mathbf{diag} \bigl\{ h^\prime(\bX^\top_{ij} \bbeta)\bigr\}_{i=1}^{n_j}=\mathbf{diag} \bigl\{ b^{\prime \prime}(\bX^\top_{ij} \bbeta)\bigr\}_{i=1}^{n_j} \in \mathbb{R}^{n_j \times n_j},
\end{align*}
where $h(\bX^\top_{i1} \bbeta)=b^{\prime}(\bX^\top_{ij} \bbeta)$ is the inverse of the link function. Define $\bDelta(\bbeta) = \{h(\bX^\top_{i1} \bbeta) - h(\bX^\top_{i1} \widehat{\bbeta}_{1})\}_{i=1}^{n_{1}}$ and $\bS(\bbeta; \lambda) = - \partial \bPsi(\bbeta; \lambda)/(\partial \bbeta)$. Then
\begin{align*}
\bS(\bbeta; \lambda)
&= \frac{1}{n_{2}} \sum \limits_{i=1}^{n_{2}} \bX_{i2} \bX^\top_{i2} h^\prime (\bX^\top_{i2} \bbeta) + \frac{\lambda}{n_{1}} \sum \limits_{i=1}^{n_{1}} \bX_{i1} \bX^\top_{i1} h^\prime (\bX^\top_{i1} \bbeta)\\
&=n_{2}^{-1} \bX_{2} \bA(\bX^\top_{2}\bbeta) \bX^\top_{2} + \lambda n_{1}^{-1} \bX_{1} \bA(\bX^\top_{1} \bbeta) \bX^\top_{1}.
\end{align*}
We assume the following conditions.
\begin{enumerate}[label=(C\arabic*)]
\item \label{c:4} $\lim_{n_{2} \rightarrow \infty} n_{2}^{-1} \sum_{i=1}^{n_{2}} \{ h(\bX^\top_{i2} \bbeta_{2}) \bX^\top_{i2} \bbeta - b(\bX^\top_{i2} \bbeta) \}$ exists and is finite.
\item \label{c:5} For any $\bbeta$ between $\bbeta_{2}$ and $\bbeta^\star_{2}(\lambda)$ inclusive, the two matrices defined below exist and are positive definite:
\[ \bv_{1}(\bbeta) = n_{1}^{-1} \bX_{1} \bA(\bX^\top_{1} \bbeta) \bX^\top_{1} \quad \text{and} \quad \bv_{2}(\bbeta) = \lim_{n_{2} \rightarrow \infty} n_{2}^{-1} \bX_{2} \bA(\bX^\top_{2} \bbeta) \bX^\top_{2}. \]
\end{enumerate}
Denote by $\bs(\bbeta; \lambda) =\lim_{n_{2} \rightarrow \infty} \E_{\bbeta_{2}} \{ \bS(\bbeta; \lambda) \} = \lim_{n_{2} \rightarrow \infty} \bS(\bbeta; \lambda)=\bv_{2}(\bbeta) + \lambda \bv_{1}(\bbeta)$. Recall that the asymptotic variance of the MLE $\widehat{\bbeta}_{2}$ is $d(\gamma_{2}) \bv^{-1}_{2}(\bbeta_{2})$. 

Lemma \ref{l:MLE-cons} is a standard result stating that the minimizer of our objective function converges to the minimizer of its expectation.

\begin{lemma}
\label{l:MLE-cons}
If Condition \ref{c:4} holds, then $\widetilde{\bbeta}_{2}(\lambda) - \bbeta^\star_{2}(\lambda) = O_p(n^{-1/2}_{2})$, for each $\lambda \geq 0$.
\end{lemma}

Since we already have that $\lim_{\lambda \rightarrow 0^+} \bbeta^\star_{2}(\lambda) = \bbeta_{2}$ as $n_{2}\rightarrow \infty$, an immediate consequence of Lemma \ref{l:MLE-cons} is that $\widetilde{\bbeta}_{2}(\lambda) \stackrel{p}{\rightarrow} \bbeta_{2}$ as $n_{2} \rightarrow \infty$ and $\lambda \rightarrow 0^+$. More can be said, however, about the local behavior of $\widetilde\bbeta_{2}(\lambda)$ depending on how quickly $\lambda$ vanishes with $n_{2}$.  

\begin{lemma}
\label{l:MLE-norm}
If Conditions \ref{c:4}--\ref{c:5} hold, and $\lambda = O(n^{-1/2}_{2})$, then as $n_{2} \to \infty$,
\begin{align*}
n_{2}^{1/2} \bJ^{1/2}(\bbeta_{2}; \lambda) \bigl\{ \widetilde{\bbeta}_{2}(\lambda) - \bbeta_{2} + \lambda n_{1}^{-1} \bS^{-1}(\bbeta_{2}; \lambda) \bX_{1} \bDelta(\bbeta_{2}) \bigr\} \stackrel{d}{\rightarrow} \mN ( \bzero, \bI_p ), 
\end{align*}
where $\bJ(\bbeta; \lambda ) = d(\gamma_{2}) \bS^\top(\bbeta; \lambda) \bv_{2}^{-1}(\bbeta) \bS(\bbeta; \lambda)$ is the observed Godambe information matrix \citep{Godambe-1991}.
Moreover, if $\lambda=o(n^{-1/2}_{2})$, then 
\[ n_{2}^{1/2} \bJ^{1/2}(\bbeta_{2}; \lambda) \{\widetilde{\bbeta}_{2}(\lambda) - \bbeta_{2} \} \stackrel{d}{\rightarrow} \mN \left( \bzero, \bI_p \right), \quad n_{2} \to \infty. \]
\end{lemma}

To summarize, if $\lambda$ vanishes not too rapidly, then there is a bias effect in the first-order asymptotic distribution approximation of $\widetilde{\bbeta}_{2}$; if $\lambda$ vanishes rapidly, then there is no bias effect in the first-order  approximation.  In either case, however, there is a reduction in the variance due to the combining of information in $\mD_{1}$ and $\mD_{2}$. This gain in efficiency can be seen, at least intuitively, by looking at the Godambe information matrix $\bJ(\bbeta_{2}; \lambda)$.  Indeed, since $\lim_{n_{2} \rightarrow \infty} \bS(\bbeta; \lambda) = \bv_{2}(\bbeta) + \lambda \bv_{1}(\bbeta)$ has eigenvalues strictly larger than those of $\bv_{2}(\bbeta)$, it follows that $\bJ(\bbeta_{2}; \lambda)$ has eigenvalues strictly smaller than those of $d(\gamma_{2}) \bv_{2}^{-1}(\bbeta_{2})$.  The following theorem, our main result, confirms the above intuition.  

\begin{theorem}
\label{t:MLE-MSE}
If Conditions \ref{c:4}--\ref{c:5} hold, then there exists a range of $\lambda>0$ values, with upper limit $O(n^{-1/2}_{2})$, on which the asymptotic mean squared error (aMSE) of $\widetilde{\bbeta}_{2}(\lambda)$ is strictly less than that of $\widehat{\bbeta}_{2}$.
\end{theorem}

The proof of Theorem \ref{t:MLE-MSE} shows that $\mbox{aMSE}\{ \widetilde{\bbeta}_{2}(\lambda)\} < \mbox{aMSE} (\widehat{\bbeta}_{2})$ for all $\lambda>0$ when $\bdelta = \bzero$; when $\bdelta \neq \bzero$, it proves that the aMSE of $\widetilde{\bbeta}_{2}(\lambda)$ is monotonically decreasing over the range $[0,\lambda^\star]$ and monotonically increasing over the range $[\lambda^\star, \infty)$, for some $\lambda^\star>0$ that does not have a closed-form. But the proof of Theorem \ref{t:MLE-MSE} does provide a bound:
\begin{align}
\lambda^\star > \frac{d(\gamma_{2})}{n_{2}} \frac{\min_{r=1, \ldots, p}\{ \kappa_{r}(\bbeta_{2}) g^{-1}_{r2}(\bbeta_{2}) \}}{ \max \mathbf{diag} \{ \bv^{-1}_{1}(\bbeta_{2}) \bX_{1} \bDelta(\bbeta_{2}) \bDelta^\top(\bbeta_{2}) \bX^\top_{1} \bv^{-1}_{1}(\bbeta_{2}) \} / n^2_{1}},
\label{e:MLE-lambda-range}
\end{align}
where $g_{r2}(\bbeta)>0$, $r=1, \ldots, p$, are the eigenvalues in increasing order of $\bv_{2}(\bbeta)$ and $\kappa_{r}(\bbeta)$, $r=1, \ldots, p$, are the eigenvalues in decreasing order of $\bv^{1/2}_{2}(\bbeta) \bv^{-1}_{1}(\bbeta) \allowbreak \bv^{1/2}_{2}(\bbeta)$.  This implies that the aMSE of $\widetilde\bbeta_{2}(\lambda)$ is less than that of $\widehat\bbeta_{2}$ for $\lambda$ no more than the expression on the right-hand side of \eqref{e:MLE-lambda-range}. The denominator of \eqref{e:MLE-lambda-range} is the nonlinear analogue of $\max_{r=1, \ldots, p} \delta^2_r$. To see this, by the mean value theorem we can write $\bX_{1} \bDelta(\bbeta_{2}) = n_{1} \bv_{1}(\bc_{2}) \bdelta$ for some vector $\bc_{2}$ between $\widehat{\bbeta}_{1}$ and $\bbeta_{2}$. Then the denominator can be rewritten as the maximum entry of
\begin{align*}
\mathbf{diag} \{ \bv^{-1}_{1} (\bbeta_{2}) \bv_{1}(\bc_{2}) \bdelta \bdelta^\top \bv_{1}(\bc_{2}) \bv^{-1}_{1}(\bbeta_{2}) \}.
\end{align*}
Recall that $\bv_{1}(\bbeta)$ is the derivative of $n_{1}^{-1} \bmu_{1}(\bbeta)$. Thus, the denominator divides the distance $\bdelta$ by the rate of change of the link function. When $\bdelta$ is small, $\bv^{-1}_{1} (\bbeta_{2}) \bv_{1}(\bc_{2}) \approx \bI_p$ and we recover the denominator in \eqref{e:lambda-range}. From this, we draw the same conclusion as in the Gaussian linear model: if $\mD_{2}$ is highly informative, then little weight should be given to $\mD_{1}$, regardless of how informative $\mD_{1}$ is.

For a data-driven choice of $\lambda^\star$, we propose 
\begin{align}
\tilde\lambda =\arg \min \limits_{\lambda > 0} \bigl[& d(\widehat{\gamma}_{2}) \mbox{trace} \{ \bS^{-2}(\widehat{\bbeta}_{2}; \lambda) \bV_{2} (\widehat{\bbeta}_{2})\} \bigr. \notag \\\bigl.& + n_{2} n_{1}^{-2} \lambda^2 \bDelta^\top(\widehat{\bbeta}_{2}) \bX^\top_{1} \bS^{-2}(\widehat{\bbeta}_{2}; \lambda) \bX_{1} \bDelta(\widehat{\bbeta}_{2}) \bigr],
\label{e:empirical-MSE}
\end{align}
the minimizer of the estimated aMSE of $\widetilde{\bbeta}_{2}(\lambda)$, where $\widehat{\gamma}_j$ the maximum likelihood estimator of $\gamma_j$ in $\mD_j$, $j=1,2$, and $\bV_{2}(\bbeta) = n_{2}^{-1} \bX_{2} \bA(\bX^\top_{2} \bbeta) \bX^\top_{2}$. Since we are assuming $n_{2}$ is large, we do not adjust for finite sample bias in $\bDelta(\widehat{\bbeta}_{2}) \bDelta^\top (\widehat{\bbeta}_{2})$ as we did in the Gaussian linear model setting. Again, the minimizer always exists since $\bDelta(\widehat{\bbeta}_{2}) \neq \bzero$ with probability 1.

As suggested in Section \ref{ss:t-linear}, Theorem \ref{t:MLE-MSE} endows our approach with robustness to model misspecification or lack of information in $\mD_{1}$ by guaranteeing (under conditions) that our approach always improves efficiency. This is far superior to traditional transfer learning which can suffer from degraded performance in $\mD_{2}$, or ``negative transfer'' \citep{Torrey-Shavlik}. As a remedy, \cite{Li-Cai-Li-2022} and \cite{Tian-Feng} develop methods that detect informative data sets $\mD_{1}$ to avoid the pitfalls of negative transfer. Their approaches can handle settings with multiple candidate data sets $\mD_{1}$: they find those most ``similar'' to $\mD_{2}$ and use these for transferring. Unfortunately, their approaches can still result in degraded performance, as illustrated in Section \ref{s:empirical}.

Interestingly, the bound in equation~\eqref{e:MLE-lambda-range} is $O(n^{-1}_{2})$, implying that the choice of $\lambda$ that maximizes the asymptotic efficiency of $\widetilde{\bbeta}_{2}(\lambda)$ is a $\lambda^\star$ strictly between $O(n^{-1/2}_{2})$ and $O(n^{-1}_{2})$; numerical results that help to confirm this are given in Section~\ref{s:empirical} below. This suggests the bias introduced by $\lambda^\star$ should be ignorable, while still affording us a small gain in efficiency. In conjunction with Lemma~\ref{l:MLE-norm}, this suggests a path to inference using the biased estimator $\widetilde{\bbeta}_{2}(\tilde\lambda)$. We propose to construct large-sample $100(1-\alpha)\%$ confidence intervals for $\bbeta_{2}$ using 
\begin{align}
\widetilde{\bbeta}_{2}(\tilde\lambda) \pm z_{\alpha/2} n^{-1/2}_{2} \bj^{-1/2}\{ \widehat{\bbeta}_{2}; \tilde\lambda \},
\label{e:MLE-CI}
\end{align}
with $z_{\alpha/2}$ the $1-\frac{\alpha}{2}$ quantile of the standard normal distribution and $\bj^{-1/2}(\bbeta; \lambda)$ the square root of the diagonal elements in 
$\{d(\widehat{\gamma}_{2}) \bS^\top(\bbeta; \lambda) \bV_{2}^{-1}(\bbeta) \bS(\bbeta; \lambda) \}^{-1}$.  Since $\bS(\bbeta; \lambda) \succeq \bV_{2}(\bbeta)$ in Loewner order, it follows that the confidence interval based on $\widetilde{\bbeta}_{2}(\tilde{\lambda})$ in equation \eqref{e:MLE-CI} is statistically more efficient than the classical confidence interval based on the sampling distribution of the MLE $\widehat{\bbeta}_{2}$. 

\section{Empirical investigation}
\label{s:empirical}

\subsection{Objectives, setup, and take-aways}

We examine the empirical performance of our proposed information-based shrinkage estimator (ISE) $\widetilde{\bbeta}_{2}(\lambda)$ through simulations in linear and logistic regression models. For all settings, we compute $\widetilde{\bbeta}_{2}(\lambda)$ for a range of $\lambda$ values, and compute $\tilde\lambda$ and $\widetilde{\bbeta}_{2}(\tilde\lambda)$. We construct large sample 95\% confidence intervals using equation \eqref{e:MLE-CI}. Finally, we investigate the robustness of our proposed approach to model misspecification in the assumed model for analyzing $\mD_{1}$. Throughout, true values $\bbeta_{1}$ and $\bdelta$ are randomly sampled from suitable uniform distributions.

In addition to comparison with the MLE $\widehat{\bbeta}_{2}$, we show that the behavior of our estimator $\widetilde{\bbeta}_{2}(\lambda)$ more closely aligns with the intuition developed in Section \ref{s:formulation} than competitor approaches, and that it is more efficient. In both the linear and logistic models, we compare our approach to the Trans-GLM estimator $\widehat{\bbeta}_{2}^{TG}$ in Algorithm 2 of \cite{Tian-Feng} with $L_1$ regularization (their default) using the R package and function \verb|glmtrans|, and the pooled MLE $\widehat{\bbeta}_{2}^p$ that estimates $\bbeta_{2}$ with $\mD_{1}$ and $\mD_{2}$ by assuming $\bbeta_{2}=\bbeta_{1}$. In the linear model, we compare our approach to the estimator $\widehat{\bbeta}_{2}(\bW_{\lambda})$ of \cite{Chen-Owen-Shi} described in Section \ref{ss:linear} over the same range of $\lambda$ values as our estimator, and to $\widehat{\bbeta}_{2}(\bW_{\widehat{\lambda}})$ with $\widehat{\lambda}$ selected by minimizing the predictive mean squared error of $\widehat{\bbeta}_{2}(\bW_{\lambda})$ using their bias-adjusted plug-in estimate (their version of our $\tilde\lambda$). We also compare our approach to the Trans-Lasso estimator $\widehat{\bbeta}_{2}^{TL}$ of \cite{Li-Cai-Li-2022} (using their software defaults): their approach assumes all covariates are Gaussian with mean zero, and so we center our outcome and estimate $\bbeta_{2}$ without its intercept. In the logistic model, we compare our approach to that of \cite{Zheng-etal}, who propose the estimator $\widehat{\bbeta}_{2}(\bW)=\bW \widehat{\bbeta}_{2} + (\bI_p - \bW) \widehat{\bbeta}_{1}$ with the matrix $\bW$ given by
\begin{align*}
\bigl[ (\widehat{\bbeta}_{1} - \widehat{\bbeta}_{2}) (\widehat{\bbeta}_{1} - \widehat{\bbeta}_{2})^\top + \bigl\{\bX_{1} \bA(\bX^\top_{1} \widehat{\bbeta}_{1}) \bX^\top_{1}\bigr\}^{-1} + \bigl\{\bX_{2} \bA(\bX^\top_{2} \widehat{\bbeta}_{2}) \bX^\top_{2}\bigr\}^{-1} \bigr]^{-1} \\
\bigl[ (\widehat{\bbeta}_{1} - \widehat{\bbeta}_{2}) (\widehat{\bbeta}_{1} - \widehat{\bbeta}_{2})^\top + \bigl\{\bX_{1} \bA(\bX^\top_{1} \widehat{\bbeta}_{1}) \bX^\top_{1}\bigr\}^{-1} \bigr].
\end{align*}

Across all simulations, we report 95\% confidence interval empirical coverage (CP) of $\widetilde{\bbeta}_{2}(\tilde{\lambda})$, mean $\tilde\lambda$ and its standard error, and empirical MSE (eMSE). \cite{Li-Cai-Li-2022, Chen-Owen-Shi, Zheng-etal} do not provide a method for constructing confidence intervals, prohibiting comparison of inferential properties of these methods to ours. While \cite{Tian-Feng} offer an alternative approach to construct confidence intervals for each element of $\bbeta_{2}$, their proposal is not designed to yield a point estimator.

We show that the mean squared error of $\widetilde{\bbeta}_{2}(\tilde\lambda)$ is smaller than that of $\widehat{\bbeta}_{2}$ and various competitors when $n_{2}$ is not too small. As expected, $\tilde\lambda$ is larger when $n_{2}$ or $\bdelta$ are smaller. We show that, when $n_{2}$ is large, $\mD_{1}$ is informative or $\bdelta$ is small, the bias of our estimator $\widetilde{\bbeta}_{2}(\tilde\lambda)$ is negligible and empirical coverage reaches nominal levels. This phenomenon is unique to our approach: our choice of $\tilde\lambda$, in contrast to competitors, guarantees that the bias is negligible across a range of practical settings. 

Finally, an additional simulation in Appendix \ref{a:multi-source} investigates the trade-offs of concatenation versus single-source transfer learning when more than one source data set is available. We summarize the key take-aways from this investigation in the discussion of Section \ref{s:conclusion}.

\subsection{Setting I: Varying sample size}
\label{ss:vi-1}

We study the performance of $\widetilde{\bbeta}_{2}(\lambda)$ in the linear model with varying degrees of likelihood information in $\mD_{1}$ relative to $\mD_{2}$. We vary $n_{1} \in \{50,100,500\}$, $n_{2} \in \{50,100,500\}$: for each $(n_{1},n_{2})$ pair, we simulate one data set $\mD_{1}=\{y_{i1}, \bX_{i1}\}_{i=1}^{n_{1}}$. For each simulated $\mD_{1}$, we generate $1000$ data sets $\mD_{2}=\{y_{i2},\allowbreak \bX_{i2}\}_{i=1}^{n_{2}}$. Features $\bX_{ij} \in \mathbb{R}^{11}$ consist of an intercept and ten continuous features independently generated from a standard Gaussian distribution. Outcomes are simulated from the Gaussian distribution with $\E(Y_{i2})= \bX^\top_{ij} \bbeta_j$ and $\V(Y_{ij})=1$. True parameter values are set to $\bbeta_{1}=(1, -1.8, 2.6, 1.4, -3.6, 3.5, 2.4, -3.3, \allowbreak 1.8, -3.4, 2.8)$, $\bdelta=(0.2,0.1,0.2,-0.1,0.1,-0.1,0.2,0.2,0.2,-0.1,0.1)$ and $\bbeta_{2}=\widehat{\bbeta}_{1}+\bdelta$. MLEs $\widehat{\bbeta}_{1}$ and corresponding values $\bbeta_{2}$ are reported in Appendix~\ref{a:estimates}. For each $\mD_{2}$, we compute $\widetilde{\bbeta}_{2}(\lambda)$ and $\widehat{\bbeta}_{2}(\bW_{\lambda})$ for $\lambda$ a sequence of 500 evenly spaced values in $[0,5]$ for $n_{2}=50,100$ and $[0,0.2]$ for $n_{2}=500$.

\begin{figure}[H]
\includegraphics[width=\textwidth]{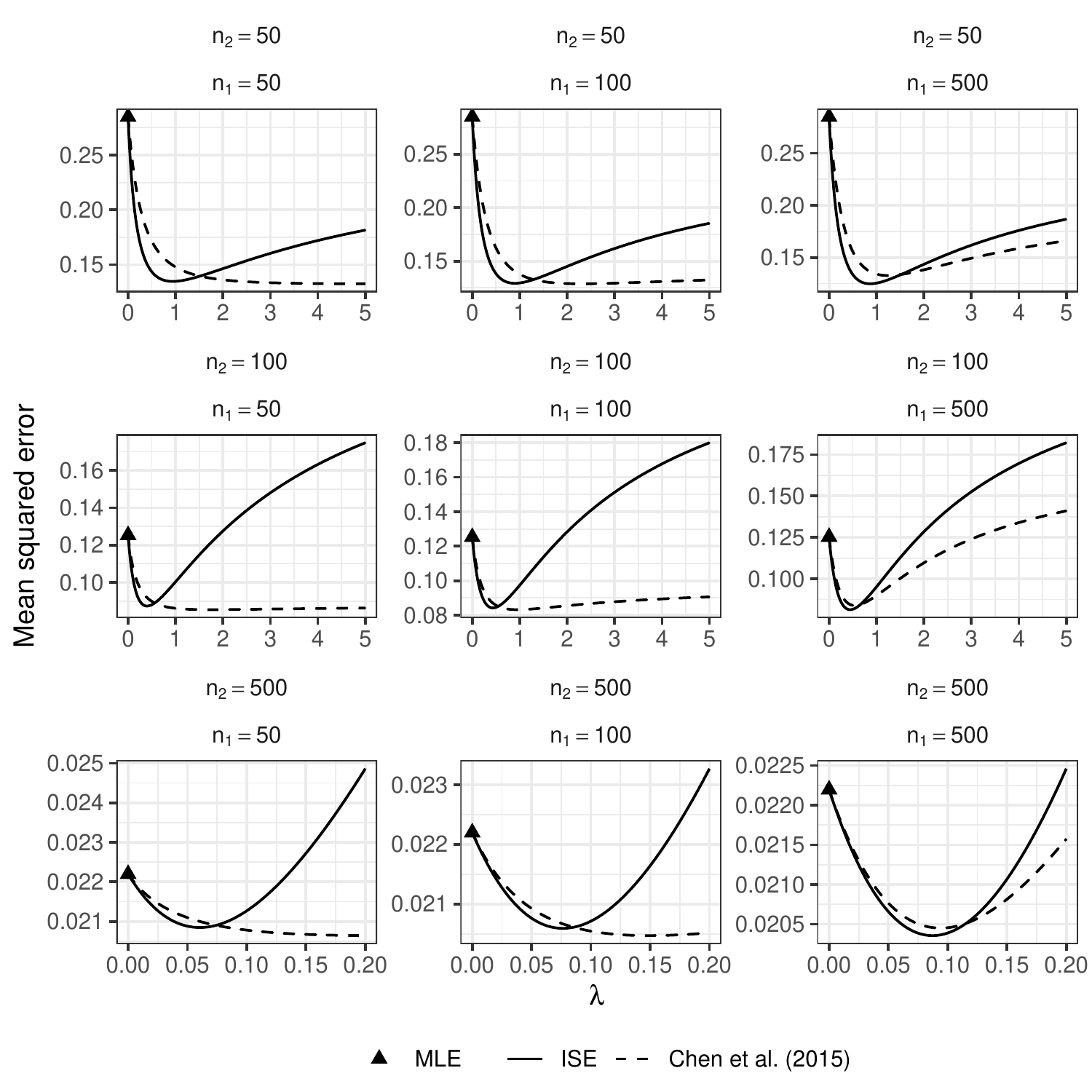}
\caption{eMSE of $\widetilde{\bbeta}_{2}(\lambda)$, $\widehat{\bbeta}_{2}(\bW_{\lambda})$ and $\widehat{\bbeta}_{2}$ over $1000$ simulated $\mD_{2}$ in Setting I.\label{f:MSE-SetI}}
\end{figure}

Empirical mean squared errors (eMSE) of $\widetilde{\bbeta}_{2}(\lambda)$, $\widehat{\bbeta}_{2}(\bW_{\lambda})$ and $\widehat{\bbeta}_{2}$ are depicted in Figure~\ref{f:MSE-SetI}. Both $\widetilde{\bbeta}_{2}(\lambda)$ and $\widehat{\bbeta}_{2}(\bW_{\lambda})$ have smaller eMSE than $\widehat{\bbeta}_{2}$ across all $(n_{1},n_{2})$ pairs. For fixed $n_{2}$, the eMSEs of $\widetilde{\bbeta}_{2}(\lambda)$ and $\widehat{\bbeta}_{2}(\bW_{\lambda})$ decrease modestly as $n_{1}$ increases. For fixed $n_{1}$, the eMSEs of $\widetilde{\bbeta}_{2}(\lambda)$ and $\widehat{\bbeta}_{2}(\bW_{\lambda})$ decrease much more drastically as $n_{2}$ increases. For fixed $n_{1}$, the minimum of $\mbox{eMSE}\{ \widetilde{\bbeta}_{2}(\lambda) \} $ is achieved at smaller $\lambda$ for larger $n_{2}$, with $\mbox{eMSE}\{ \widetilde{\bbeta}_{2}(\lambda)\} < \mbox{eMSE}(\widehat{\bbeta}_{2})$ for shrinking ranges of $\lambda$ values as $n_{2}$ increases. This is consistent with the intuition developed in Sections \ref{s:formulation} and \ref{s:theory}: little weight should be placed on $\mD_{1}$, regardless of how informative it is, when $\mD_{2}$ is highly informative. In contrast, $\widehat{\bbeta}_{2} (\bW_{\lambda})$ continues to place a large weight on $\mD_{1}$ even when $n_{2}$ is large. Theoretically, this results in substantial bias of the most efficient $\widehat{\bbeta}_{2}(\bW_{\lambda})$, while the bias of the most efficient $\widetilde{\bbeta}_{2}(\lambda)$ will become negligible, allowing us to construct confidence intervals that reach nominal levels (discussed below). Practically, this suggests $\widehat{\bbeta}_{2}(\bW_{\lambda})$ struggles with predicting how much efficiency gain can be expected from incorporating inference in $\mD_{1}$ when $n_{2}$ is large. Our approach is advantageous when proposing practical guidelines for improving inference in data integration.

A relevant question that the theory in Section~\ref{s:theory} is unable to answer is how large $\lambda^\star$ is as a function of $n_{2}$.  That is, we have an upper bound that is $O(n_{2}^{-1/2})$ and a lower bound that is $O(n_{2}^{-1})$, but what about $\lambda^\star$ itself?  For an empirical check, this simulation provides realizations of $\tilde\lambda$ for a range of $n_{2}$ values, so if we regress $\log\tilde\lambda$ against $\log n_{2}$, then the estimated slope coefficient would provide some information about how fast $\lambda^\star$ vanishes with $n_{2}$.  Based on the results summarized in Table~\ref{t:SetI}, the estimated slope is $-0.77$, with a 95\% confidence interval $(-0.78, -0.76)$, which is well within the range $[-1.0, -0.5]$ that we were looking in. This provides empirical support of the claim that $\lambda^\star$ is strictly between our lower and upper bounds, which are $O(n_{2}^{-1})$ and $O(n_{2}^{-1/2})$, respectively.

We show in Table \ref{t:SetI} that our proposed $\tilde\lambda$ results in a reduced eMSE by reporting eMSE of $\widetilde{\bbeta}_{2}(\tilde\lambda)$, $\widehat{\bbeta}_{2}(\bW_{\widehat{\lambda}})$, $\widehat{\bbeta}_{2}^{TG}$, $\widehat{\bbeta}_{2}^{TL}$, $\widehat{\bbeta}_{2}^p$ and $\widehat{\bbeta}_{2}$ (Monte Carlo standard errors, MCse, of eMSE are reported in Table \ref{t:SetI-sd} of the Appendix). Aside from the pooled MLE, our estimator achieves the smallest eMSE when $n_{2} \leq n_{1}$. The Trans-GLM approach of \cite{Tian-Feng} apparently gives too much weight to $\mD_{1}$ and suffers from negative transfer in most settings. We report the mean value of $\tilde\lambda$ and the median (over the $11$ features) CP of $\widetilde{\bbeta}_{2}(\tilde\lambda)$, $\widehat{\bbeta}_{2}^p$ and $\widehat{\bbeta}_{2}$ over the 1000 simulated $\mD_{2}$ in Table \ref{t:SetI-CP}. Coverage of $\widetilde{\bbeta}_{2}(\tilde\lambda)$ reaches the nominal 95\% coverage when $n_{2}=500$. As discussed above, this is a consequence of the negligible bias when $n_{2}$ is
\noindent large or one of $\tilde\lambda$, $\bdelta$ is sufficiently small. The bias incurred by assuming $\bbeta_{1}=\bbeta_{2}$ leads to inflated eMSE of the pooled MLE and severe under-coverage when $n_{2} > n_{1} >50$. Of course, the value of $\bdelta$ is not known in practice and the use of the pooled analysis runs the substantial risk of introducing a large bias that is not offset by a reduction in variance.

\begin{table}[t]
\centering
\begin{tabular}{rrrrrrrr}
\toprule
$n_{1}$ & $n_{2}$ & \multicolumn{6}{c}{eMSE$\cdot 10^{2}$} \\
& & $\widetilde{\bbeta}_{2}(\tilde\lambda)$ & $\widehat{\bbeta}_{2}( \bW_{\widehat{\lambda}})$ & $\widehat{\bbeta}_{2}^{TG}$ & $\widehat{\bbeta}_{2}^{TL}$ & $\widehat{\bbeta}_{2}^p$ & $\widehat{\bbeta}_{2}$ \\ 
\midrule
\multirow{3}{*}{$50$} & $50$ & ${17.1}$ & ${17.2}$ & $32.4$ & $27.7$ & ${13.5}$ & $28.5$ \\ 
& 100 & ${9.69}$ & ${9.56}$ & $20.8$ & $12.6$ & ${8.80}$ & $12.5$ \\ 
& $500$ & ${2.16}$ & ${2.11}$ & ${2.23}$ & ${2.21}$ & ${2.13}$ & ${2.22}$ \\ 
\midrule
\multirow{3}{*}{$100$} & $50$ & ${16.5}$ & ${16.8}$ & $27.3$ & $24.2$ & ${14.5}$ & $28.5$ \\ 
& $100$ & ${9.34}$ & ${9.33}$ & $15.8$ & ${11.0}$ & ${9.74}$ & ${12.5}$ \\ 
& $500$ & ${2.12}$ & ${2.09}$ & ${2.26}$ & ${2.10}$ & ${2.33}$ & ${2.22}$ \\ 
\midrule
\multirow{3}{*}{$500$} & $50$ & ${16.0}$ & ${16.6}$ & ${23.6}$ & $24.2$ & ${21.6}$ & $28.5$ \\ 
& $100$ & ${9.06}$ & ${9.24}$ & ${12.7}$ & ${11.0}$ & $18.2$ & ${12.5}$ \\ 
& $500$ & ${2.10}$ & ${2.10}$ & ${2.38}$ & ${2.12}$ & $6.93$ & ${2.22}$ \\ 
\bottomrule
\end{tabular}
\caption{Setting I: eMSE of $\widetilde{\bbeta}_{2}(\tilde\lambda)$, $\widehat{\bbeta}_{2}(\bW_{\widehat{\lambda}})$, $\widehat{\bbeta}_{2}^{TG}$, $\widehat{\bbeta}_{2}^{TL}$, $\widehat{\bbeta}_{2}^p$, $\widehat{\bbeta}_{2}$.\label{t:SetI}}
\end{table}

\begin{table}[t]
\centering
\begin{tabular}{rrrrrr}
\toprule
$n_{1}$ & $n_{2}$ & $\tilde\lambda$ $(\mbox{s.e.})$ & \multicolumn{3}{c}{CP} \\
& & & $\widetilde{\bbeta}_{2}(\tilde\lambda)$ & $\widehat{\bbeta}_{2}^p$ & $\widehat{\bbeta}_{2}$ \\ 
\midrule
\multirow{3}{*}{$50$} & $50$ & $4.87\cdot 10^{-1}$ ($3.81\cdot 10^{-1}$) & $88.4$ & $94.9$ & $94.5$ \\ 
& 100 & $3.05\cdot 10^{-1}$ ($1.93\cdot 10^{-1}$) & $90.8$ & $93.5$ & $94.4$ \\ 
& $500$ & $6.16\cdot 10^{-2}$ ($1.90\cdot 10^{-2}$) & $94.5$ & $95.4$ & $94.9$ \\ 
\midrule
\multirow{3}{*}{$100$} & $50$ & $4.97\cdot 10^{-1}$ ($3.45\cdot 10^{-1}$) & $88.1$ & $93.3$ & $94.5$ \\ 
& $100$ & $3.31\cdot 10^{-1}$ ($1.74\cdot 10^{-1}$) & $88.7$ & $90.3$ & $94.4$ \\ 
& $500$ & $7.62\cdot 10^{-2}$ ($1.92\cdot 10^{-2}$) & $93.5$ & $93.2$ & $94.9$ \\ 
\midrule
\multirow{3}{*}{$500$} & $50$ & $4.89\cdot 10^{-1}$ ($3.07\cdot 10^{-1}$) & $89.7$ & $24.7$ & $94.5$ \\ 
& $100$ & $3.37\cdot 10^{-1}$ ($1.62\cdot 10^{-1}$) & $91.7$ & $39.6$ & $94.4$ \\ 
& $500$ & $8.50\cdot 10^{-2}$ ($1.87\cdot 10^{-2}$) & $93.9$ & $69.1$ & $94.9$ \\ 
\bottomrule
\end{tabular}
\caption{Setting I: mean $\tilde\lambda$ (s.e.), \%CP of $\widetilde{\bbeta}_{2}(\tilde\lambda)$, $\widehat{\bbeta}_{2}^{TL}$, $\widehat{\bbeta}_{2}^p$, $\widehat{\bbeta}_{2}$.\label{t:SetI-CP}}
\end{table}

\subsection{Setting II: Varying feature information}
\label{ss:vi-2}

We study the performance of $\widetilde{\bbeta}_{2}(\lambda)$ in the logistic model with varying degrees of likelihood information in $\mD_{1}$ relative to $\mD_{2}$ when $\bdelta=\bzero$ and when $\bdelta \neq \bzero$. We fix $(n_{1},n_{2})=(500,500)$ and vary $\bX_{1}\bX^\top_{1}$ by simulating two data sets $\mD_{1}$: one with large $\bX_{1} \bX^\top_{1}$ and one with small $\bX_{1} \bX^\top_{1}$. For each simulated $\mD_{1}$, we generate $1000$ data sets $\mD_{2}=\{y_{i2},\bX_{i2}\}_{i=1}^{n_{2}}$. Features $\bX_{ij} \in \mathbb{R}^{5}$ consist of an intercept and four continuous features independently generated from $\mN(0,0.75^2)$ and $\mN(0,3^2)$ distributions for $\bX_{1} \bX^\top_{1}$ small and large respectively. Outcomes are simulated with mean $\E(Y_{i2})=\expit ( \bX^\top_{ij} \bbeta_j)$ from the Bernoulli distribution. True parameter values are set to $\bbeta_{1}=(1, -1.8, -1.2, 1.6, 0.2)$ and $\bbeta_{2}=\widehat{\bbeta}_{1}+\bdelta$. We let $\bdelta$ take two values: $\bdelta=(0, 0.25, 0, -0.25, 0.25)$ ($\bdelta \neq \bzero)$ and $\bdelta = \bzero$. MLEs $\widehat{\bbeta}_{1}$ and corresponding values $\bbeta_{2}$ are reported in Appendix~\ref{a:estimates}. For each $\mD_{2}$, we compute $\widetilde{\bbeta}_{2}(\lambda)$ for $\lambda$ a sequence of 100 evenly spaced values in $[0,7]$.

\begin{figure}[H]
\includegraphics[width=\textwidth]{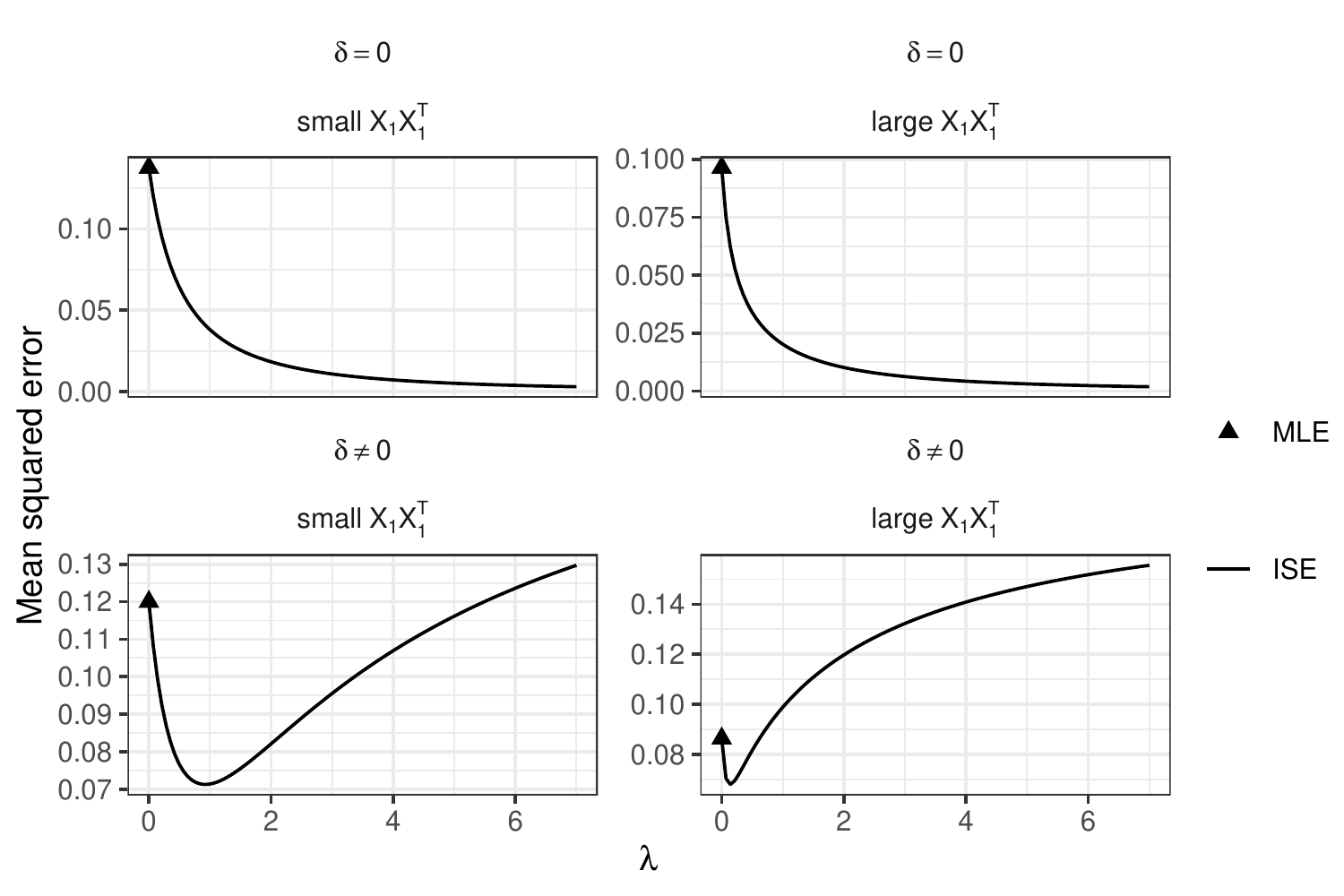}
\caption{eMSE of $\widetilde{\bbeta}_{2}(\lambda)$ and $\widehat{\bbeta}_{2}$ over $1000$ simulated $\mD_{2}$ in Setting II.\label{f:MSE-SetII}}
\end{figure}

Empirical mean squared error (eMSE) of $\widetilde{\bbeta}_{2}(\lambda)$ is depicted in Figure \ref{f:MSE-SetII}. When $\bdelta = \bzero$, the smallest $\mbox{eMSE}\{ \widetilde{\bbeta}_{2}(\lambda)\}$ is achieved for $\lambda$ only marginally smaller when $\mD_{1}$ is less informative ($\bX_{1} \bX^\top_{1}$ is small): our estimator uses more or less the same amount of information from $\mD_{1}$
\noindent when $\bdelta=\bzero$. When $\bdelta \neq \bzero$, however, the smallest $\mbox{eMSE}\{ \widetilde{\bbeta}_{2}(\lambda)\}$ is achieved for smaller $\lambda$ when $\mD_{1}$ is informative:  when the two data sets are different and $\mD_{1}$ provides sufficient information to discriminate between the two, our approach does not use as much information from $\mD_{1}$. 

We show in Table \ref{t:SetII} that $\tilde\lambda$ results in a reduced eMSE by reporting eMSE of $\widetilde{\bbeta}_{2}(\tilde\lambda)$, $\widehat{\bbeta}_{2}(\bW)$, $\widehat{\bbeta}_{2}^{TG}$, $\widehat{\bbeta}_{2}^p$ and $\widehat{\bbeta}_{2}$ (MCse of eMSE are reported in Table \ref{t:SetII-sd} of the Appendix). Aside from the pooled MLE, our estimator achieves the smallest eMSE across all settings. As a sample mean, the eMSE is approximately normally distributed given the large number of simulations ($1000$ Monte Carlo replicates). An approximate 95\% confidence interval for the MSE can be constructed following the formula $\mbox{eMSE} \pm 1.96 ~\mbox{MCse}$ and two-sample z-tests show that, for example, the MSEs for $\widehat{\bbeta}^{TG}_2$ and $\widehat{\bbeta}_2$ are not statistically significantly different at level $0.05$ when $\bdelta=\bzero$. When $\bX_{1} \bX^\top_{1}$ is large, $\mD_{1}$ provides a substantial amount of information on $\bbeta_{2}$. When $\bdelta=\bzero$, this information is used by turning the dial parameter $\lambda$ to a large value, whereas when $\bdelta \neq \bzero$, the dial should not be turned very far. In other words, the likelihood in $\mD_{1}$ is ``sharp'', and the distance between $f^{(n_{1})}_{1}(\by_{1}; \widehat{\bbeta}_{1}, \gamma_{1})$ and $f^{(n_{1})}_{1}(\by_{1}; \bbeta_{2}, \gamma_{1})$ is either small (when $\bdelta = \bzero$) or large (when $\bdelta \neq \bzero$). On the other hand, when $\bX_{1} \bX^\top_{1}$ is small, there is insufficient information in $\mD_{1}$ to distinguish $\bbeta_{2}$ and $\widehat{\bbeta}_{1}$, and more weight can be placed on $\mD_{1}$ with little loss of efficiency when $\bdelta \neq \bzero$. Our estimator uses the abundance of information in $\mD_{1}$ differently based on the value of $\bdelta$. In Table \ref{t:SetII-CP}, we also report the mean value of $\tilde\lambda$ and the median (over the $5$ features) CP of $\widetilde{\bbeta}_{2}(\lambda)$, $\widehat{\bbeta}_{2}^p$ and $\widehat{\bbeta}_{2}$ over the 1000 simulated $\mD_{2}$. Empirical coverage of $\widetilde{\bbeta}_{2}(\tilde\lambda)$ reaches the nominal level in all but the most difficult setting, when $\mD_{1}$ and $\mD_{2}$ are dissimilar and $\mD_{1}$ is less informative. The CP and eMSE of the pooled MLE reveals over- and under-coverage when $\bdelta=\bzero$ ($\bbeta_{2}=\widehat{\bbeta}_{1}$) and $\bdelta \neq \bzero$ ($\bbeta_{2}\neq \widehat{\bbeta}_{1}$), respectively.

\begin{table}[t]
\centering
\begin{tabular}{rrrrrrr}
\toprule
$\bdelta$ & $\bX_{1} \bX^\top_{1}$ & \multicolumn{5}{c}{eMSE$\cdot 10$}\\
& & $\widetilde{\bbeta}_{2}(\tilde\lambda)$ & $\widehat{\bbeta}_{2}(\bW)$ & $\widehat{\bbeta}_{2}^{TG}$ & $\widehat{\bbeta}_{2}^p$ & $\widehat{\bbeta}_{2}$ \\ 
\midrule
\multirow{2}{*}{$=\bzero$} & large & ${0.310}$ & $0.751$ & $0.971$ & ${0.202}$ & $0.964$ \\ 
& small & ${0.464}$ & $1.11$ & $1.32$ & ${0.381}$ & $1.38$ \\ 
\midrule
\multirow{2}{*}{$\neq \bzero$} & large & ${0.773}$ & ${0.805}$ & ${0.930}$ & ${0.989}$ & ${0.864}$ \\ 
& small & ${0.849}$ & ${1.06}$ & $1.20$ & ${0.714}$ & $1.20$ \\
\bottomrule
\end{tabular}
\caption{Setting II: eMSE of $\widetilde{\bbeta}_{2}(\tilde\lambda)$, $\widehat{\bbeta}_{2}(\bW)$, $\widehat{\bbeta}_{2}^{TG}$, $\widehat{\bbeta}_{2}^p$, $\widehat{\bbeta}_{2}$.\label{t:SetII}}
\end{table}

\begin{table}[t]
\centering
\begin{tabular}{rrrrrr}
\toprule
$\bdelta$ & $\bX_{1} \bX^\top_{1}$ & $\tilde\lambda$ $(\mbox{s.e.})$ & \multicolumn{3}{c}{CP}\\
& & & $\widetilde{\bbeta}_{2}(\tilde\lambda)$ & $\widehat{\bbeta}_{2}^p$ & $\widehat{\bbeta}_{2}$ \\ 
\midrule
\multirow{2}{*}{$=\bzero$} & large & $2.83$ ($6.42$) & $95.1$ & $99.7$ & $95.3$ \\ 
& small & $2.27$ ($3.20$) & $95.3$ & $99.0$ & $95.5$ \\ 
\midrule
\multirow{2}{*}{$\neq \bzero$} & large & $0.120$ ($0.266$) & $93.1$ & $77.1$ & $94.7$ \\ 
& small & $0.720$ ($0.476$) & $80.5$ & $93.8$ & $94.9$ \\
\bottomrule
\end{tabular}
\caption{Setting II: mean $\tilde\lambda$ (s.e.), \%CP of $\widetilde{\bbeta}_{2}(\tilde\lambda)$, $\widehat{\bbeta}_{2}^p$, $\widehat{\bbeta}_{2}$.\label{t:SetII-CP}}
\end{table}

\subsection{Setting III: Varying parameter distance}
\label{ss:distance}

In Setting III, we study the performance of $\widetilde{\bbeta}_{2}(\lambda)$ in the logistic model as a function of $\bdelta$ with both independent and correlated features. With independent features, we fix $(n_{1},n_{2})=(500,500)$ and generate one data set $\mD_{1}=\{y_{i1},\bX_{i1}\}_{i=1}^{n_{1}}$ with $\bbeta_{1}=(1, -0.5, 0.5)$, yielding $\widehat{\bbeta}_{1}=(0.966, -0.563, 0.551)$. Source features $\bX_{i1} \in \mathbb{R}^{3}$ consist of an intercept and two continuous features independently generated from a normal distribution with mean 0 and variance $4$. Target features $\bX_{i2} \in \mathbb{R}^{3}$ consist of an intercept and two continuous features independently generated from a standard normal distribution. The correlation between features in source and target data sets is $\rho=0$. With correlated features, we fix $(n_{1},n_{2})=(500,100)$ and generate one data set $\mD_{1}=\{y_{i1},\bX_{i1}\}_{i=1}^{n_{1}}$ with $\bbeta_{1}=(1, -0.5, 0.5)$, yielding $\widehat{\bbeta}_{1}=(0.881,-0.516,0.571)$. Source and target features $\bX_{ij} \in \mathbb{R}^{3}$ consist of an intercept and two continuous features generated from a multivariate Gaussian distribution with mean $0$, correlation $\rho=0.4$ and variance $1$. With independent and correlated features, we set $\bbeta_{2}=\widehat{\bbeta}_{1}+\bdelta$ and let $\bdelta$ take values $\bdelta_{1}=(0,0,0)$, $\bdelta_{2}=(0,1,0)$ and $\bdelta_3=(0,2,0)$. For each value $\bdelta_j$, we generate $1000$ data sets $\mD_{2}=\{y_{i2},\bX_{i2}\}_{i=1}^{n_{2}}$. Outcomes are simulated with $\E(Y_{ij})=\expit( \bX^\top_{ij} \bbeta_j)$ from the Bernoulli distribution, with features generated as in the source data set for independent and correlated features, respectively. For each $\mD_{2}$, we compute $\widetilde{\bbeta}_{2}(\lambda)$ for $\lambda$ a sequence of 100 evenly spaced values in $[0,10]$, $[0,0.02]$ and $[0,0.01]$ for $\bdelta_{1}, \bdelta_{2}, \bdelta_3$ respectively with independent features, and $\lambda$ a sequence of 100 evenly spaced values in $[0,10]$, $[0,0.6]$ and $[0,0.2]$ for $\bdelta_{1}, \bdelta_{2}, \bdelta_3$ respectively with correlated features. 

\begin{figure}[H]
\includegraphics[width=\textwidth]{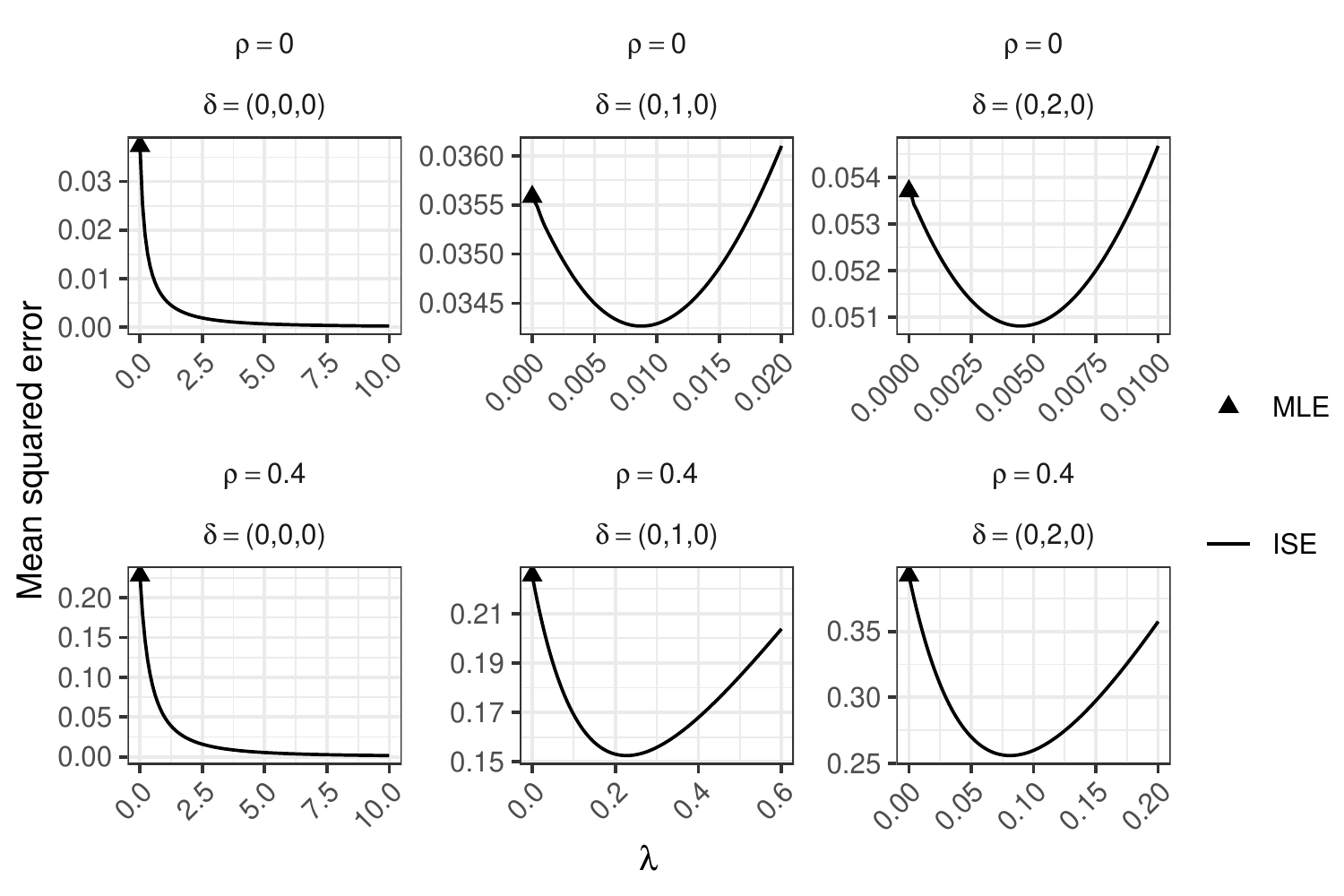}
\caption{eMSE of $\widetilde{\bbeta}_{2}(\lambda)$ and $\widehat{\bbeta}_{2}$ over $1000$ simulated $\mD_{2}$ in Setting III.\label{f:MSE-SetIII}}
\end{figure}

Empirical mean squared error (eMSE) of $\widetilde{\bbeta}_{2}(\lambda)$ is depicted in Figure \ref{f:MSE-SetIII}. The smallest 
\noindent $\mbox{eMSE}\{\widetilde{\bbeta}_{2}(\lambda)$ is achieved for smaller values of $\lambda$ when $\bdelta$ is larger. We show in Table \ref{t:SetIII} that $\tilde\lambda$ results in a reduced eMSE by reporting eMSE of $\widetilde{\bbeta}_{2}(\tilde\lambda)$, $\widehat{\bbeta}_{2}(\bW)$, $\widehat{\bbeta}_{2}^{TG}$, $\widehat{\bbeta}_{2}^p$ and $\widehat{\bbeta}_{2}$ (Monte Carlo standard errors of eMSE are reported in Table \ref{t:SetIII-sd} of the Appendix). Our estimator's efficiency gain remains robust to substantial differences between $\mD_{1}$ and $\mD_{2}$, and our proposed $\tilde\lambda$ consistently minimizes eMSE over competitors. In Table \ref{t:SetIII-CP}, we report mean value of $\tilde\lambda$ and the median (over the $2$ features) CP of $\widetilde{\bbeta}_{2}(\tilde\lambda)$, $\widehat{\bbeta}_{2}^p$ and $\widehat{\bbeta}_{2}$ over the 1000 simulated $\mD_{2}$. Empirical coverage of $\widetilde{\bbeta}_{2}(\tilde\lambda)$ reaches the nominal level in all settings. The pooled MLE shows inflated eMSE when $\bdelta \neq \bzero$. With independent features, $\bbeta_{2}$ is over- and under-covered when $\bdelta=\bzero$ and $\bdelta \neq \bzero$, respectively, whereas it is over-covered for all values of $\bdelta$ with correlated features.

\begin{table}[t]
\centering
\begin{tabular}{rrrrrr}
\toprule
$\bdelta$ & \multicolumn{5}{c}{eMSE$\cdot 10^{2}$}\\
& $\widetilde{\bbeta}_{2}(\tilde\lambda)$ & $\widehat{\bbeta}_{2}(\bW)$ & $\widehat{\bbeta}_{2}^{TG}$ & $\widehat{\bbeta}_{2}^p$ & $\widehat{\bbeta}_{2}$ \\ 
\midrule
\multicolumn{5}{c}{(i) independent features, $\rho=0$}\\
$(0,0,0)$ & $1.45$ & $2.55$ & $4.21$ & $0.572$ & $3.73$ \\ 
$(0,1,0)$ & $3.53$ & $3.84$ & $3.64$ & $58.3$ & ${3.56}$ \\ 
$(0,2,0)$ & $5.21$ & $30.2$ & ${5.22}$ & $259$ & $5.37$ \\
\multicolumn{5}{c}{(ii) correlated features, $\rho=0.4$}\\
$(0,0,0)$ & ${8.34}$ & $16.9$ & $28.1$ & $0.546$ & $22.8$ \\ 
$(0,1,0)$ & ${18.9}$ & $29.5$ & $21.0$ & $67.1$ & $22.5$ \\ 
$(0,2,0)$ & $31.6$ & $146$ & ${31.0}$ & $287$ & $39.2$ \\ 
\bottomrule
\end{tabular}
\caption{Setting III: eMSE of $\widetilde{\bbeta}_{2}(\tilde\lambda)$, $\widehat{\bbeta}_{2}(\bW)$, $\widehat{\bbeta}_{2}^{TG}$, $\widehat{\bbeta}_{2}^p$, $\widehat{\bbeta}_{2}$.\label{t:SetIII}}
\end{table}

\begin{table}[t]
\centering
\begin{tabular}{rrrrr}
\toprule
$\bdelta$ & $\tilde\lambda$ $(\mbox{s.e.})$ & \multicolumn{3}{c}{CP}\\
& & $\widetilde{\bbeta}_{2}(\tilde\lambda)$ & $\widehat{\bbeta}_{2}^p$ & $\widehat{\bbeta}_{2}$ \\ 
\midrule
\multicolumn{5}{c}{(i) independent features, $\rho=0$}\\
$(0,0,0)$ & $3.43$ ($26.2$) & $94.3$ & $100$ & $94.8$ \\ 
$(0,1,0)$ & $5.56 \cdot 10^{-3}$ ($1.18\cdot 10^{-3}$) & $94.7$ & $89.4$ & $95.5$ \\ 
$(0,2,0)$ & $2.05\cdot 10^{-3}$ ($1.93\cdot 10^{-4}$) & $94.9$ & $7.20$ & $95.0$ \\
\multicolumn{5}{c}{(ii) correlated features, $\rho=0.4$}\\
$(0,0,0)$ & $3.22$ ($9.96$) & $94.9$ & $100$ & $95.1$ \\ 
$(0,1,0)$ & $0.197$ ($0.162$) & $95.2$ & $99.9$ & $95.6$ \\ 
$(0,2,0)$ & $0.0483$ ($0.0183$) & $95.2$ & $99.3$ & $95.5$ \\ 
\bottomrule
\end{tabular}
\caption{Setting III: mean $\tilde\lambda$ (s.e.), \%CP of $\widetilde{\bbeta}_{2}(\tilde\lambda)$, $\widehat{\bbeta}_{2}^p$, $\widehat{\bbeta}_{2}$.\label{t:SetIII-CP}}
\end{table}

\subsection{Setting IV: Robustness to model misspecification}
\label{ss:robustness}

In Setting IV, we study the robustness of the gain in MSE of $\widetilde{\bbeta}_{2}(\lambda)$ in the linear model under model misspecification. We fix $(n_{1},n_{2})=(500,500)$ and generate three data sets $\mD_{1}=\{y_{i1},\bX_{i1}\}_{i=1}^{n_{1}}$ constructed as follows:
\begin{enumerate}[label=(\roman*),topsep=0pt,itemsep=-1ex,partopsep=1ex,parsep=1ex]
\item (Cauchy) outcomes are generated as $y_{i1}=\bX_{i1} \bbeta_{1} + \epsilon_{ij}$ with $\epsilon_{ij}$ independent standard Cauchy random variables;
\item (dropped $Z$) outcomes $y_{i1}$ are generated from a Gaussian distribution with mean $\bX_{i1} \bbeta_{1} + Z_{i1}$ and variance $\V(Y_{i1})=1$, where $Z_{i1}$ is generated from a standard Gaussian distribution and acts as an unmeasured confounder; 
\item ($X^2$) outcomes are generated as $y_{i1}=\mathbf{diag} (\bX_{i1} \bX^\top_{i1}) \bbeta_{1} + \epsilon_{ij}$, i.e., the relationship between outcome and squared features is linear, with $\epsilon_{ij}$ independent standard Gaussian random variables.
\end{enumerate}
True parameter values are set to $\bbeta_{1}=(1, -1.8, 2.6, 1.4, -3.6, 3.5, 2.4, -3.3, 1.8, \allowbreak -3.4, 2.8, 1)$ and $\bbeta_{2}=\bbeta_{1}+(0, 0.1, 0, -0.1, 0.1, -0.1, 0, 0, 0, -0.1, 0.1)$, which differs from previous settings; we prefer to use $\bbeta_{1}$ to define $\bbeta_{2}$ since $\widehat{\bbeta}_{1}$ will be substantially biased in these misspecified settings. MLEs $\widehat{\bbeta}_{1}$ and corresponding values $\bdelta$ are reported in Appendix~\ref{a:estimates}. For each simulated $\mD_{1}$, we generate $1000$ data sets $\mD_{2}=\{y_{i2},\bX_{i2}\}_{i=1}^{n_{2}}$ as in Setting I. For each $\mD_{2}$, we compute $\widetilde{\bbeta}_{2}(\lambda)$ and $\widehat{\bbeta}_{2}(\bW_{\lambda})$ for $\lambda$ a sequence of 100 evenly spaced values in $[0,0.02]$, $[0,0.4]$ and $[0,5\times 10^{-4}]$ for misspecifications (i), (ii) and (iii) respectively.

\begin{figure}[H]
\includegraphics[width=\textwidth]{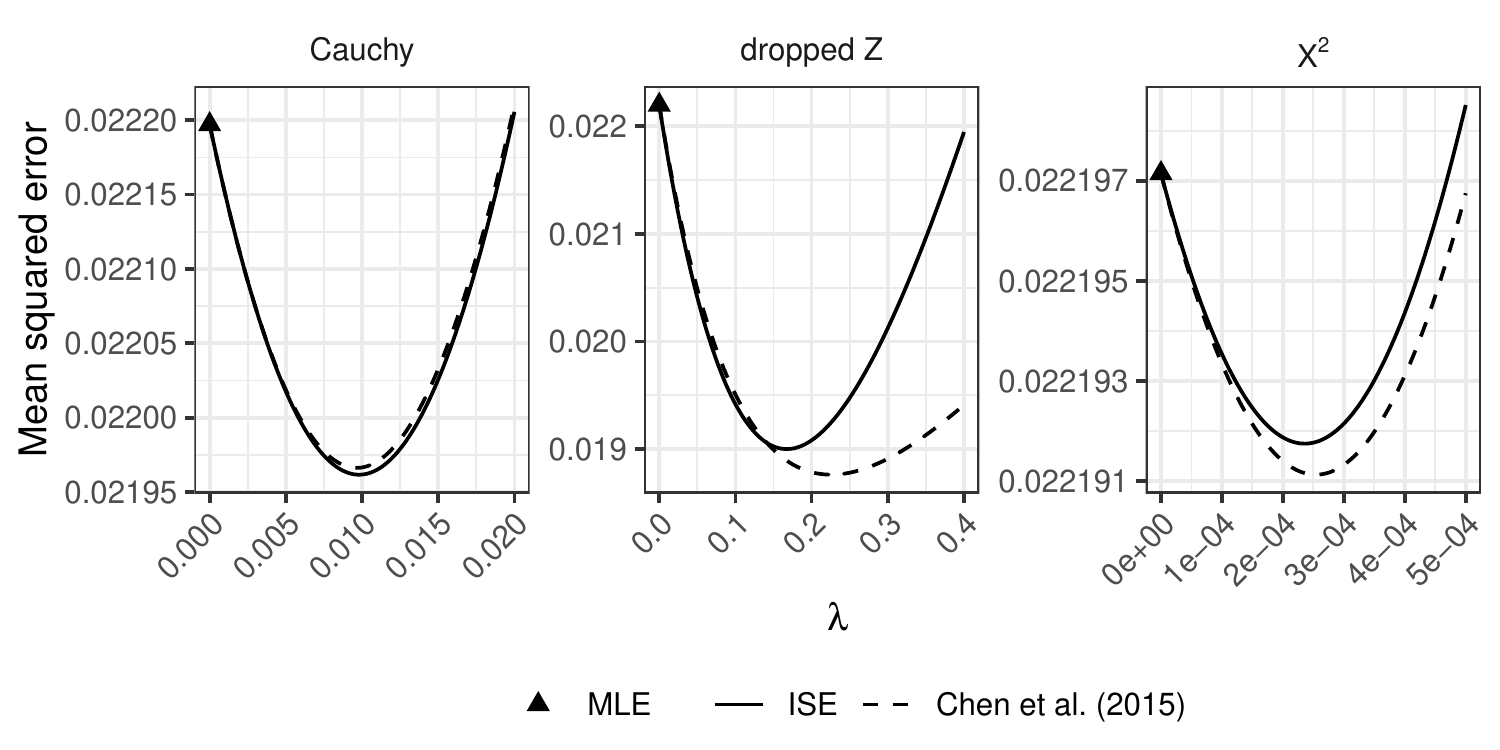}
\caption{eMSE of $\widetilde{\bbeta}_{2}(\lambda)$, $\widehat{\bbeta}_{2}(\bW_{\lambda})$ and $\widehat{\bbeta}_{2}$ over $1000$ simulated $\mD_{2}$ in Setting IV.\label{f:MSE-SetIV}}
\end{figure}

Empirical mean squared errors (eMSE) of $\widetilde{\bbeta}_{2}(\lambda)$, $\widehat{\bbeta}_{2}(\bW_{\lambda})$ and $\widehat{\bbeta}_{2}$ are depicted in Figure~\ref{f:MSE-SetIV}. 
\noindent The estimator proposed by \cite{Chen-Owen-Shi} appears slightly more efficient than our approach when the mean model is misspecified (dropped $Z$ and $X^2$). We show in Table \ref{t:SetIV} that $\tilde\lambda$ results in a reduced eMSE by reporting eMSE of $\widetilde{\bbeta}_{2}(\tilde\lambda)$, $\widehat{\bbeta}_{2}(\bW_{\widehat{\lambda}})$, $\widehat{\bbeta}_{2}^{TG}$, $\widehat{\bbeta}_{2}^{TL}$, $\widehat{\bbeta}_{2}^p$ and $\widehat{\bbeta}_{2}$ (Monte Carlo standard errors of eMSE are reported in Table \ref{t:SetI-sd} of the Appendix). Our estimator's gain in efficiency is robust when $f^{(n_{1})}_{1}$ is misspecified (Cauchy misspecification): as with Stein shrinkage, our use of the KL divergence in the shrinkage is robust to the misspecification of the density $f^{(n_{1})}_{1}$. Moreover, our approach is robust to unmeasured confounders in $\mD_{1}$ (dropped $Z$ misspecification), encouraging the use of $\widetilde{\bbeta}_{2}(\tilde\lambda)$ when $\mD_{2}$ only measures a subset of the features measured in $\mD_{1}$. Finally, when mean models in $\mD_{1}$ and $\mD_{2}$ are sizeably different ($X^2$ misspecification), our approach reverts to the MLE in $\mD_{2}$ with a very small amount of shrinkage. In Table \ref{t:SetIV-CP}, we also report mean value of $\tilde\lambda$ and the median (over the $11$ features) CP of $\widetilde{\bbeta}_{2}(\tilde\lambda)$, $\widehat{\bbeta}_{2}^p$ and $\widehat{\bbeta}_{2}$ over the 1000 simulated $\mD_{2}$. Empirical coverage of $\widetilde{\bbeta}_{2}(\tilde\lambda)$ reaches the nominal level in all settings. The Trans-Lasso estimator occasionally has smaller eMSE than our estimator; it does not, however, suggest a path to inference, so that the reduction in the MSE may give false confidence in the strength of a result that may not in fact replicate. The CP and eMSE of the pooled MLE show inflated eMSE in all settings, with over-coverage, nominal coverage and under-coverage in the Cauchy, dropped $Z$, and $X^2$ cases, respectively.

\begin{table}[t]
\centering
\begin{tabular}{rrrrrrr}
\toprule
Case & \multicolumn{6}{c}{eMSE$\cdot 10^{2}$}\\
& $\widetilde{\bbeta}_{2}(\tilde\lambda)$ & $\widehat{\bbeta}_{2}(\bW_{\widehat{\lambda}})$ & $\widehat{\bbeta}_{2}^{TG}$ & $\widehat{\bbeta}_{2}^{TL}$ & $\widehat{\bbeta}_{2}^p$ & $\widehat{\bbeta}_{2}$ \\ 
\midrule
Cauchy & ${2.20}$ & ${2.20}$ & ${2.22}$ & ${2.15}$ & $60.0$ & ${2.22}$ \\ 
dropped $Z$ & ${1.99}$ & ${1.96}$ & ${2.37}$ & ${1.93}$ & $3.68$ & ${2.22}$ \\ 
$X^2$ & ${2.22}$ & ${2.22}$ & ${2.22}$ & ${2.29}$ & $230 \cdot 10$ & ${2.22}$ \\ 
\bottomrule
\end{tabular}
\caption{Setting IV: eMSE of $\widetilde{\bbeta}_{2}(\tilde\lambda)$, $\widehat{\bbeta}_{2}(\bW_{\widehat{\lambda}})$, $\widehat{\bbeta}_{2}^{TG}$, $\widehat{\bbeta}_{2}^{TL}$, $\widehat{\bbeta}_{2}^p$, $\widehat{\bbeta}_{2}$. \label{t:SetIV}}
\end{table}

\begin{table}[t]
\centering
\begin{tabular}{rrrrr}
\toprule
Case & $\tilde\lambda$ $(\mbox{s.e.})$ & \multicolumn{3}{c}{CP}\\
& & $\widetilde{\bbeta}_{2}(\tilde\lambda)$ & $\widehat{\bbeta}_{2}^p$ & $\widehat{\bbeta}_{2}$ \\ 
\midrule
Cauchy & $9.49\cdot 10^{-3}$ ($1.44\cdot 10^{-3}$) & $94.7$ & $100$ & $94.7$ \\ 
dropped $Z$ & $1.58 \cdot 10^{-1}$ ($4.66\cdot 10^{-2}$) & $93.6$ & $95.3$ & $93.6$ \\ 
$X^2$ & $2.44\cdot 10^{-4}$ ($3.45\cdot 10^{-5}$) & $95.1$ & $0$ & $95.1$ \\ 
\bottomrule
\end{tabular}
\caption{Setting IV: mean $\tilde\lambda$ (s.e.), \%CP of $\widetilde{\bbeta}_{2}(\tilde\lambda)$, $\widehat{\bbeta}_{2}^p$, $\widehat{\bbeta}_{2}$. \label{t:SetIV-CP}}
\end{table}

\section{Real data analysis}
\label{s:data-analysis}

We present a real-data illustration of our information-based shrinkage estimator $\widetilde{\bbeta}_{2}(\lambda)$ in an analysis of data from the multi-center eICU Collaborative Research Database \citep{eICU} maintained by the Philips eICU Research Institute. The database consists of data from patients admitted to one of several intensive care units (ICUs) throughout the continental United States in 2014 and 2015. Information on data access and pre-processing is provided in Appendix~\ref{S:data}.

Our analysis focuses on estimating the association between death and baseline features for patients suffering from cardiac arrest upon admission to the ICU. Inclusion criteria are described in Appendix~\ref{S:data}. Our first population consists of ICU admissions at hospitals located in the western United States, and the data set $\mD_{1}$ consists of cardiac arrest ICU admissions at these 34 hospitals, $n_{1}=575$.  (The source data set $\mD_{1}$ is a concatenation of source data from 34 hospitals, as discussed in Section~\ref{ss:setup}; our justification for this concatenation is that these records are from hospitals in the same geographic region, so substantial heterogeneity between hospitals would not be expected.) The data set $\mD_{2}$ consists of cardiac arrest ICU admissions at one hospital in the southern United States, $n_{2}=145$. The probability of death $\mu_{ij}$ for participant $i$ in data set $\mD_j$ is modeled by
\begin{align*}
\log\frac{\mu_{ij}}{1-\mu_{ij}} &= \beta_{0j} + \beta_{1j} \, \text{sex}_{ij} + \beta_{2j} \, \text{age}_{ij} + \beta_{3j} \, \text{ethnicity}_{ij} + \beta_{4j} \, \text{BMI}_{ij},
\end{align*}
with $\text{sex}_{ij}$, $\text{age}_{ij}$, $\text{ethnicity}_{ij}$ and $\text{BMI}_{ij}$ the sex ($1$ for male, $0$ for female), age (in years), ethnicity ($1$ for African American, $0$ for Caucasian) and body mass index (in $\text{kg}/\text{m}^2$) of participant $i$ in data set $\mD_j$, respectively.

Maximum likelihood estimation based on $\mD_{1}$ reveals that sex and age are significantly associated with death following cardiac arrest at hospitals in the west at level $0.05$, with estimated effects $\widehat{\beta}_{1j}=-0.39$ (standard error, se $=0.18$) and $\widehat{\beta}_{2j}=0.022$ (se $=0.0073$). The MLEs and approximate standard errors for $\mD_{1}$ and $\mD_{2}$ are displayed in Table~\ref{t:data-MLE}. While associations between the outcome and age and sex are in the same direction in $\mD_{1}$ and $\mD_{2}$, the estimates in the latter are not significant. 

\begin{table}[t]
\centering
\begin{tabular}{rrrr}
\toprule
Covariate & West hospitals $\mD_{1}$ & South hospital $\mD_{2}$\\
\midrule
intercept & $-1.3$ ($0.65)$ & $-1.1$ ($1.3$) \\
sex & $\mathbf{-0.39}$ ($0.18)$ & $-0.17$ ($0.36$) \\
age & $\mathbf{0.022}$ ($0.0073)$ & $0.017$ ($0.015$) \\
ethnicity & $-0.29$ ($0.39)$ & $0.27$ ($0.36$) \\
BMI & $-0.00069$ ($0.011)$ & $0.019$ ($0.024$) \\
\bottomrule
\end{tabular}
\caption{Maximum likelihood estimates (standard errors) of feature effects in the data sets $\mD_{1}$ and $\mD_{2}$ of cardiac arrest ICU admissions at the hospitals in the west and the hospital in the south, respectively. Bolded estimates are significant at level $0.05$. \label{t:data-MLE}}
\end{table}

\begin{figure}[H]
\centering
\includegraphics[width=\textwidth]{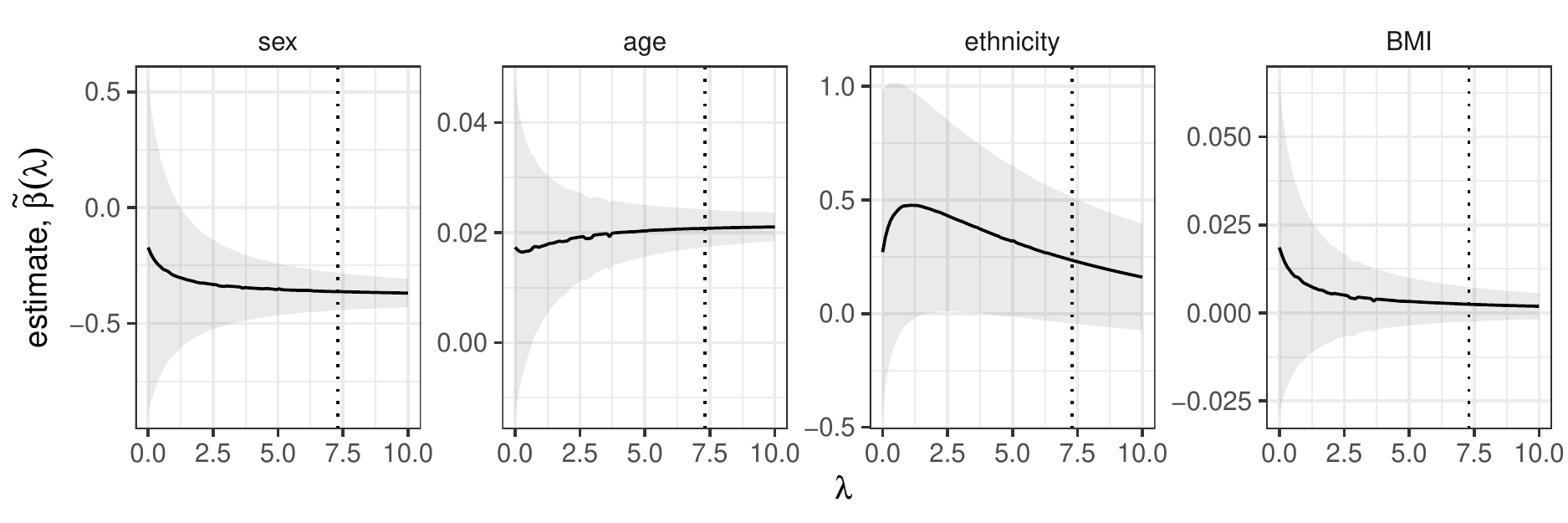}
\caption{ISE $\widetilde{\bbeta}_{2}(\lambda)$ of the sex, age, ethnicity and BMI effects in $\mD_{2}$, the data set of cardiac arrest ICU admissions at the southern U.S.~hospital. The dotted vertical line gives the value of the minimizer of the estimated aMSE, $\tilde{\lambda}$. \label{f:eICU-estimates}}
\end{figure}

Our proposed information-shrinkage approach can be used to borrow information from $\mD_{1}$ to improve the efficiency of estimates in $\mD_{2}$. Trace plots of the ISE $\widetilde{\bbeta}_{2}(\lambda)$ with shaded 95\% (pointwise) confidence bands are plotted in Figure \ref{f:eICU-estimates} for a sequence of $\lambda \in [0,10]$. The minimizer $\tilde{\lambda}$ of the estimated aMSE is also depicted there and shows that our proposed estimator $\widetilde{\bbeta}_{2}(\tilde{\lambda})$ leads to statistically significant estimates of the sex and age effects in $\mD_{2}$, at level $0.05$. The 95\% confidence intervals based on equation \eqref{e:MLE-CI} for the sex and age effects are $(-0.44,-0.28)$ and $(0.017,0.024)$ respectively.

We also report estimates of the intercept, sex, age, ethnicity and BMI effects for the estimators $\widetilde{\bbeta}_{2}(\tilde{\lambda})$, $\widehat{\bbeta}_{2}(\bW)$, $\widehat{\bbeta}_{2}^{TG}$, $\widehat{\bbeta}_{2}^p$ and $\widehat{\bbeta}_{2}$ in Table \ref{t:data-comparison}. The estimator $\widehat{\bbeta}_{2}^{TG}$ uses the source detection algorithm in \cite{Tian-Feng} and finds that data from all 34 hospitals in the west are transferable. Our proposed ISE appears close to the pooled and target-only estimates. In contrast, $\widehat{\bbeta}_{2}(\bW)$ exhibits substantial bias in the age and BMI effects, whereas the debiasing step in $\widehat{\bbeta}_{2}^{TG}$ appears to fail and returns effect estimates that are zero for the ethnicity and BMI covariates. In fairness, $\widehat{\bbeta}_{2}^{TG}$ is designed for a different setting than ours, as discussed in Section \ref{ss:t-GLM}. Specifically, the estimator is designed to improve target data set predictions in high-dimensional transfer learning problems with multiple heterogeneous sources, and so the algorithm may be over-designed for our setting. We can nonetheless compare predictive metrics of the estimators. Area under the receiver operating characteristic curve (AUC) (standard error) are $57.6\%$ ($4.79\%$), $58.6\%$ ($4.74\%$), $56.7\%$ ($4.81\%$), $57.4\%$ ($4.81\%$) and $59.6\%$ ($4.77\%$) using estimates $\widetilde{\bbeta}_{2}(\tilde{\lambda})$, $\widehat{\bbeta}_{2}(\bW)$, $\widehat{\bbeta}_{2}^{TG}$, $\widehat{\bbeta}_{2}^p$ and $\widehat{\bbeta}_{2}$, respectively. While not designed for prediction, our approach does not give worse predictive performance---as measured by AUC---in comparison to $\widehat{\bbeta}_{2}(\bW)$, $\widehat{\bbeta}_{2}^p$, and  $\widehat{\bbeta}_{2}^{TG}$, despite the latter being designed specifically to improve prediction. Further, our information shrinkage estimate $\widetilde{\bbeta}_{2}(\tilde{\lambda})$ appears less biased than $\widehat{\bbeta}_{2}(\bW)$ and $\widehat{\bbeta}_{2}^{TG}$, facilitating estimation and inference.

\begin{table}[t]
\centering
\begin{tabular}{rrrrrr}
\toprule
& $\widetilde{\bbeta}_{2}(\tilde{\lambda})$ & $\widehat{\bbeta}_{2}(\bW)$ & $\widehat{\bbeta}_{2}^{TG}$ & $\widehat{\bbeta}_{2}^p$ & $\widehat{\bbeta}_{2}$ \\ 
\midrule
intercept & $-1.3$ & $-0.94$ & $0.15$ & $-1.2$ & $-1.1$ \\ 
sex & $-0.36$ & $-0.17$ & $-0.085$ & $-0.35$ & $-0.17$ \\ 
age & $0.021$ & $19$ & $0.0070$ & $0.020$ & $0.017$ \\ 
ethnicity & $0.23$ & $0.063$ & $0$ & $0.36$ & $0.27$ \\ 
BMI & $0.0025$ & $9.1$ & $0$ & $0.0038$ & $0.019$ \\
\bottomrule
\end{tabular}
\caption{Estimates of feature effects in the target data set $\mD_{2}$ of cardiac arrest ICU admissions at the hospital in the south. \label{t:data-comparison}}
\end{table}

\section{Conclusion}
\label{s:conclusion}

This paper proposes a new approach to data integration, focusing not on {\em if} two data sets should be integrated, but on the {\em extent} to which inference based on the second data set could be made more efficient by leveraging information in the first.  This {\em extent} is controlled by a dial parameter $\lambda$. We proposed a $\lambda$-dependent estimator in generalized linear models, offered a data-driven choice of $\lambda$, and established theoretical support for our claims that this new estimator is more efficient than that which ignores the first data set, even when the underlying populations differ.  Our new, more nuanced data integration framework not only matches statistical intuition on notions of informativeness but empirically out-performs the state-of-the-art techniques. Our proposed approach yields relatively simple parameter estimates, yet performs powerfully in practice and is intuitively related to a broad scope of inferential frameworks. 

A unique feature of our approach is the ability to essentially ignore $\mD_{1}$ when $n_{2}$ is large. This is important. We have already discussed that our approach is protected from negative transfer. Almost equally important, our approach says something useful about when to expect efficiency gains by incorporating inference from another data set. Intuition tells us that the contribution from another data set should vanish as the sample size $n_{2}$ grows when $\bdelta \neq \bzero$, which is typical in practice; this intuition underpins the foundations of efficiency results for maximum likelihood estimation. This is borne out by our information-shrinkage approach because we are accounting for the relative {\em information} in $\mD_{1}$ with respect to $\mD_{2}$. It is somewhat surprising, however, that this is not the case for all shrinkage estimators, as evidenced in Section \ref{s:empirical}. While the estimator in \cite{Chen-Owen-Shi} has a smaller asymptotic variance than our proposed estimator when $n_{2}$ is large, theirs is asymptotically biased and does not offer a path to inference. In contrast, we guarantee that, asymptotically, the bias is vanishing and approximately valid confidence intervals are available and can be computed. 

Another key point is that the bias is controlled by choosing $\lambda^\star$ {\em at the right rate}, namely between $O(n^{-1}_{2})$ and $O(n^{-1/2}_{2})$. This rate guarantees a gain in efficiency, even when the bias is negligible. The  need for data integration is driven by the high cost of data collection, and a gain in efficiency, however small, can make the difference between a null finding and a new discovery.

In contrast to data fusion discussed in Section \ref{ss:fusion}, our approach limits its consideration to data sets $\mD_{1}$ and $\mD_{2}$ as the units of integration, rather than features. While elements of $\widetilde{\bbeta}_{2}(\lambda)$ will shrink unevenly depending on feature information, we do not allow the user to differentially shrink estimates using feature-specific dial parameters. It is unclear how to incorporate feature-specific dial parameters because the penalty in \eqref{e:const-opt} is data-dependent, i.e. a summation over independent units $i=1, \ldots, n_{1}$ rather than parameters.

Our focus on generalized linear models is motivated by a desire to balance practical utility and the insights we can obtain into the efficiency gains expected from integrating data sets.  Our approach can surely be applied in problems that fall outside the generalized linear model framework, but the theory may be substantially complicated by, e.g., the lack of a closed form for the KL divergence. The substantial and practically useful intuition developed in the present paper, for example on the rate of $\lambda^\star$, should be helpful in guiding extensions to more general models in future work.

We envision that our proposed method can be extended to the setting with high-dimensional predictors ($p > n_{2}$). A special case of interest considers the setting where the correctly specified model in the source $\mD_{1}$ depends on a set of $q>n_{1}$ features, but the correctly specified model in the target $\mD_{2}$ depends only on a subset of $p<n_{2}$ of these features. Then, our information-based shrinkage estimation can be carried out using only these $p$ features without modification. This is due to the fact that the model in $\mD_{1}$ need not be correctly specified for us to borrow information from $\mD_{1}$. Theorems 1 and 2 and the simulation results in Setting IV of Section \ref{s:empirical} support this solution, although further investigation is required to determine how much efficiency can be gained when more than one feature differs between $\mD_{2}$ and $\mD_{1}$. More generally, extensions to our work may consider the setting where the correctly specified models in both data sets $\mD_{1}, \mD_{2}$ depend on $p$ features, $p>n_{2} \vee n_{1}$. We envision that our approach can be extended by regularizing the estimator $\widetilde{\bbeta}_{2}(\lambda)$ of $\bbeta_{2}$. It is doubtful that the bias introduced by this regularization decreases quickly enough to yield valid inference, and so we expect to lose many of the appealing properties we derived for our estimator $\widetilde{\bbeta}_{2}(\tilde{\lambda})$. Additional theoretical study is required to investigate the implementation of debiasing strategies in our data integrative setting.

As seen in Section \ref{s:data-analysis}, the multiple source setting is very realistic. When the number of source data sets is small, an investigator may compare our proposed source concatenation approach to, for example, one that uses each source individually, one at a time. Appendix \ref{a:multi-source} shows that our suggested concatenation strategy is safe in the sense that it is theoretically guaranteed to protect against negative transfer, and efficient in that it generally outperforms the integration of a single source data set. We also argue that concatenation is a reasonable \textit{de facto} approach in all but the most extreme cases, e.g., where one source data set obviously matches the target data much better than the other source data sets.  A further question centers around the optimal source configuration to minimize the MSE of the estimator. In Appendix \ref{a:multi-source}, we consider the idea of choosing a source data configuration by minimizing our estimated MSE.  That is, in the linear model case, define the function 
\begin{align*}
\mS \mapsto n_{\mT}^{-1} \widehat{\sigma}^2_{\mT} \, \mbox{trace} \{ \bS^{-2}(\tilde{\lambda}) \bG_{\mT} \} + \tilde{\lambda}^2 \mbox{trace} \{ \bG_{\mS} \bS^{-2}(\tilde{\lambda}) \bG_{\mS} \widehat{\bdelta^2} \},
\end{align*}
where the input $\mS$ could be the full concatenation of source data sets, any one of the individual source data sets, or some intermediate configuration consisting of a concatenation of only some of the source data sets, and the right-hand side is the minimized estimated MSE for this source configuration. The same idea can be applied in the generalized linear model case, with an expression slightly more complicated than that above. Then the idea is to choose $\widehat\mS$ as the minimizer of the above objective function. 
We show in Appendix \ref{a:multi-source} that this strategy reliably selects the most efficient source configuration. The minimized estimated MSE can form the basis for this comparison by quantifying the effectiveness of the data integration: a smaller value signals that a large amount of information can be safely borrowed and should therefore be preferred. As with all model selection approaches, the downstream inference does not account for the model selection and so an investigator should be wary of over-interpreting confidence intervals/standard errors. A natural extension of our concatenation proposal would incorporate features of data-driven heterogeneous source detection methods \citep{Li-Cai-Li-2022, Tian-Feng} to perform inference in the multi-source setting.

\section*{Acknowledgments}

The authors thank the reviewer and associate editor for their valuable feedback, which led to an improved manuscript.  The second author's work is partially supported by the U.~S.~National Science Foundation, grant SES--2051225. 

\appendix

\section{Proofs}
\label{S:proofs}

\subsection{Proof of Theorem 1}

We start with the mean squared error of $\widetilde{\bbeta}_{2}(\lambda)$. First,
\begin{align*}
\{ \widetilde{\bbeta}_{2}(\lambda) - \bbeta_{2} \}^\top \{ \widetilde{\bbeta}_{2}(\lambda) - \bbeta_{2} \} 
=
\{ \bG_{2} ( \widehat{\bbeta}_{2} - \bbeta_{2} ) + \lambda \bG_{1} ( \widehat{\bbeta}_{1} - \bbeta_{2} ) \}^\top ( \bG_{2} + \lambda \bG_{1} )^{-2}\\
\{ \bG_{2} ( \widehat{\bbeta}_{2} - \bbeta_{2} ) +\lambda \bG_{1} ( \widehat{\bbeta}_{1} - \bbeta_{2} ) \}.
\end{align*}
Define $\bM=\bG^{-1/2}_{2} \bG_{1} \bG^{-1/2}_{2}$ and $\bR=\bG^{-1}_{1} \bG_{2}$. Then
\begin{equation}
\begin{split}
\mbox{MSE}\{\widetilde{\bbeta}_{2}(\lambda)\}
&=\E_{\bbeta_{2}} [ \{ \widetilde{\bbeta}_{2}(\lambda) - \bbeta_{2} \}^\top \{ \widetilde{\bbeta}_{2}(\lambda) - \bbeta_{2} \} ] \\
&=
n_{2}^{-1} \sigma_{2}^2 \mbox{trace} \{ \bS^{-2}(\lambda) \bG_{2} \} + \lambda^2 \bdelta^\top \bG_{1} \bS^{-2}(\lambda) \bG_{1}  \bdelta\\
&=
n_{2}^{-1} \sigma_{2}^2 \mbox{trace} \{ ( \bI_p + \lambda \bM )^{-2} \bG^{-1}_{2}\} + \lambda^2 \mbox{trace} \{ (\bR + \lambda \bI_p)^{-2} \bdelta \bdelta^\top \}\\
&=v(\lambda) + b(\lambda).
\label{e:MSE1}
\end{split}
\end{equation}
The term $v(\lambda)$ is the sum of the variances of the (weighted) least squares estimates from $\mD_{1}$ and $\mD_{2}$. On the other hand, $b(\lambda)$ is the squared distance from $\widehat{\bbeta}_{1}$ to $\bbeta_{2}$ and will be $\bzero$ when $\lambda=0$ or $\widehat{\bbeta}_{1} = \bbeta_{2}$. Clearly, $v(\lambda)$ is monotonically decreasing and $b(\lambda)$ is monotonically increasing. Note that, as expected,
\begin{align*}
v(0)+b(0)&=n_{2}^{-1} \sigma_{2}^2 \mbox{trace} (\bG^{-1}_{2}) 
=\mbox{MSE}(\widehat{\bbeta}_{2}).
\end{align*}
When $\bdelta = \bzero$, we therefore have $\mbox{MSE} \{ \widetilde{\bbeta}_{2}(\lambda)\} < \mbox{MSE}(\widehat{\bbeta}_{2})$ for all $\lambda>0$. Throughout the rest of the proof, we consider the case when $\bdelta \neq \bzero$. 

Since $\bG_{2}$ is symmetric positive definite, it has an invertible and symmetric square root denoted by $\bG^{1/2}_{2}$. From
\begin{align*}
\bG^{1/2}_{2} \bR \bG^{-1/2}_{2}
&= \bG^{1/2}_{2} \bG^{-1}_{1} \bG^{1/2}_{2}=\bM^{-1},
\end{align*}
we have that $\bR$ and $\bM^{-1}$ are similar matrices. By symmetry and, therefore, orthogonal diagonalizability of the latter, the former is diagonalizable, and $\bR$ and $\bM^{-1}$ share eigenvalues. Let $\kappa_{r}>0$, $r=1,\ldots,p$, the eigenvalues of $\bM^{-1}$ and $\bR$ in decreasing order. Denote by $\bKappa=\mathbf{diag} \{\kappa_{r}\}_{r=1}^p$ and $\bP$ the matrices of eigenvalues and eigenvectors of $\bM^{-1}$, with $\bP^\top=\bP^{-1}$ such that $\bM^{-1}=\bP^\top \bKappa \bP$. Then the mean squared error of $\widetilde{\bbeta}_{2}(\lambda)$ can be rewritten as
\begin{align*}
\mbox{MSE} \{ \widetilde{\bbeta}_{2}(\lambda) \} &= n_{2}^{-1} \sigma_{2}^2 \mbox{trace} \{ ( \bI_p + \lambda \bKappa^{-1} )^{-2} \bP \bG^{-1}_{2} \bP^\top \} + \mbox{trace} \{ (\lambda^{-1} \bR + \bI_p )^{-2} \bdelta \bdelta^\top \} .
\end{align*}
We see that $\mbox{MSE} \{ \widetilde{\bbeta}_{2}(\lambda) \}$ is monotonically decreasing on an interval $[0, \lambda^\star]$ and monotonically increasing on an interval $[\lambda^\star,\infty)$ for some $\lambda^\star>0$. Our aim is to find a bound on $\lambda^\star$, so that $\mbox{MSE} \{ \widetilde{\bbeta}_{2}(\lambda) \} < \mbox{MSE} (\widehat{\bbeta}_{2})$ for all $\lambda$ less than that bound.

Denote by $(\delta^2_r)_{r=1}^p$ the values of $\mathbf{diag} (\bdelta \bdelta^\top)$ sorted in increasing order, i.e. $0 \leq \delta^2_{1} \leq \ldots \leq \delta^2_p $. Let $g_{r2}>0$, $r=1, \ldots, p$, the eigenvalues of $\bG_{2}$ in increasing order. Then the bias and variance functions satisfy
\begin{align*}
b(\lambda) &\leq \lambda^2 \sum \limits_{r=1}^p \frac{\delta^2_r}{(\kappa_{r}+\lambda)^2}\\
v(\lambda) &\leq \frac{\sigma^2_{2}}{n_{2}} \sum \limits_{r=1}^p \frac{g^{-1}_{r2}}{( 1 + \lambda \kappa^{-1}_{r} )^2} =\frac{\sigma^2_{2}}{n_{2}} \sum \limits_{r=1}^p \frac{g^{-1}_{r2} \kappa^2_{r} }{ (\kappa_{r} + \lambda)^2},
\end{align*}
using Von Neumann's trace inequality and the fact that the eigenvalues of $\bA^\top \bA$ are the squared eigenvalues of $\bA$ for a positive definite matrix $\bA$. Therefore
\begin{align*}
\mbox{MSE}\{\widetilde{\bbeta}_{2}(\lambda)\}&=v(\lambda) + b(\lambda)
\leq \sum \limits_{r=1}^p \frac{\sigma^2_{2} g^{-1}_{r2} \kappa^2_{r} + \lambda^2 n_{2} \delta^2_r}{n_{2} (\kappa_{r} +\lambda)^2}.
\end{align*}
Denote the upper bound on the right-hand side by $U(\lambda)$. We proceed to show that there exists a $\lambda >0$ such that $\mbox{MSE}\{ \widetilde{\bbeta}_{2}(\lambda)\} \leq U(\lambda) < \mbox{MSE}(\widehat{\bbeta}_{2})$. 

Towards this, we examine the monotonicity of $U(\lambda)$. First,
\begin{align*}
&\frac{\partial}{\partial \lambda} U(\lambda)
=2 \sum \limits_{r=1}^p  \frac{\kappa_{r} (\lambda n_{2} \delta^2_r - \sigma^2_{2} g^{-1}_{r2} \kappa_{r})}{n_{2}(\kappa_{r} + \lambda)^3}.
\end{align*}
From
\begin{align*}
\lim \limits_{\lambda \rightarrow 0^+} \frac{\partial}{\partial \lambda} U(\lambda) &= - \frac{2\sigma^2_{2}}{n_{2}} \sum \limits_{r=1}^p \frac{ g^{-1}_{r2} }{\kappa_{r}} <0,
\end{align*}
the upper bound $U(\lambda)$ immediately decreases as $\lambda$ moves away from $0$. To find a range of $\lambda$ such that $\mbox{MSE}\{ \widetilde{\bbeta}_{2}(\lambda)\} \leq U(\lambda) < \mbox{MSE}(\widehat{\bbeta}_{2})$, it is therefore sufficient to find the range of $\lambda$ over which $U(\lambda)$ is decreasing, due to the fact that, at $\lambda=0$,
\begin{align*}
\mbox{MSE}\{\widetilde{\bbeta}_{2}(0)\} = U(0) = \mbox{MSE}(\widehat{\bbeta}_{2}).
\end{align*}
The range of $\lambda$ such that $U(\lambda)$ is decreasing corresponds to the range of $\lambda$ values for which the derivative of $U(\lambda)$ is negative. This, in turn, corresponds to the range of $\lambda$ satisfying
\begin{align*}
\lambda \sum_{r=1}^p \frac{\kappa_{r} \delta^2_r}{(\kappa_{r}+\lambda)^3} < \frac{\sigma^2_{2}}{n_{2}} \sum_{r=1}^p \frac{g^{-1}_{r2} \kappa^2_{r}}{(\kappa_{r}+\lambda)^3} .
\end{align*}
This is satisfied by
\begin{align}
\label{e:lambda-bound1}
\lambda < \frac{\sigma^2_{2}}{n_{2}} \frac{\min_{r=1, \ldots, p}( \kappa_{r} g^{-1}_{r2} )}{ \max_{r=1, \ldots, p}( \delta^2_r )}.
\end{align}
Equation \eqref{e:lambda-bound1} gives a range of $\lambda$ values such that $\mbox{MSE}\{ \widetilde{\bbeta}_{2}(\lambda)\} < \mbox{MSE}(\widehat{\bbeta}_{2})$.

\subsection{Proof of Lemma 1}

First, observe that $-O(\bbeta; \lambda)$ is a convex function of $\bbeta \in \Theta$, where $\Theta \in \mathbb{R}^p$ is an open convex set. Define
\begin{align*}
o_i(\bbeta; \lambda)&=y_{i2} \bX^\top_{i2} \bbeta - b( \bX^\top_{i2} \bbeta) \\
&~~~~~~~~ - \frac{ \lambda }{n_{1}} \sum \limits_{i=1}^{n_{1}}\left\{ b^\prime(\bX^\top_{i1} \widehat{\bbeta}_{1}) (\bX^\top_{i1} \widehat{\bbeta}_{1}-\bX^\top_{i1}\bbeta) + b(\bX^\top_{i1}\bbeta) - b(\bX^\top_{i1} \widehat{\bbeta}_{1}) \right\},
\end{align*}
so that $-O(\bbeta; \lambda)=n_{2}^{-1}\sum_{i=1}^{n_{2}} -o_i(\bbeta; \lambda)$. By the law of large numbers, for each fixed $\bbeta \in \Theta$, $O(\bbeta; \lambda)$ converges in probability to
\begin{align*}
\lim \limits_{n_{2} \rightarrow \infty} \frac{1}{n_{2}} \sum \limits_{i=1}^{n_{2}} \E_{\bbeta_{2}} o_i(\bbeta; \lambda) &=\lim \limits_{n_{2} \rightarrow \infty} \frac{1}{n_{2}} \sum \limits_{i=1}^{n_{2}} \left\{ h(\bX^\top_{i2} \bbeta_{2}) \bX^\top_{i2} \bbeta - b(\bX^\top_{i2} \bbeta)\right\} \\
&\hspace*{-5em} - \frac{ \lambda }{n_{1}} \sum \limits_{i=1}^{n_{1}}\left\{ b^\prime(\bX^\top_{i1} \widehat{\bbeta}_{1}) (\bX^\top_{i1} \widehat{\bbeta}_{1}-\bX^\top_{i1}\bbeta) + b(\bX^\top_{i1}\bbeta) - b(\bX^\top_{i1} \widehat{\bbeta}_{1}) \right\}.
\end{align*}
By convexity, it follows from Lemma 1 in \cite{Hjort-Pollard} that the above convergence is uniform in $\bbeta$ over compact subsets $K \subset \Theta$.  Proofs are given in \cite{Andersen-Gill} or \cite{Pollard}. Consistency of $\widetilde{\bbeta}_{2}(\lambda)$ and the rate $\widetilde{\bbeta}_{2}(\lambda) - \bbeta^\star_{2}(\lambda) = O_p(n^{-1/2}_{2})$ follow directly from the conditions and Theorem 2.2 in \cite{Hjort-Pollard}.

\subsection{Proof of Lemma 2}

By a Taylor's expansion of $\bzero=\bPsi\{ \widetilde{\bbeta}_{2}(\lambda); \lambda\}$ around $\bbeta_{2}$, we obtain
\begin{align}
\bzero&=\bPsi(\bbeta_{2}; \lambda) - \bS(\bbeta_{2}; \lambda) \{ \widetilde{\bbeta}_{2}(\lambda) - \bbeta_{2} \} + O_p \{ \| \widetilde{\bbeta}_{2}(\lambda) - \bbeta_{2} \|^2 \} \nonumber \\
&=\bPsi(\bbeta_{2}; \lambda) - \bS(\bbeta_{2}; \lambda) \{ \widetilde{\bbeta}_{2}(\lambda) - \bbeta_{2} \} + O_p \{ \| \widetilde{\bbeta}_{2}(\lambda) - \bbeta^\star_{2}(\lambda) \|^2 + \| \bbeta^\star_{2}(\lambda) - \bbeta_{2} \|^2 \nonumber \\
&\hspace*{10em} + 2 \| \widetilde{\bbeta}_{2}(\lambda) - \bbeta^\star_{2}(\lambda) \| \| \bbeta^\star_{2}(\lambda) - \bbeta_{2} \| \}. \label{e:MLE-Taylor}
\end{align}
We bound the higher-order terms in the above expansion. By continuity of $\bPsi(\bbeta; \lambda)$, there exists a vector $\bc_{\lambda} \in \mathbb{R}^p$ between $\bbeta_{2}$ and $\bbeta^\star_{2}(\lambda)$ such that
\begin{align*}
\bbeta_{2} - \bbeta^\star_{2}(\lambda)&= -\bS^{-1}(\bc_{\lambda}; \lambda) \{ \bPsi(\bbeta_{2}; \lambda) - \bPsi(\bbeta^\star_{2}(\lambda); \lambda) \}.
\end{align*}
Recall that $\bbeta^\star_{2}(\lambda)$ is the unique solution to $\E_{\bbeta_{2}} \{\bPsi(\bbeta^\star_{2}(\lambda); \lambda)\} = \bzero$. Taking expectations,
\begin{align}
\label{e:MLE-min-1}
\bbeta_{2} - \bbeta^\star_{2}(\lambda) = -\bS^{-1}(\bc_{\lambda}; \lambda) \E_{\bbeta_{2}} \{ \bPsi(\bbeta_{2}; \lambda) \}.
\end{align}
We decompose $\bPsi(\bbeta; \lambda)=\bU_{2}(\bbeta) - \lambda \bP_{1}(\bbeta)$ into the difference of the score function $\bU_{2}(\bbeta)$ in $\mD_{2}$ and a (non-random) penalty term $\bP_{1}(\bbeta)$:
\begin{align*}
\bU_{2}(\bbeta) &= n_{2}^{-1} \bX_{2} \{ \by_{2} - \bmu_{2}(\bbeta) \}, \quad \bP_{1}(\bbeta) = n_{1}^{-1} \bX_{1} \{ \bmu_{1}(\bbeta) - \bmu_{1} (\widehat{\bbeta}_{1}) \}.
\end{align*}
Since $\bbeta_{2}$ is the true value of $\bbeta$ in $\mD_{2}$, i.e., $\E_{\bbeta_{2}} (\bY_{2})=\bX^\top_{2} \bbeta_{2}=\bmu_{2}(\bbeta_{2})$, clearly $\E_{\bbeta_{2}} \{ \bU_{2}(\bbeta_{2})\}=\bzero$. This implies 
\begin{equation}
\begin{split}
\label{e:MLE-min-2}
\E_{\bbeta_{2}} \left\{ \bPsi(\bbeta_{2}; \lambda) \right\}&=-\lambda n_{1}^{-1} \bX_{1} \{ h ( \bX^\top_{i1} \bbeta_{2} ) - h ( \bX^\top_{i1} \widehat{\bbeta}_{1} ) \}_{i=1}^{n_{1}}\\
&=-\lambda n_{1}^{-1} \bX_{1} \bDelta(\bbeta_{2})\\
&=-\lambda n_{1}^{-1} \bX_{1} \bA ( \bX^\top_{1} \bc_{2} ) \bX^\top_{1} \bdelta\\
&=-\lambda \bv_{1}(\bc_{2}) \bdelta,
\end{split}
\end{equation}
where the third line is by the mean value theorem with $\bc_{2} \in \mathbb{R}^p$ a vector between $\bbeta_{2}$ and $\widehat{\bbeta}_{1}$. Plugging \eqref{e:MLE-min-2} into \eqref{e:MLE-min-1} gives $\bbeta_{2} - \bbeta^\star_{2}(\lambda) = \lambda \bS^{-1}(\bc_{\lambda}; \lambda) \bv_{1}(\bc_{2}) \bdelta$. Taking the limit as $n_{2} \rightarrow \infty$ on both sides yields
\begin{align*}
\bbeta_{2} - \bbeta^\star_{2}(\lambda) = \lambda \left\{ \bv_{2}(\bc_{\lambda}) + \lambda \bv_{1}(\bc_{\lambda}) \right\}^{-1} \bv_{1}( \bc_{2}) \bdelta.
\end{align*}
By condition \ref{c:5}, the eigenvalues of $\bv_j(\bc_{\lambda})$ are bounded away from $0$, $j=1,2$. If $\lambda=O(n^{-1/2}_{2})$, then we obtain $\bbeta^\star_{2}(\lambda) - \bbeta_{2} = O(n^{-1/2}_{2})$. Plugging this rate into equation \eqref{e:MLE-Taylor} and using Lemma \ref{l:MLE-cons},
\begin{align*}
\bzero
&=\bPsi(\bbeta_{2}; \lambda) - \bS(\bbeta_{2}; \lambda) \{ \widetilde{\bbeta}_{2}(\lambda) - \bbeta_{2} \} + O_p ( n^{-1}_{2}) + O(n^{-1}_{2}) + O_p (n^{-1/2}_{2}) O(n^{-1/2}_{2}) \\
&=\bPsi(\bbeta_{2}; \lambda) - \bS(\bbeta_{2}; \lambda) \{ \widetilde{\bbeta}_{2}(\lambda) - \bbeta_{2} \} + O_p(n^{-1}_{2}) + O(n^{-1}_{2}).
\end{align*}
Rearranging gives
\begin{align}
\label{e:true-rate}
n_{2}^{1/2}\{ \widetilde{\bbeta}_{2}(\lambda) - \bbeta_{2} \}&= \bS^{-1}(\bbeta_{2}; \lambda) n_{2}^{1/2} \bPsi(\bbeta_{2}; \lambda) + O_p(n^{-1/2}_{2}) + O(n^{-1/2}_{2}).
\end{align}
We examine the asymptotic behavior of $n_{2}^{1/2} \bPsi(\bbeta_{2}; \lambda)$. For $i=1, \ldots, n_{2}$, define
\begin{align*}
\bpsi_i(\bbeta; \lambda)&=  \bX_{i2} \bigl\{ y_{i2} - h ( \bX^\top_{i2} \bbeta ) \bigr\} - \frac{\lambda}{n_{1}} \sum \limits_{j=1}^{n_{1}} \bX_{j1} \bigl\{ h ( \bX^\top_{j1} \bbeta ) - h ( \bX^\top_{j1} \widehat{\bbeta}_{1} ) \bigr\},
\end{align*}
such that $\bPsi(\bbeta;  \lambda)=n_{2}^{-1} \sum_{i=1}^{n_{2}} \bpsi_i(\bbeta; \lambda)$. Since
\begin{align*}
\lim \limits_{n_{2} \rightarrow \infty} \frac{1}{n_{2}} \sum \limits_{i=1}^{n_{2}} \V_{\bbeta_{2}} \{ \bpsi_i ( \bbeta; \lambda) \}&=\lim \limits_{n_{2} \rightarrow \infty} \frac{1}{n_{2}} \sum \limits_{i=1}^{n_{2}} \bX_{i2} \V_{\bbeta_{2}}\left\{ Y_{i2} - h\left( \bX^\top_{i2} \bbeta \right) \right\} \bX^\top_{i2}\\
&=\lim \limits_{n_{2} \rightarrow \infty} \frac{1}{n_{2}} d(\gamma_{2}) \bX_{2} \bA(\bX^\top_{2} \bbeta_{2}) \bX^\top_{2}\\
&=d(\gamma_{2}) \bv_{2} (\bbeta_{2}),
\end{align*}
by equation \eqref{e:MLE-min-2} and conditions \ref{c:4}--\ref{c:5},
\begin{align*}
n_{2}^{1/2} \{ \bPsi(\bbeta_{2}; \lambda) + \lambda n_{1}^{-1} \bX_{1} \bDelta(\bbeta_{2}) \} \stackrel{d}{\rightarrow} \mN \{ \bzero, d(\gamma_{2}) \bv_{2}(\bbeta_{2}) \}.
\end{align*}
In other words,
\begin{align*}
n_{2}^{1/2} \bPsi(\bbeta_{2}; \lambda) = -\lambda n_{1}^{-1} n_{2}^{1/2} \bX_{1} \bDelta(\bbeta_{2}) + d(\gamma_{2})^{1/2} \bv^{1/2}_{2}(\bbeta_{2}) \bZ + o_p(1),
\end{align*}
where $\bZ \sim \mN(\bzero, \bI_p)$. By equation \eqref{e:true-rate},
\begin{equation}
\begin{split}
\label{e:norm-bias}
n_{2}^{1/2} \{ \widetilde{\bbeta}_{2}(\lambda) - \bbeta_{2} \}
&=\bS^{-1}(\bbeta_{2}; \lambda) d(\gamma_{2})^{1/2} \bv^{1/2}_{2}(\bbeta_{2}) \bZ \\
&\hspace*{-6em}  - \lambda n_{1}^{-1} n_{2}^{1/2} \bS^{-1}(\bbeta_{2}; \lambda) \bX_{1} \bDelta(\bbeta_{2})  + o_p(1) + O_p(n^{-1/2}_{2}) + O(n^{-1/2}_{2}).
\end{split}
\end{equation}
As $n_{2}\rightarrow \infty$, using symmetry of $\bS(\bbeta; \lambda)$, it follows that
\begin{align*}
 n_{2}^{1/2} d(\gamma_{2})^{-1/2} \bS^\top(\bbeta_{2}; \lambda) \bv^{-1/2}_{2}(\bbeta_{2}) \bigl\{ \widetilde{\bbeta}_{2}(\lambda) - \bbeta_{2} + \lambda n_{1}^{-1} \bS^{-1}(\bbeta_{2}; \lambda) \bX_{1} \bDelta(\bbeta_{2}) \bigr\} 
\end{align*}
converges in distribution to $\mN ( \bzero, \bI_p )$, which proves the first claim. When $\lambda=o(n_{2}^{-1/2})$, the above display can be re-expressed as  
\[ n_{2}^{1/2} d(\gamma_{2})^{-1/2} \bS^\top(\bbeta_{2}; \lambda) \bv^{-1/2}_{2}(\bbeta_{2}) \bigl\{ \widetilde{\bbeta}_{2}(\lambda) - \bbeta_{2} \bigr\} + o(1). \]
Then the second claim follows from the first together with Slutsky's theorem.

\subsection{Proof of Theorem 2}

We start with the asymptotic mean squared error of $\widetilde{\bbeta}_{2}(\lambda)$. By \eqref{e:norm-bias} in the proof of Lemma~\ref{l:MLE-norm}, 
\begin{align*}
\mbox{aMSE} \{ \widetilde{\bbeta}_{2}(\lambda)\} &=d(\gamma_{2}) \mbox{trace} \{ \bs^{-2}( \bbeta_{2}; \lambda) \bv_{2}(\bbeta_{2}) \} + n_{2} \lambda^2 \bdelta^\top \bv_{1}(\bc_{2}) \bs^{-2}(\bbeta_{2}; \lambda) \bv_{1}(\bc_{2}) \bdelta,
\end{align*}
where $\bc_{2} \in \mathbb{R}^p$ is a vector between $\bbeta_{2}$ and $\widehat{\bbeta}_{1}$ such that $\bv_{1}(\bc_{2}) \bdelta = n^{-1}_{1} \bX_{1} \bDelta(\bbeta_{2})$. Notice that $\bs(\bbeta_{2}; \lambda) = \bv_{2}(\bbeta_{2}) + \lambda \bv_{1}(\bbeta_{2})$. Analogous to the Gaussian linear model case, define $\bM(\bbeta) = \bv^{-1/2}_{2}(\bbeta) \bv_{1}(\bbeta) \bv^{-1/2}_{2}(\bbeta)$, $\bR(\bbeta) = \bv^{-1}_{1}(\bbeta) \bv_{2}(\bbeta)$, and $\bC=\bv^{-1}_{1}(\bbeta_{2}) \bv_{1}(\bc_{2})$. Then the $\mbox{aMSE}$ can be rewritten as
\begin{align*}
\mbox{aMSE} \{ \widetilde{\bbeta}_{2}(\lambda)\} &= v(\lambda) + b(\lambda), 
\end{align*}
where 
\begin{align*}
v(\lambda) & = d(\gamma_{2}) \mbox{trace} [ \{ \bI_p + \lambda \bM(\bbeta_{2}) \}^{-2} \bv^{-1}_{2}(\bbeta_{2}) ] \\
b(\lambda) & = n_{2} \lambda^2 \bdelta^\top \bC^\top \{ \bR(\bbeta_{2}) + \lambda \bI_p \}^{-2} \bC \bdelta. 
\end{align*}
The $v(\lambda)$ term is the sum of the variances of the (weighted) least squares estimates from $\mD_{1}$ and $\mD_{2}$, whereas $b(\lambda)$ is the squared distance from $\widehat{\bbeta}_{1}$ to $\bbeta_{2}$ and will be $0$ when $\lambda=0$ or $\bdelta=\bzero$. Clearly, $v(\lambda)$ is monotonically decreasing, $b(\lambda)$ is monotonically increasing, and  
\begin{align*}
v(0) + b(0) &=d(\gamma_{2}) \mbox{trace} \{ \bv^{-1}_{2}(\bbeta_{2})\} = \mbox{aMSE}(\widehat{\bbeta}_{2}).
\end{align*}
When $\bdelta = \bzero$, we therefore have $\mbox{aMSE} \{ \widetilde{\bbeta}_{2}(\lambda)\} < \mbox{aMSE}(\widehat{\bbeta}_{2})$ for all $\lambda>0$. Throughout the rest of the proof, we consider the case when $\bdelta \neq \bzero$. 

Since $\bv_{2}(\bbeta)$ is symmetric positive definite, it has an invertible and symmetric square root denoted by $\bv^{1/2}_{2}(\bbeta)$. From
\begin{align*}
\bv^{1/2}_{2}(\bbeta) \bR(\bbeta) \bv^{-1/2}_{2}(\bbeta) = \bv^{1/2}_{2}(\bbeta) \bv^{-1}_{1}(\bbeta) \bv^{1/2}_{2}(\bbeta)=\bM^{-1}(\bbeta),
\end{align*}
we have that $\bR(\bbeta)$ and $\bM^{-1}(\bbeta)$ are similar matrices. By symmetry, and therefore orthogonal diagonalizability, of the latter, the former is diagonalizable, and $\bR(\bbeta)$ and $\bM^{-1}(\bbeta)$ share eigenvalues. Let $\kappa_{r}(\bbeta)>0$, $r=1, \ldots, p$ the (common) eigenvalues of $\bM^{-1}(\bbeta)$ and $\bR(\bbeta)$ in decreasing order. Denote by $\bKappa(\bbeta) = \mathbf{diag} \{ \kappa_{r}(\bbeta)\}_{r=1}^p$ and $\bP(\bbeta)$ the matrices of eigenvalues and eigenvectors of $\bM^{-1}(\bbeta)$, with $\bP^\top(\bbeta) = \bP^{-1}(\bbeta)$ such that $\bM^{-1}(\bbeta) = \bP^\top(\bbeta) \bKappa(\bbeta) \bP(\bbeta)$. Then 
\begin{align*}
v(\lambda) & = d(\gamma_{2}) \mbox{trace} [ \{ \bI_p + \lambda \bKappa^{-1}(\bbeta_{2}) \}^{-2} \bP(\bbeta_{2}) \bv^{-1}_{2}(\bbeta_{2}) \bP^\top(\bbeta_{2}) ] \\
b(\lambda) & = n_{2} \bdelta^\top \bC^\top \{ \lambda^{-1} \bR(\bbeta_{2}) + \bI_p \}^{-2} \bC \bdelta.
\end{align*}

Recall that $\bv_{1}(\bc_{2}) \bdelta = n^{-1}_{1} \bX_{1} \bDelta(\bbeta_{2})$. Denote by $\{ \delta^2_r(\bbeta_{2})\}_{r=1}^p$ the values of
\begin{align*}
\mathbf{diag} (\bC \bdelta \bdelta^\top \bC^\top) &= \mathbf{diag} \{ \bv^{-1}_{1}(\bbeta_{2}) \bv_{1} (\bc_{2}) \bdelta \bdelta^\top \bv_{1}(\bc_{2}) \bv^{-1}_{1}(\bbeta_{2}) \}\\
&=n_{1}^{-2} \mathbf{diag} \{ \bv^{-1}_{1}(\bbeta_{2}) \bX_{1} \bDelta(\bbeta_{2}) \bDelta^\top(\bbeta_{2}) \bX^\top_{1} \bv^{-1}_{1}(\bbeta_{2}) \},
\end{align*}
sorted in increasing order, i.e. $0 \leq \delta^2_{1}(\bbeta) \leq \ldots \leq \delta^2_p(\bbeta)$. Let $g_{r2}(\bbeta)$, $r=1, \ldots, p$, the eigenvalues of $\bv_{2}(\bbeta)$ in increasing order. Then the bias and variance functions satisfy
\begin{align*}
n_{2} \lambda^2 \sum \limits_{r=1}^p \frac{\delta^2_{p-r+1}(\bbeta_{2})}{\{ \kappa_{r}(\bbeta_{2}) +\lambda \}^2 } \leq b(\lambda)& \leq n_{2} \lambda^2 \sum \limits_{r=1}^p \frac{\delta^2_r (\bbeta_{2})}{\{\kappa_{r}(\bbeta_{2})+\lambda\}^2}\\
d(\gamma_{2}) \sum \limits_{r=1}^p \frac{g^{-1}_{p-r+1~2}(\bbeta_{2}) \kappa^2_{p-r+1}(\bbeta_{2})}{\{ \kappa_r(\bbeta_{2}) +\lambda \}^2 } \leq v(\lambda) &\leq 
d(\gamma_{2}) \sum \limits_{r=1}^p \frac{g^{-1}_{r2}(\bbeta_{2}) \kappa^2_{r}(\bbeta_{2})}{\{ \kappa_{r}(\bbeta_{2})+\lambda\}^2},
\end{align*}
using Von Neumann's trace inequality and the fact that the eigenvalues of $\bA^2$ are the squared eigenvalues of $\bA$ for a positive definite matrix $\bA$. Then we have a bound on the aMSE, 
\[ L(\lambda) \leq \mbox{aMSE}\{ \widetilde{\bbeta}_{2}(\lambda)\} \leq U(\lambda), \]
where 
\begin{align*}
L(\lambda) &= \sum \limits_{r=1}^p \frac{d(\gamma_{2}) g^{-1}_{p-r+1~2}(\bbeta_{2}) \kappa^2_{p-r+1}(\bbeta_{2}) + n_{2}\lambda^2 \delta^2_{p-r+1}(\bbeta_{2})}{\{ \kappa_{r}(\bbeta_{2}) +\lambda\}^2}\\
U(\lambda) &=  \sum \limits_{r=1}^p \frac{d(\gamma_{2}) g^{-1}_{r2}(\bbeta_{2}) \kappa^2_{r}(\bbeta_{2}) + n_{2}\lambda^2 \delta^2_r(\bbeta_{2})}{\{ \kappa_{r}(\bbeta_{2}) +\lambda\}^2},
\end{align*}
are the sums of the lower and upper bounds on $v(\lambda)$ and $b(\lambda)$ above, respectively. We calculate the derivatives:
\begin{align*}
\frac{\partial}{\partial \lambda} L(\lambda)&= 2\sum \limits_{r=1}^p \frac{2 \delta_{p-r+1}(\bbeta_{2}) \{ \delta_{p-r+1}(\bbeta_{2}) \kappa_{r}(\bbeta_{2}) n_{2} \lambda -g^{-1}_{p-r+1~2} \kappa^2_{p-r+1}(\bbeta_{2}) \}}{\{ \kappa_{r}(\bbeta_{2}) + \lambda \}^3}\\
\frac{\partial}{\partial \lambda} U(\lambda)&= 2\sum \limits_{r=1}^p \frac{\kappa_{r}(\bbeta_{2}) \{ \lambda n_{2} \delta^2_r(\bbeta_{2}) - d(\gamma_{2}) g^{-1}_{r2}(\bbeta_{2}) \kappa_{r}(\bbeta_{2}) \}}{\{\kappa_{r}(\bbeta_{2}) + \lambda\}^3}.
\end{align*}
Since the derivative of the lower bound $L(\lambda)$ is positive for large enough $\lambda$, $L(\lambda)$ is increasing for large enough $\lambda$. From
\begin{align*}
\lim \limits_{\lambda \rightarrow 0^+} \frac{\partial}{\partial \lambda}U(\lambda) &= -2d(\gamma_{2}) \sum \limits_{r=1}^p \frac{g^{-1}_{r2}(\bbeta_{2})}{\kappa_{r}(\bbeta_{2})} < 0,
\end{align*}
we find that $U(\lambda)$ is decreasing in a neighborhood of the origin.  All together, we find that $\mbox{aMSE} \{ \widetilde{\bbeta}_{2}(\lambda)\}$ is monotonically decreasing on an interval $[0, \lambda^\star]$ and monotonically increasing on an interval $[\lambda^\star,\infty)$ for some $\lambda^\star>0$. 

Next we show that the upper limit on the range of $\lambda$ on which there is an efficiency gain is $O(n_{2}^{-1/2})$.  Let $\lambda>0$ be such that $\mbox{aMSE}\{ \widetilde{\bbeta}_{2}(\lambda)\} = \mbox{aMSE}(\widehat{\bbeta}_{2})$. Then
\begin{align*}
0&=v(\lambda) + b(\lambda) - \{v(0) + b(0) \} \\
&=d(\gamma_{2}) \mbox{trace} [ \{ \bI_p + \lambda \bM(\bbeta_{2}) \}^{-2} \bv^{-1}_{2} (\bbeta_{2}) - \bv^{-1}_{2}(\bbeta_{2}) ]  \\
& \qquad + n_{2} \bdelta^\top \bC^\top \{ \lambda^{-1} \bR(\bbeta_{2}) + \bI_p \}^{-2} \bC \bdelta
\end{align*}
Recall that $\bC \bdelta$ is a constant with respect to $\lambda$. Rearranging, $\lambda$ satisfies
\begin{align*}
0&=d(\gamma_{2}) \mbox{trace} \big( [ \{ \bI_p + \lambda \bM(\bbeta_{2}) \}^{-2} -\bI_p ] \bv^{-1}_{2} (\bbeta_{2}) \big) \\
&\qquad + n_{2} \bdelta^\top \bC^\top \{ \lambda^{-1} \bR(\bbeta_{2}) + \bI_p \}^{-2} \bC \bdelta
\end{align*}
The first term on the right-hand side is a decreasing function of $\lambda$ with limit $0$ as $\lambda \rightarrow 0^+$ and limit $-\mbox{aMSE}(\widehat{\bbeta}_{2})$ as $\lambda \rightarrow \infty$; the first term is, therefore, effectively (a negative) constant in $\lambda$. The second term is a strictly positive increasing function of $\lambda$ for all $\lambda > 0$.  For this, too, to be a constant, so that the above equality is satisfied, we need $n_{2} \lambda^2$ to be effectively a constant.  Therefore, the $\lambda$ at which equality of aMSEs is achieved is $O(n_{2}^{-1/2})$.

Finally, we can get a lower bound on $\lambda^\star$, so that $\mbox{aMSE} \{ \widetilde{\bbeta}_{2}(\lambda)\} < \mbox{aMSE} (\widehat{\bbeta}_{2})$ for all $\lambda$ less than that bound. Since $U(\lambda)$ is decreasing in a neighborhood of the origin, it is sufficient to find the $\lambda$ values for which this derivative is negative, i.e., 
it is sufficient to find $\lambda$ such that
\begin{align*}
\lambda n_{2} \sum_{r=1}^p \frac{\kappa_{r}(\bbeta_{2}) \delta^2_r(\bbeta_{2})}{\{\kappa_{r}(\bbeta_{2}) +\lambda\}^3} < d(\gamma_{2}) \sum_{r=1}^p \frac{g^{-1}_{r2}(\bbeta_{2}) \kappa^2_{r}(\bbeta_{2})}{\{\kappa_{r} (\bbeta_{2}) +\lambda\}^3}.
\end{align*}
This is satisfied by
\begin{align}
\label{e:MLE-lambda-bound1}
\lambda < \frac{d(\gamma_{2})}{n_{2}} \frac{\min_{r=1, \ldots, p}\{ \kappa_{r}(\bbeta_{2}) g^{-1}_{r2}(\bbeta_{2}) \}}{ \max_{r=1, \ldots, p}\{ \delta^2_r (\bbeta_{2}) \}}.
\end{align}
Equation \eqref{e:MLE-lambda-bound1} gives a range of $\lambda$ values such that $\mbox{aMSE}\{ \widetilde{\bbeta}_{2}(\lambda)\} < \mbox{aMSE}(\widehat{\bbeta}_{2})$, so the right-hand side is a lower bound on $\lambda^\star$.

\section{Additional numerical results}

\subsection{Further results from Sections \ref{ss:vi-1}-\ref{ss:robustness}}
\label{a:estimates}

\begin{figure}[H]
\includegraphics[width=\textwidth]{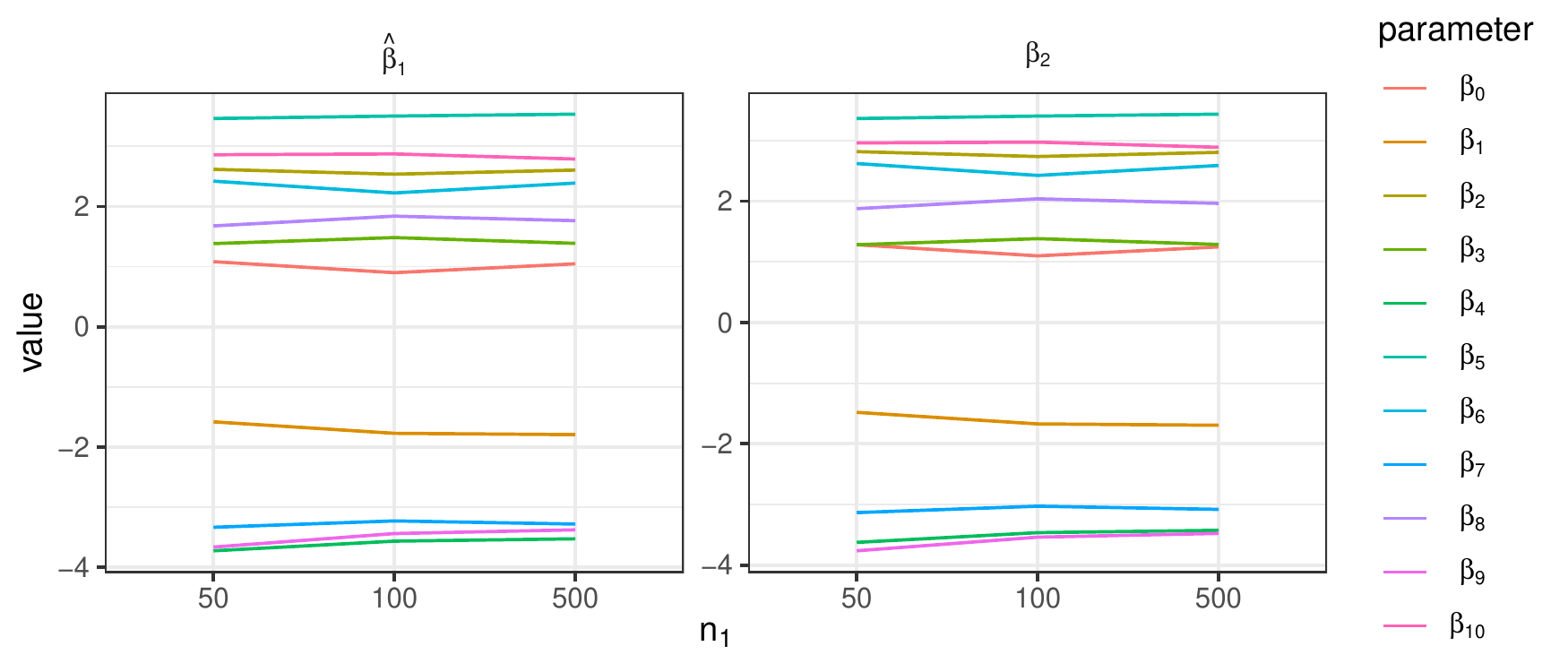}
\caption{MLE $\widehat{\bbeta}_{1}$ (left) with corresponding value of $\bbeta_{2}$ (right) for each $\mD_{1}$ in Setting I. \label{f:MLE-SetI}}
\end{figure}

\begin{figure}[H]
\includegraphics[width=\textwidth]{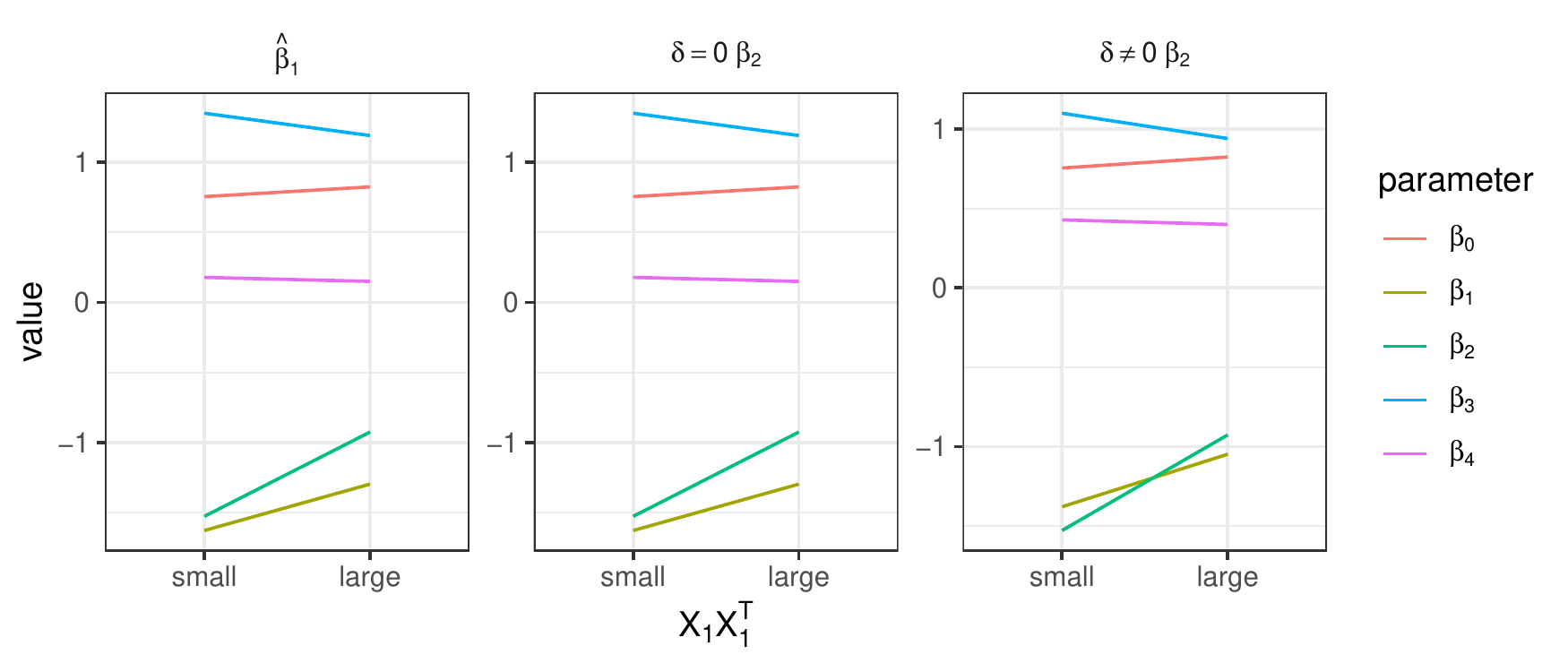}
\caption{MLE $\widehat{\bbeta}_{1}$ (left) with corresponding value of $\bbeta_{2}$ (right) for each $\mD_{1}$ in Setting II. \label{f:MLE-SetII}}
\end{figure}

\begin{figure}[H]
\includegraphics[width=\textwidth]{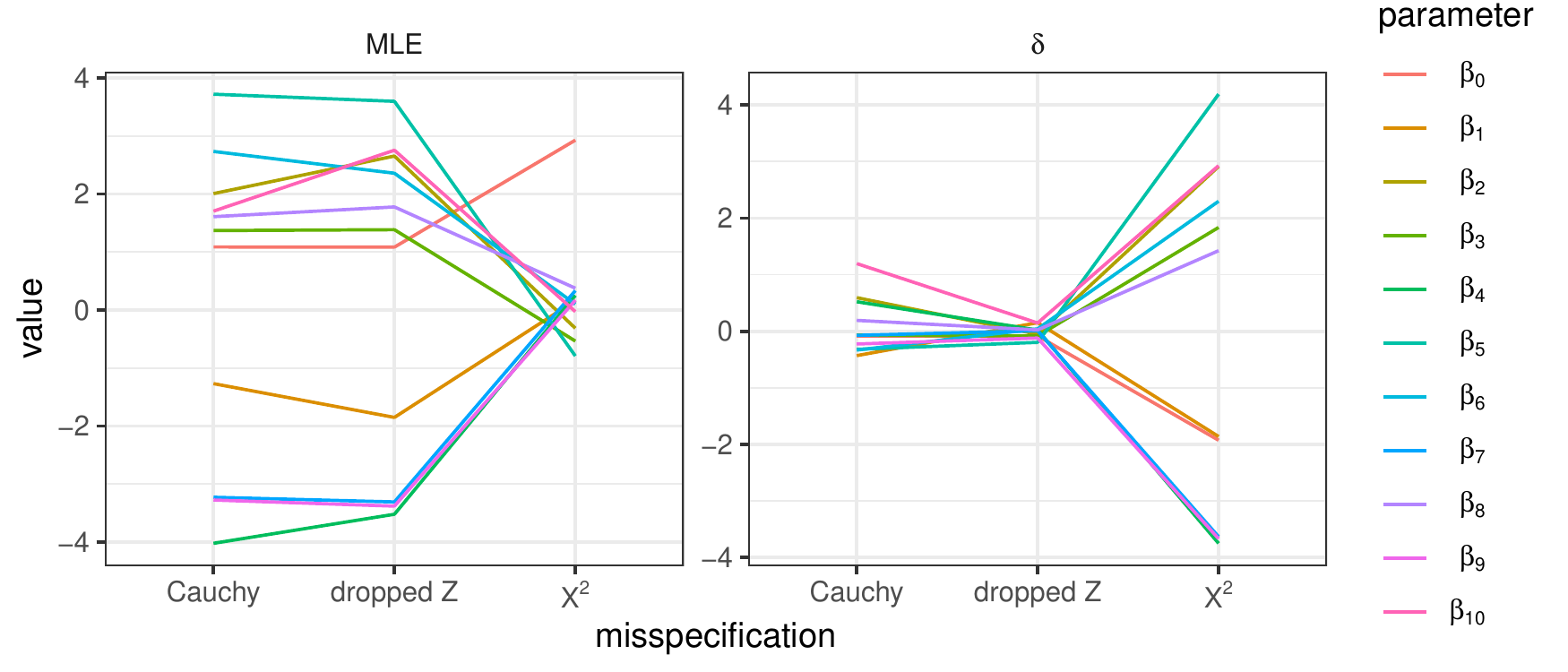}
\caption{MLE $\widehat{\bbeta}_{1}$ (left) with corresponding value of $\bdelta$ (right) for each $\mD_{1}$ in Setting IV. \label{f:MLE-SetIV}}
\end{figure}

Figures~\ref{f:MLE-SetI}--\ref{f:MLE-SetII} plot $\widehat{\bbeta}_{1}$ and $\bbeta_{2}$ in Settings I and II, respectively. Figure~\ref{f:MLE-SetIV} plots $\widehat{\bbeta}_{1}$ and the corresponding value of $\bdelta$ in Setting IV.

Tables \ref{t:SetI-sd}--\ref{t:SetIV-sd} give the Monte Carlo standard error of the empirical mean squared error for $\widetilde{\bbeta}_{2}(\tilde\lambda)$ and comparison estimators in Settings I--IV respectively.

\begin{table}[H]
\centering
\begin{tabular}{rrrrrrrr}
\toprule
$n_{1}$ & $n_{2}$ & \multicolumn{6}{c}{MCse~$\cdot 10^{3}$ of eMSE} \\
& & $\widetilde{\bbeta}_{2}(\tilde\lambda)$ & $\widehat{\bbeta}_{2}( \bW_{\widehat{\lambda}})$ & $\widehat{\bbeta}_{2}^{TG}$ & $\widehat{\bbeta}_{2}^{TL}$ & $\widehat{\bbeta}_{2}^p$ & $\widehat{\bbeta}_{2}$ \\ 
\midrule
\multirow{3}{*}{$50$} & $50$ & $2.73$ & $2.73$ & $3.80$ & $4.48$ & $1.49$ & $4.38$ \\ 
& $100$ & $1.31$ & $1.35$ & $2.94$ & $1.87$ & $1.06$ & $1.81$ \\ 
& $500$ & $0.295$ & $0.289$ & $0.316$ & $0.321$ & $0.284$ & $0.31$ \\ 
\multirow{3}{*}{$100$} & $50$ & $2.64$ & $2.63$ & $3.48$ & $3.98$ & $1.15$ & $4.38$ \\ 
& $100$ & $1.27$ & $1.31$ & $2.68$ & $1.60$ & $0.945$ & $1.81$ \\ 
& $500$ & $0.289$ & $0.286$ & $0.327$ & $0.304$ & $0.291$ & $0.310$ \\ 
\multirow{3}{*}{$500$} & $50$ & $2.60$ & $2.57$ & $3.11$ & $3.98$ & $0.438$ & $4.38$ \\ 
& $100$ & $1.25$ & $1.28$ & $2.15$ & $1.58$ & $0.526$ & $1.81$ \\ 
& $500$ & $0.284$ & $0.285$ & $0.313$ & $0.304$ & $0.384$ & $0.310$ \\ 
\bottomrule
\end{tabular}
\caption{Setting I: Monte Carlo standard error (MCse) of eMSE of $\widetilde{\bbeta}_{2}(\tilde\lambda)$, $\widehat{\bbeta}_{2}(\bW_{\widehat{\lambda}})$, $\widehat{\bbeta}_{2}^{TG}$, $\widehat{\bbeta}_{2}^{TL}$, $\widehat{\bbeta}_{2}^p$, $\widehat{\bbeta}_{2}$.\label{t:SetI-sd}}
\end{table}

\begin{table}[H]
\centering
\begin{tabular}{rrrrrrr}
\toprule
$\bdelta$ & $\bX_{1} \bX^\top_{1}$ & \multicolumn{5}{c}{MCse~$\cdot 10^3$ of eMSE}\\
& & $\widetilde{\bbeta}_{2}(\tilde\lambda)$ & $\widehat{\bbeta}_{2}(\bW)$ & $\widehat{\bbeta}_{2}^{TG}$ & $\widehat{\bbeta}_{2}^p$ & $\widehat{\bbeta}_{2}$ \\ 
\midrule
\multirow{2}{*}{$=\bzero$} & large & $1.37$ & $2.24$ & $2.54$ & $0.586$ & $2.35$ \\ 
& small & $2.29$ & $3.66$ & $3.82$ & $0.865$ & $3.78$ \\ 
\midrule
\multirow{2}{*}{$\neq \bzero$} & large & $1.68$ & $1.62$ & $2.37$ & $1.03$ & $1.98$ \\ 
& small & $1.89$ & $2.30$ & $3.28$ & $1.44$ & $3.05$ \\ 
\bottomrule
\end{tabular}
\caption{Setting II: Monte Carlo standard error (MCse) of eMSE of $\widetilde{\bbeta}_{2}(\tilde\lambda)$, $\widehat{\bbeta}_{2}(\bW)$, $\widehat{\bbeta}_{2}^{TG}$, $\widehat{\bbeta}_{2}^p$, $\widehat{\bbeta}_{2}$.\label{t:SetII-sd}}
\end{table}

\begin{table}[H]
\centering
\begin{tabular}{rrrrrr}
\toprule
$\bdelta$ & \multicolumn{5}{c}{MCse~$\cdot 10^{3}$ of eMSE}\\
& $\widetilde{\bbeta}_{2}(\tilde\lambda)$ & $\widehat{\bbeta}_{2}(\bW)$ & $\widehat{\bbeta}_{2}^{TG}$ & $\widehat{\bbeta}_{2}^p$ & $\widehat{\bbeta}_{2}$ \\ 
\midrule
\multicolumn{5}{c}{(i) independent features, $\rho=0$}\\
$(0,0,0)$ & $0.679$ & $0.894$ & $1.23$ & $0.190$ & $1.03$ \\ 
$(0,1,0)$ & $0.950$ & $0.980$ & $1.03$ & $1.13$ & $0.978$ \\ 
$(0,2,0)$ & $1.50$ & $4.30$ & $1.48$ & $1.87$ & $1.58$ \\ 
\multicolumn{5}{c}{(ii) correlated features, $\rho=0.4$}\\
$(0,0,0)$ & $5.16$ & $8.94$ & $6.60$ & $0.166$ & $8.59$ \\ 
$(0,1,0)$ & $5.77$ & $8.64$ & $6.07$ & $2.48$ & $7.76$ \\ 
$(0,2,0)$ & $11.6$ & $58.1$ & $9.82$ & $5.07$ & $18.1$ \\
\bottomrule
\end{tabular}
\caption{Setting III: Monte Carlo standard error (MCse) of eMSE of $\widetilde{\bbeta}_{2}(\tilde\lambda)$, $\widehat{\bbeta}_{2}(\bW)$, $\widehat{\bbeta}_{2}^{TG}$, $\widehat{\bbeta}_{2}^p$, $\widehat{\bbeta}_{2}$.\label{t:SetIII-sd}}
\end{table}

\begin{table}[H]
\centering
\begin{tabular}{rrrrrrr}
\toprule
Case & \multicolumn{6}{c}{MCse~$\cdot 10^{3}$ of eMSE}\\
& $\widetilde{\bbeta}_{2}(\tilde\lambda)$ & $\widehat{\bbeta}_{2}(\bW_{\widehat{\lambda}})$ & $\widehat{\bbeta}_{2}^{TG}$ & $\widehat{\bbeta}_{2}^{TL}$ & $\widehat{\bbeta}_{2}^p$ & $\widehat{\bbeta}_{2}$ \\ 
\midrule
Cauchy & $0.308$ & $0.308$ & $0.310$ & $0.315$ & $1.70$ & $0.310$ \\ 
dropped $Z$ & $0.272$ & $0.269$ & $0.362$ & $0.285$ & $0.264$ & $0.310$ \\ 
$X^2$ & $0.310$ & $0.310$ & $0.310$ & $0.334$ & $47.0$ & $0.310$ \\ 
\bottomrule
\end{tabular}
\caption{Setting IV: Monte Carlo standard error (MCse) of eMSE of $\widetilde{\bbeta}_{2}(\tilde\lambda)$, $\widehat{\bbeta}_{2}(\bW_{\widehat{\lambda}})$, $\widehat{\bbeta}_{2}^{TG}$, $\widehat{\bbeta}_{2}^{TL}$, $\widehat{\bbeta}_{2}^p$, $\widehat{\bbeta}_{2}$. \label{t:SetIV-sd}}
\end{table}

\subsection{Multi-source simulation}
\label{a:multi-source}

For the purposes of this section, we denote by $\mD^{\mS}_1, \ldots, \mD^{\mS}_M$ $M$ source data sets and $\mD^{\mT}$ the target data set. Each data set $\mD^{\mS}_j$ is composed of features $\bX^{\mS}_{ij}$ and outcomes $y^{\mS}_{ij}$, $i=1, \ldots, n^{\mS}_j$, $j=1,\ldots, M$, and the target data set $\mD^{\mT}$ is composed of features $\bX^{\mT}_i$ and outcomes $y^{\mT}_i$, $i=1, \ldots, n^{\mT}$.

In the notation of the main manuscript, the concatenation approach considers $\bX_1=(\bX^{\mS}_1, \ldots, \bX^{\mS}_M)$, $\by_1=(\by^{\mS~\top}_1, \ldots, \by^{\mS~\top}_M)\top$, $\mD_1=\{\bX_1, \by_1\}$ and $\mD_2=\mD^{\mT}$, with $n_1=n^{\mS}_1 + \ldots + n^{\mS}_M$ and $n_2=n^{\mT}$, $\bbeta_2=\bbeta_{\mT}$. We have suggested that this concatenation approach will work well when the source data sets are mostly homogeneous or mostly heterogeneous. It is possible, as described in Section 2.1, that there are groups of source data sets that are homogeneous within and heterogeneous across. When the number of source data sets is small, as is the case below, it may be possible to consider integrating the target data set with only one of the source data sets, and to compare performance across these separate integrative estimates in order to select the best approach. In this section, we investigate this question by examining the trade-offs of concatenation versus single-source transfer learning when $M=3$ source data sets are available. 

\begin{figure}[H]
\includegraphics[width=\textwidth]{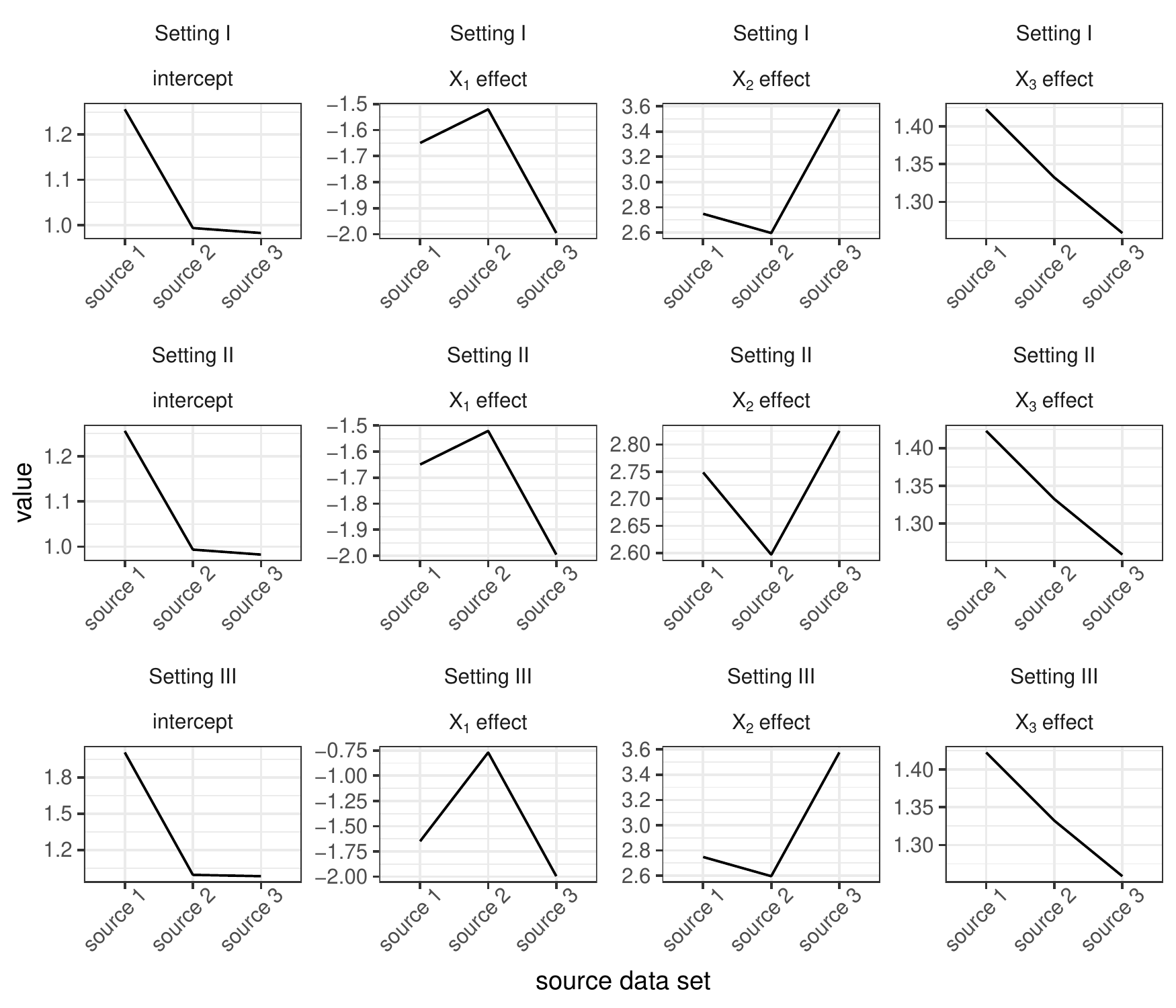}
\caption{MLEs $\widehat{\bbeta}^{\mS}_1$, $\widehat{\bbeta}^{\mS}_2$, $\widehat{\bbeta}^{\mS}_3$ and $\widehat{\bbeta}^{\mT}$ in Settings I, II and III. \label{f:supp-MLE-multi}}
\end{figure}

In both source and target data sets, features consist of an intercept and three continuous variables independently generated from a standard Gaussian distribution. Source outcomes $y^{\mS}_{ij}$ are simulated from the Gaussian distribution with $\E(Y^{\mS}_{ij})= (\bX^{\mS}_{ij})^\top \bbeta^{\mS}_j$ and $\V(Y^{\mS}_{ij})=1$, $j=1,\ldots, M$, and target outcomes $y^{\mT}_i$ are simulated from the Gaussian distribution with $\E(Y^{\mT}_i)= (\bX^{\mT}_i)^\top \bbeta_{\mT}$ and $\V(Y^{\mT}_i)=1$. We fix $n^{\mS}_1=n^{\mS}_2=n^{\mS}_3=100$, $n_{\mT}=50$. We simulate three source data sets $\mD^{\mS}_j=\{y^{\mS}_{ij}, \bX^{\mS}_{ij}\}_{i=1}^{n^{\mS}_j}$. We consider three settings of values for $\bbeta^{\mS}_j$ and $\bbeta^{\mT}$ that characterize the heterogeneity/homogeneity of the sources relative to the target. Specifically, defining $\bc=(1, -1.8, 2.6, 1.4)$:
\begin{itemize}
\item In Setting I, $\bbeta^{\mS}_1 = \bc + (1/4, 0, 0, 0)$, $\bbeta^{\mS}_2 = \bc + (0, 1/4, 0, 0)$, $\bbeta^{\mS}_3 = \bc + (0, 0, 1, 0)$ and $\bbeta^{\mT}=\bc + (0, 0, 5/4, 0)$. Here, the first and second source data sets are closer to each other than to the third source data set, which is itself closer to the target data set.
\item In Setting II, $\bbeta^{\mS}_1$ and $\bbeta^{\mS}_2$ are as in Setting I, $\bbeta^{\mS}_3=\bc+(0, 0, 1/4, 0)$ and $\bbeta^{\mT}=\bc+(0, 0, 0, 1/4)$. All source data sets are close to each other and to the target data set.
\item In Setting III, $\bbeta^{\mS}_1 = \bc + (1, 0, 0, 0)$, $\bbeta^{\mS}_2 = \bc + (0, 1, 0, 0)$, $\bbeta^{\mS}_3 = \bc + (0, 0, 1, 0)$ and $\bbeta^{\mT}=\bc + (0, 0, 0, 1)$. All source data sets are far from each other and the target data set.
\end{itemize}
The MLEs $\widehat{\bbeta}^{\mS}_j$ are reported in Figure \ref{f:supp-MLE-multi}. For each setting, we generate $1000$ data sets $\mD_{\mT}=\{y^{\mT}_i, \bX^{\mT}_i\}_{i=1}^{n^{\mT}}$.

For each $\mD_{\mT}$, we compute $\widetilde{\bbeta}_2(\tilde{\lambda})$ using each source separately and using the concatenation of the sources. We also compute the target-only MLE $\widehat{\bbeta}_2$, which corresponds to setting the source data set to be $\varnothing$. Empirical mean squared error (eMSE) of $\widetilde{\bbeta}_2(\tilde{\lambda})$ averaged across the 1000 target data sets in each setting are reported in Table \ref{t:supp-multi}. In all cases, our information-shrinkage estimator is more efficient than the target-only MLE. In Setting I, the estimator with the smallest eMSE uses $\mD^{\mS}_3$ only, which is to be expected given that $\mD_3^{\mS}$ is objectively closer to the target data than the others.  In Setting II, the estimator with the smallest eMSE uses the concatenation $\{\mD^{\mS}_1, \mD^{\mS}_2, \mD^{\mS}_3\}$. Again, this makes intuitive sense given that the three source data sets are similar and similar to the target---no reason not to concatenate.  In Setting III, the estimator with the smallest eMSE uses the concatenation $\{\mD^{\mS}_1, \mD^{\mS}_2, \mD^{\mS}_3\}$.  This, too, is not unexpected, since the concatenation of three source data sets that differ from each other and the target can be no worse than a single source data set that differs from the target. 

The concatenation approach appears to be the optimal choice in most settings. Setting I is a somewhat pathological example, in that it is rare that a single source out of many is much more similar to the target than any other source. Usually, such as setting is easy to identify \textit{a priori} based on external information such as study characteristics. Nonetheless, it is reassuring that, even in this extreme setting, the concatenation approach remains a reasonable choice. This is in fact very much by design: the selection of $\tilde{\lambda}$ theoretically guarantees that the concatenation outperforms the target-only estimation (in terms of MSE). In other words, regardless of the heterogeneity/homogeneity structure of the sources, the concatenation guarantees protection against negative transfer and is a safe approach.

\begin{table}[t]
\centering
\begin{tabular}{rrrrrrr}
\toprule
source data set & Setting I & Setting II & Setting III \\ 
\midrule
$\mD^{\mS}_1$ & $8.68$ ($1.96$) & $6.17$ ($1.38$) & $8.53$ ($1.94$) \\ 
$\mD^{\mS}_2$ & $8.69$ ($1.96$) & $7.24$ ($1.50$) & $8.68$ ($1.99$) \\ 
$\mD^{\mS}_3$ & $\mathbf{6.73}$ ($1.36$) & $7.17$ ($1.55$) & $8.60$ ($1.94$) \\ 
$\{\mD^{\mS}_1, \mD^{\mS}_2, \mD^{\mS}_3\}$ & $8.63$ ($1.93$) & $5.90$ ($1.28$) & $\mathbf{8.48}$ ($1.94$) \\ 
$\varnothing$ & $8.65$ ($1.98$) & $8.65$ ($1.98$) & $8.65$ ($1.98$) \\ 
\bottomrule
\end{tabular}
\caption{Multi-source simulation: eMSE $\times 10^2$ (Monte Carlo standard error $\times 10^3$) of $\widetilde{\bbeta}_{\mT}(\tilde\lambda)$. Minimum eMSE in each Setting is bolded. \label{t:supp-multi}}
\end{table}

Beyond guaranteeing robustness to heterogeneity/homogeneity of multiple sources, an investigator may wish to select the approach that is ``best'' in the sense of minimizing MSE.  Specifically, treating $\widehat{\mbox{MSE}}(\tilde{\lambda})$ as a function of the source data set configuration
\begin{align*}
\mS \mapsto \underbrace{n_{\mT}^{-1} \widehat{\sigma}^2_{\mT} \, \mbox{trace} \{ \bS^{-2}(\tilde{\lambda}) \bG_{\mT} \} + \tilde{\lambda}^2 \mbox{trace} \{ \bG_{\mS} \bS^{-2}(\tilde{\lambda}) \bG_{\mS} \widehat{\bdelta^2} \}}_{\widehat{\mbox{MSE}}(\tilde{\lambda})},
\end{align*}
\noindent one may select the source data set configuration $\widehat\mS$ to minimize the right-hand side.  As $\widehat{\mbox{MSE}}(\tilde{\lambda})$ balances bias and variance to evaluate the most efficient estimator's inferential performance, an investigator may choose the single source or concatenated source that gives the most efficient estimator. Across the 1000 simulations, we compare the value of $\widehat{\mbox{MSE}}(\tilde{\lambda})$ across the five source configurations in each setting. In Setting I, the estimator that uses $\mD^{\mS}_3$ has smaller $\widehat{\mbox{MSE}}(\tilde{\lambda})$ in $100\%$, $100\%$ and $99.7\%$ of simulations compared to the estimator that uses $\mD^{\mS}_1$, $\mD^{\mS}_2$ and $\{\mD^{\mS}_1, \mD^{\mS}_2, \mD^{\mS}_3\}$, respectively. In Setting II, the estimator that uses $\{\mD^{\mS}_1, \mD^{\mS}_2, \mD^{\mS}_3\}$ has smaller $\widehat{\mbox{MSE}}(\tilde{\lambda})$ in $61.4\%$, $83.0\%$ and $83.5\%$ of simulations compared to the estimator that uses $\mD^{\mS}_1$, $\mD^{\mS}_2$ and $\mD^{\mS}_3$, respectively. In Setting III, the estimator that uses $\{\mD^{\mS}_1, \mD^{\mS}_2, \mD^{\mS}_3\}$ has smaller $\widehat{\mbox{MSE}}(\tilde{\lambda})$ in $67.8\%$, $98.4\%$ and $89.4\%$ of simulations compared to the estimators that use $\mD^{\mS}_1$, $\mD^{\mS}_2$ and $\mD^{\mS}_3$, respectively. In other words, across all settings, the criterion $\widehat{\mbox{MSE}}(\tilde{\lambda})$ does reasonably well in helping an investigator select the most useful source configuration in terms of minimizing MSE, showing less certainty in settings for which the source configurations perform similarly. We therefore suggest that an investigator rely on $\widehat{\mbox{MSE}}(\tilde{\lambda})$ to decide among several multi-source integrative estimators and a concatenation approach. 

\section{eICU Database information}
\label{S:data}

The eICU Collaborative Research Database \citep{eICU} data are publicly available upon completion of training and authentication. Users may begin their data access request at \verb|https://eicu-crd.mit.edu/gettingstarted/access/|.

Our analysis focuses on African American and Caucasian patients between the ages of 40 and 89 with body mass index (BMI) between 14 and 60 that were admitted through any means other than another intensive care unit (ICU). 

\bibliographystyle{apalike}
\bibliography{mhd_meta-bib-20211209}

\begin{thebibliography}{}

\bibitem[Amari and Nagaoka, 2000]{Amari-Nagaoka}
Amari, S. and Nagaoka, H. (2000).
\newblock {\em Methods of Information Geometry}, volume 191.
\newblock American Mathematical Society.

\bibitem[Andersen and Gill, 1982]{Andersen-Gill}
Andersen, P. and Gill, R. (1982).
\newblock Cox's regression model for counting processes: a large sample study.
\newblock {\em The Annals of Statistics}, 10(4):1100--1120.

\bibitem[Baldi and Itti, 2010]{Baldi-Itti}
Baldi, P. and Itti, L. (2010).
\newblock Of bits and wows: A {B}ayesian theory of surprise with applications
  to attention.
\newblock {\em Neural Networks}, 23(5):649--666.

\bibitem[Bondell and Reich, 2009]{Bondell-Reich}
Bondell, H.~D. and Reich, B.~J. (2009).
\newblock Simultaneous factor selection and collapsing levels in {ANOVA}.
\newblock {\em Biometrics}, 65(1):169--177.

\bibitem[Cahoon and Martin, 2020]{Cahoon-Martin}
Cahoon, J. and Martin, R. (2020).
\newblock A generalized inferential model for meta-analyses based on few
  studies.
\newblock {\em Statistics and Applications}, 18(2):299--316.

\bibitem[Cai et~al., 2019]{cai2019individual}
Cai, T., Liu, M., and Xia, Y. (2019).
\newblock Individual data protected integrative regression analysis of
  high-dimensional heterogeneous data.
\newblock {\em arXiv preprint arXiv:1902.06115}.

\bibitem[Cai and Wei, 2021]{Cai-Wei}
Cai, T.~T. and Wei, J. (2021).
\newblock Transfer learning for nonparametric classification: Minimax rate and
  adaptive classifier.
\newblock {\em The Annals of Statistics}, 49(1):100--128.

\bibitem[Chatterjee et~al., 2016]{chatterjee2016constrained}
Chatterjee, N., Chen, Y.-H., Maas, P., and Carroll, R.~J. (2016).
\newblock Constrained maximum likelihood estimation for model calibration using
  summary-level information from external big data sources.
\newblock {\em Journal of the American Statistical Association},
  111(513):107--117.

\bibitem[Chen et~al., 2015]{Chen-Owen-Shi}
Chen, A., Owen, A.~B., and Shi, M. (2015).
\newblock Data enriched linear regression.
\newblock {\em Electronic Journal of Statistics}, 9(1):1078--1112.

\bibitem[Chen et~al., 2019]{chen2019distributed}
Chen, X., Liu, W., Mao, X., and Yang, Z. (2019).
\newblock Distributed high-dimensional regression under a quantile loss
  function.
\newblock {\em arXiv preprint arXiv:1906.05741}.

\bibitem[Cheng et~al., 2020]{Cheng-etal}
Cheng, C., Zhou, B., Ma, G., Wu, D., and Yuan, Y. (2020).
\newblock Wasserstein distance based deep adversarial transfer learning for
  intelligent fault diagnosis.
\newblock {\em Neurocomputing}, 409:35--45.

\bibitem[Claggett et~al., 2014]{Claggett-Xie-Tian}
Claggett, B., Xie, M., and Tian, L. (2014).
\newblock Meta-analysis with fixed, unknown, study-specific parameters.
\newblock {\em Journal of the American Statistical Association},
  109(508):1660--1671.

\bibitem[Duan et~al., 2019]{duan2019heterogeneity}
Duan, R., Ning, Y., and Chen, Y. (2019).
\newblock Heterogeneity-aware and communication-efficient distributed
  statistical inference.
\newblock {\em arXiv preprint arXiv:1912.09623}.

\bibitem[Dube et~al., 2020]{Dube-etal}
Dube, P., Bhattacharjee, B., Petit-Bois, E., and Hill, M. (2020).
\newblock Improving transferability of deep neural networks.
\newblock {\em Domain Adaptation for Visual Understanding}, pages 51--64.

\bibitem[Efron and Morris, 1973]{Efron-Morris-JRSSB}
Efron, B. and Morris, C. (1973).
\newblock Combining possibly related estimation problems.
\newblock {\em Journal of the Royal Statistical Society Series B},
  35(3):379--421.

\bibitem[Efron and Morris, 1977]{Efron-Morris-1977}
Efron, B. and Morris, C. (1977).
\newblock Stein's paradox in statistics.
\newblock {\em Scientific American}, 236(5):119--127.

\bibitem[Farebrother, 1976]{Farebrother}
Farebrother, R.~W. (1976).
\newblock Further results on the mean square error of ridge regression.
\newblock {\em Journal of the Royal Statistical Society Series B},
  38(3):248--250.

\bibitem[Godambe, 1991]{Godambe-1991}
Godambe, V.~P. (1991).
\newblock {\em Estimating functions}.
\newblock Oxford University Press.

\bibitem[Hector and Song, 2020]{Hector-Song-JMLR}
Hector, E.~C. and Song, P. X.-K. (2020).
\newblock Doubly distributed supervised learning and inference with
  high-dimensional correlated outcomes.
\newblock {\em Journal of Machine Learning Research}, 21:1--35.

\bibitem[Hemmerle, 1975]{Hemmerle}
Hemmerle, W.~J. (1975).
\newblock An explicit solution for generalized ridge regression.
\newblock {\em Technometrics}, 17(3):309--314.

\bibitem[Higgins and Thompson, 2002]{Higgins-Thompson}
Higgins, J. P.~T. and Thompson, S.~G. (2002).
\newblock Quantifying heterogeneity in a meta-analysis.
\newblock {\em Statistics in Medicine}, 21(11):1539--1558.

\bibitem[Hjort and Pollard, 1993]{Hjort-Pollard}
Hjort, N.~L. and Pollard, D. (1993).
\newblock Asymptotics for minimisers of convex processes.
\newblock {\em arXiv}, arXiv:1107.3806.

\bibitem[Hoerl and Kennard, 1970]{Hoerl-Kennard}
Hoerl, A.~E. and Kennard, R.~W. (1970).
\newblock Ridge regression: biased estimation for nonorthogonal problems.
\newblock {\em Technometrics}, 12(1):55--67.

\bibitem[James and Stein, 1960]{James-Stein}
James, W. and Stein, C. (1960).
\newblock Estimation with quadratic loss.
\newblock {\em Fourth Berkeley Symposium on Mathematical Statistics and
  Probability}, 4.1:361--379.

\bibitem[Jordan et~al., 2018]{jordan2018communication}
Jordan, M.~I., Lee, J.~D., and Yang, Y. (2018).
\newblock Communication-efficient distributed statistical inference.
\newblock {\em Journal of the American Statistical Association}.

\bibitem[Ke et~al., 2015]{ke2015homogeneity}
Ke, Z.~T., Fan, J., and Wu, Y. (2015).
\newblock Homogeneity pursuit.
\newblock {\em Journal of the American Statistical Association},
  110(509):175--194.

\bibitem[Li et~al., 2020]{Li-Cai-Li-2020}
Li, S., Cai, T.~T., and Li, H. (2020).
\newblock Transfer learning in large-scale gaussian graphical models with false
  discovery rate control.
\newblock {\em arXiv}, arXiv:2010.11037.

\bibitem[Li et~al., 2022]{Li-Cai-Li-2022}
Li, S., Cai, T.~T., and Li, H. (2022).
\newblock Transfer learning for high-dimensional linear regression: Prediction,
  estimation and minimax optimality.
\newblock {\em Journal of the Royal Statistical Society Series B},
  84(1):149--173.

\bibitem[Lin and Lu, 2019]{lin2019race}
Lin, L. and Lu, J. (2019).
\newblock A race-dc in big data.
\newblock {\em arXiv preprint arXiv:1911.11993}.

\bibitem[Liu et~al., 2015]{Liu-Liu-Xie-2015}
Liu, D., Liu, R.~Y., and Xie, M. (2015).
\newblock Multivariate meta-analysis of heterogeneous studies using only
  summary statistics: efficiency and robustness.
\newblock {\em Journal of the American Statistical Association},
  110(509):326--340.

\bibitem[Luo and Song, 2021]{Luo-Song-2021}
Luo, L. and Song, P. X.-K. (2021).
\newblock Multivariate online regression analysis with heterogeneous streaming
  data.
\newblock {\em Canadian Journal of Statistics}, 0(0):1--23.

\bibitem[Michael et~al., 2019]{Michael-Thornton-Xie-Tian}
Michael, H., Thornton, S., Xie, M., and Tian, L. (2019).
\newblock Exact inference on the random-effects model for meta-analyses with
  few studies.
\newblock {\em Biometrics}, 75(2):485--493.

\bibitem[Nielsen, 2020]{Nielsen}
Nielsen, F. (2020).
\newblock An elementary introduction to information geometry.
\newblock {\em Entropy}, 22(10):1100.

\bibitem[Obenchain, 1977]{Obenchain}
Obenchain, R.~L. (1977).
\newblock Classical {F}-tests and confidence regions for ridge regression.
\newblock {\em Technometrics}, 19(4):429--439.

\bibitem[Olkin and Pukelsheim, 1982]{Olkin-Pukelsheim}
Olkin, I. and Pukelsheim, F. (1982).
\newblock The distance between two random vectors with given dispersion
  matrices.
\newblock {\em Linear Algebra and its Applications}, 48:257--263.

\bibitem[Pan and Yang, 2010]{Pan-Yang}
Pan, S.~J. and Yang, Q. (2010).
\newblock A survey on transfer learning.
\newblock {\em IEEE Transactions on knowledge and data engineering},
  22(10):1345--1359.

\bibitem[Pollard, 1991]{Pollard}
Pollard, D. (1991).
\newblock Asymptotics for least absolute deviation regression estimators.
\newblock {\em Econometric Theory}, 7(2):186--199.

\bibitem[Pollard et~al., 2018]{eICU}
Pollard, T.~J., Johnson, A. E.~W., Raffa, J.~D., Celi, L.~A., Mark, R.~G., and
  Badawi, O. (2018).
\newblock The {eICU Collaborative Research Database}, a freely available
  multi-center database for critical care research.
\newblock {\em Scientific Data}, 5(1):1--13.

\bibitem[Raghu et~al., 2019]{Raghu-etal}
Raghu, M., Zhang, C., Kleinberg, J., and Bengio, S. (2019).
\newblock Transfusion: understanding transfer learning for medical imaging.
\newblock {\em Advances in NEural Information Processing Systems 33}.

\bibitem[Reeve et~al., 2021]{Reeve-Cannings-Samworth}
Reeve, H.~W., Cannings, T.~I., and Samworth, R.~J. (2021).
\newblock Adaptive transfer learning.
\newblock {\em The Annals of Statistics}, 49(6):3618--3649.

\bibitem[Schifano et~al., 2016]{Schifano-etal}
Schifano, E.~D., Wu, J., Wang, C., Yan, J., and Chen, M.-H. (2016).
\newblock Online updating of statistical inference in the big data setting.
\newblock {\em Technometrics}, 58(3):393--403.

\bibitem[Shen et~al., 2020]{shen2020fusion}
Shen, J., Liu, R.~Y., and Xie, M.-g. (2020).
\newblock i fusion: Individualized fusion learning.
\newblock {\em Journal of the American Statistical Association},
  115(531):1251--1267.

\bibitem[Shen et~al., 2018]{Shen-etal}
Shen, J., Qu, Y., Zhang, W., and Yu, Y. (2018).
\newblock Wasserstein distance guided representation learning for domain
  adaptation.
\newblock In {\em Thirty-Second AAAI Conference on Artificial Intelligence},
  pages 3--9.

\bibitem[Shen and Huang, 2010]{shen2010grouping}
Shen, X. and Huang, H.-C. (2010).
\newblock Grouping pursuit through a regularization solution surface.
\newblock {\em Journal of the American Statistical Association},
  105(490):727--739.

\bibitem[Stein, 1956]{Stein}
Stein, C.~M. (1956).
\newblock Inadmissibility of the usual estimator for the mean of a multivariate
  normal distribution.
\newblock {\em Proceedings of the Third Berkeley Symposium}, 1:197--206.

\bibitem[Stein, 1981]{Stein-1981}
Stein, C.~M. (1981).
\newblock Estimation of the mean of a multivariate normal distribution.
\newblock {\em The Annals of Statistics}, 9(6):1135--1151.

\bibitem[Tan et~al., 2018]{Tan-etal}
Tan, C., Sun, F., Kong, T., Zhang, W., Yang, C., and Liu, C. (2018).
\newblock A survey on deep transfer learning.
\newblock In K{\r{u}}rkov{\'a}, V., Manolopoulos, Y., Hammer, B., Iliadis, L.,
  and Maglogiannis, I., editors, {\em Artificial Neural Networks and Machine
  Learning -- ICANN 2018}, pages 270--279, Cham. Springer International
  Publishing.

\bibitem[Tang and Song, 2016]{Tang-Song}
Tang, L. and Song, P. X.-K. (2016).
\newblock Fused lasso approach in regression coefficients clustering --
  learning parameter heterogeneity in data integration.
\newblock {\em Journal of Machine Learning Research}, 17(113):1--23.

\bibitem[Tang and Song, 2020]{tang2020poststratification}
Tang, L. and Song, P. X.-K. (2020).
\newblock Poststratification fusion learning in longitudinal data analysis.
\newblock {\em Biometrics}.

\bibitem[Tang et~al., 2020]{Tang-Zhou-Song-2020}
Tang, L., Zhou, L., and Song, P. X.-K. (2020).
\newblock Distributed simultaneous inference in generalized linear models via
  confidence distribution.
\newblock {\em Journal of Multivariate Analysis}, 176:104567.

\bibitem[Theobald, 1974]{Theobald}
Theobald, C.~M. (1974).
\newblock Generalizations of mean square error applied to ridge regression.
\newblock {\em Journal of the Royal Statistical Society Series B},
  36(1):103--106.

\bibitem[Tian and Feng, 2022]{Tian-Feng}
Tian, Y. and Feng, Y. (2022).
\newblock Transfer learning under high-dimensional generalized linear models.
\newblock {\em Journal of the American Statistical Association}, doi:
  10.1080/01621459.2022.2071278:1--14.

\bibitem[Tibshirani and Taylor, 2011]{tibshirani2011solution}
Tibshirani, R.~J. and Taylor, J. (2011).
\newblock The solution path of the generalized lasso.
\newblock {\em The Annals of Statistics}, 39(3):1335--1371.

\bibitem[Torrey and Shavlik, 2009]{Torrey-Shavlik}
Torrey, L. and Shavlik, J. (2009).
\newblock {\em Handbook of research on machine learning applications and
  trends: algorithms, methods, and techniques}, chapter Transfer learning,
  pages 242--264.
\newblock IGI Global.

\bibitem[Toulis and Airoldi, 2017]{Toulis-Airold}
Toulis, P. and Airoldi, E.~M. (2017).
\newblock Asymptotic and finite-sample properties of estimators based on
  stochastic gradients.
\newblock {\em The Annals of Statistics}, 45(4):1694--1727.

\bibitem[van Wieringen, 2021]{vanWieringen}
van Wieringen, W.~N. (2021).
\newblock Lecture notes on ridge regression.
\newblock {\em arXiv}, arXiv:1509.09169v7.

\bibitem[Vinod, 1987]{Vinod}
Vinod, H.~D. (1987).
\newblock Confidence intervals for ridge regression parameters.
\newblock {\em Time Series and Econometric Modelling}, 36:279--300.

\bibitem[Wang et~al., 2018]{Wang-Kim-Yang}
Wang, Z., Kim, J.~K., and Yang, S. (2018).
\newblock Approximate bayesian inference under informative sampling.
\newblock {\em Biometrika}, 105(1):91--102.

\bibitem[Weiss et~al., 2016]{Weiss-etal}
Weiss, K., Khoshgoftaar, T.~M., and Wang, D. (2016).
\newblock A survey of transfer learning.
\newblock {\em Journal of Big Data}, 3(9):1--40.

\bibitem[Xie et~al., 2011]{Xie-Singh-Strawderman}
Xie, M., Singh, K., and Strawderman, W.~E. (2011).
\newblock Confidence distributions and a unifying framework for meta-analysis.
\newblock {\em Journal of the American Statistical Association},
  106(493):320--333.

\bibitem[Yang and Ding, 2019]{yang2019combining}
Yang, S. and Ding, P. (2019).
\newblock Combining multiple observational data sources to estimate causal
  effects.
\newblock {\em Journal of the American Statistical Association},
  115(531):1540--1554.

\bibitem[Yoon et~al., 2020]{Yoon-etal}
Yoon, T., Lee, J., and Lee, W. (2020).
\newblock Joint transfer of model knowledge and fairness over domains using
  wasserstein distance.
\newblock {\em IEEE Access}, 8:123783--123798.

\bibitem[Yosinski et~al., 2014]{Yosinski-etal}
Yosinski, J., Clune, J., Bengio, Y., and Lipson, H. (2014).
\newblock How transferable are features in deep neural networks?
\newblock {\em Advances in Neural Information Processing Systems 27}, pages
  3320--3328.

\bibitem[Zheng et~al., 2019]{Zheng-etal}
Zheng, C., Dasgupta, S., Xie, Y., Haris, A., and Chen, Y.~Q. (2019).
\newblock On data enriched logistic regression.
\newblock {\em arXiv}, arXiv:1911.06380.

\bibitem[Zhu et~al., 2021]{Zhu-Wang-Peng-Li}
Zhu, Z., Wang, L., Peng, G., and Li, S. (2021).
\newblock {WDA}: an improved {W}asserstein distance-based transfer learning
  fault diagnosis method.
\newblock {\em Sensors}, 21(13):4394.

\bibitem[Zhuang et~al., 2021]{Zhuang-etal}
Zhuang, F., Qi, Z., Duan, K., Xi, D., Zhu, Y. Z.~H., Xiong, H., and He, Q.
  (2021).
\newblock A comprehensive survey on transfer learning.
\newblock {\em Proceedings of the IEEE}, 109(1):43--76.

\end{thebibliography}

\end{document}